\documentclass[11pt]{article}

\usepackage[english]{babel}
\usepackage[intlimits]{amsmath}
\usepackage{setspace}
\usepackage[dviwin]{graphicx}
\usepackage{amsfonts}
\usepackage{amssymb}
\usepackage{mathrsfs}
\usepackage{amsmath}
\usepackage{epsfig}
\usepackage{graphicx}
\usepackage{mathabx}
\usepackage{xcolor}
\usepackage{dsfont}
\usepackage{mathrsfs}
\usepackage{esint}
\usepackage{textpos}
\usepackage{wrapfig}
\usepackage{hyperref}
\usepackage{mathtools}
\usepackage{mathtext}  
\usepackage{wasysym}

\usepackage{bm}
\DeclareBoldMathCommand\boldlangle{\left\langle}
\DeclareBoldMathCommand\boldrangle{\right\rangle}

\DeclareSymbolFont{lettersA}{U}{txmia}{m}{it}
\DeclareMathSymbol{\epsilonup}{\mathord}{lettersA}{15}

\makeatletter
\def\blx@maxline{77}
\makeatother

\newcommand{\bea}{\begin{equation}}
\newcommand{\eea}{\end{equation}}
\newcommand{\bear}{\begin{eqnarray}}
\newcommand{\eear}{\end{eqnarray}}
\newcommand{\bearr}{\begin{eqnarray*}}
\newcommand{\eearr}{\end{eqnarray*}}

\newcommand{\beal}{\begin{align}}
\newcommand{\eeal}{\end{align}}
\newcommand{\beall}{\begin{align*}}
\newcommand{\eeall}{\end{align*}}

\newcommand{\D}{\mathscr{D}}

\newcommand{\tr}{\mathrm{tr}\,}
\newcommand{\CP}{\mathds{C}\mathds{P}}
\newcommand{\CC}{\mathds{C}}

\newcommand{\dd}{\partial}

\newcommand{\comment}[1]{}

\newcommand{\fl}{\mathcal{F}}
\newcommand{\Zz}{\mathbf{Z}}
\newcommand{\Hh}{\mathrm{H}}

\newcommand{\Tr}{\mathrm{Tr}}

\usepackage{array}
\newcolumntype{L}[1]{>{\raggedright\let\newline\\\arraybackslash\hspace{0pt}}m{#1}}
\newcolumntype{C}[1]{>{\centering\let\newline\\\arraybackslash\hspace{0pt}}m{#1}}
\newcolumntype{R}[1]{>{\raggedleft\let\newline\\\arraybackslash\hspace{0pt}}m{#1}}

\begin{document}

\title{\vspace{-1.5cm}Flag manifold $\sigma$-models: \\The $1\over N$-expansion and the anomaly two-form}
\author{Dmitri Bykov\footnote{Emails:
bykov@mpp.mpg.de, bykov@mi-ras.ru}  \\ \\ 
{\small $\bullet$ Max-Planck-Institut f\"ur Physik, F\"ohringer Ring 6, D-80805 Munich, Germany}\\
{\small $\bullet$ Steklov
Mathematical Institute of Russ. Acad. Sci.,}\\ {\small Gubkina str. 8, 119991 Moscow, Russia \;}}
\date{}

\maketitle
\vspace{-1cm}
\begin{center}
\line(1,0){380}
\end{center}
\vspace{-0.2cm}
\textbf{Abstract.} We construct a gauged linear sigma-model representation and develop a $1\over N$-expansion for flag manifold $\sigma$-models previously proposed by the author. Classically there exists a zero-curvature representation for the equations of motion of these models, which leads in particular to the existence of a conserved non-local charge. We show that at the quantum level this charge is no longer conserved and calculate explicitly the anomaly in its conservation law. 

\vspace{-0.5cm}
\begin{center}
\line(1,0){380}
\end{center}

The subject of integrability in two-dimensional field theory has a long history. After the initial success with the Korteweg-de Vries equation (the main developments being summarized in~\cite{Pitaevskiy}), similar technology, based on the zero-curvature representation, was applied to relativistic $\sigma$-models with symmetric target spaces~\cite{Pohlmeyer, ZM}. In classical theory the mathematical groundwork was laid in~\cite{Uhlenbeck, Hitchin}. In quantum theory the initial advances in the sine-Gordon theory~\cite{Arefeva, Zam} were succeeded by the calculation of the S-matrix of the $S^{N-1}$ sigma-model~\cite{ZZ}. For an introduction to the subject of quantum-integrable sigma-models with symmetric target spaces see the lectures~\cite{ZaremboSigma}. 

 \vspace{0.3cm}
Around the same time the $1\over N$-expansion was formulated for the $S^{N-1}$-model in~\cite{ZZ} and for the $\CP^{N-1}$-model in~\cite{DAdda1}. The latter model was found to be non-integrable due to anomalies in the higher local conservation laws~\cite{Goldschmidt}, whose potential appearance had been discussed in~\cite{Polyakov77}. The Hamiltonian structure of the flat connections of the model was discussed in~\cite{Maillet1, Maillet2}, where in particular involutivity of the local charges was verified. Another approach to the question of integrability was developed in~\cite{Luscher} and was based on the non-local conserved charges, which follow from the zero-curvature representation. These charges, if unobstructed by anomalies, generate non-abelian algebraic structures such as the Yangian, that underlie the integrable structure of the theory, as was elaborated in~\cite{Bernard}. In theories like the $\CP^{N-1}$-model, however, the conservation equation of the first non-local charge is spoiled by an anomaly~\cite{AbdallaAnomaly}. This anomaly is also present in various sigma-models with symmetric target spaces~\cite{AbdallaGeneral} but is canceled in certain fermionic generalizations (see~\cite{AbdallaFermions} and references therein), for example in the supersymmetric $\CP^{N-1}$-model~\cite{Cremmer, Adda, AbdallaCancel}.

 \vspace{0.3cm}
What fell outside of the scope of these classical papers is the case of homogeneous but \emph{non-symmetric} target spaces. These models are in general believed to be not integrable even at the classical level. The technical reason is that the Noether current of such models is in general not flat, which makes it impossible to construct a zero-curvature representation by a standard procedure of~\cite{Pohlmeyer}. To start with, symmetric spaces, which may also be called $\mathbb{Z}_2$-symmetric spaces, allow for a direct generalization, the so-called $\mathbb{Z}_m$-symmetric spaces. These arise in various contexts in differential geometry, for example in the theory of nearly K\"ahler and twistor spaces, see Section~\ref{gradedsec} below for more details. In certain cases sigma-models with such target spaces might be integrable~\cite{Bena, Young}. The discussion of $\mathbb{Z}_m$-symmetric spaces in the context of integrable sigma-models can be found in the mathematical literature in the book~\cite{Guest}.

\vspace{0.3cm}
Other important advances in the direction of integrable non-symmetric space sigma-models are related to integrable deformations of the principal chiral model, in particular in the work~\cite{Cherednik} and in~\cite{Fateev}. The deformation of the latter paper is now known as the Yang-Baxter or  $\eta$-deformation. The algebraic structure behind such deformations was understood in~\cite{Klimcik}, and the Hamiltonian structure of the corresponding models was extensively analyzed in~\cite{Delduc, Lacroix}. Another approach (proposed by the author) is based on integrable complex structures on the target spaces~\cite{BykovZeroCurv, BykovFlag1, BykovFlag2, BykovGLSM, BykovCyclic}. Some of the most interesting complex homogeneous spaces are the flag manifolds. These manifolds are coincidentally also $\mathbb{Z}_m$-graded spaces, and the corresponding sigma-models may be understood, in certain cases, as limits of the $\eta$-deformed models~\cite{BykovCSind}. Generally speaking, flag manifolds are ubiquitous objects in representation theory, due to the Borel-Weil-Bott theorem (we dedicate Appendix~\ref{appBWB} to the discussion of this point), and also feature in certain supersymmetric constructions, such as harmonic superspace~\cite{Ivanov1} or supersymmetric quantum mechanics~\cite{Ivanov2}.

 \vspace{0.3cm}
Although this is not directly related to the main line of the present paper, it should be noted that in recent years there is a strong interest in systems with $SU(N)$-symmetry ($N>2$) from the point of view of condensed matter physics. This is related to the fact that such systems have now been experimentally realized in systems of cold atoms (see~\cite{Zoller}, for example). Some of the most well studied systems in statistical physics are the one-dimensional spin chains, which in appropriate limits can be described by flag manifold sigma-models, as shown in~\cite{BykovHaldane, BykovAF, Affleck}. The phase structure of such models, in particular the IR effective field theory for various values of parameters,  has been studied in~\cite{Sulejmanpasic, Seiberg}. It should be noted that these models in general are likely to be non-integrable and differ from the sigma-models studied in the present paper by a choice of $B$-field. One interesting common feature, though, is the role played by the $\mathbb{Z}_m$-symmetry. In flag manifold sigma-models $m$ is the number of steps in the flag. From the point of view of spin chains, $m$ is the length of the elementary cell, and the $\mathbb{Z}_m$-symmetry arises from shifting the elementary cell by one site. In the context of integrable models discussed in this paper, the group $\mathbb{Z}_m$ acts on the complex structure defining the $B$-field of the model, see 
the proposition in Section~\ref{GLSMsect}.

\vspace{0.3cm}
The paper is organized as follows. We start in Section~\ref{geomsec} by reminding facts about the geometry of flag manifolds, pertinent to the discussion of the sigma-models in question. In particular, we describe the general procedure of constructing invariant symplectic forms and complex structures in Sections~\ref{symplman} and~\ref{compstructsec0} respectively. Cohomology is discussed in Section~\ref{cohsec}. We briefly introduce the nonlinear action of the relevant sigma-models in Section~\ref{flagmanintrosec}, for more details the reader is referred to our earlier papers, for example~\cite{BykovZeroCurv, BykovFlag1}. Subsequently in section~\ref{gradedsec} we prove the relation between our models and the models with $\mathbb{Z}_m$-graded target spaces introduced in~\cite{Young}. It turns out that the latter models are essentially independent of the grading, which enters only through topological terms. In Section~\ref{GLSMsect} we construct the gauged linear sigma-model (GLSM) representation for our models (here the $\mathbb{Z}_m$-symmetry plays an important role). For the case of K\"ahler metrics on flag manifolds the GLSM-representations were constructed in~\cite{Donagi, Ginzburg}, however the peculiarity of our case is that the chosen metric is not K\"ahler. In Section~\ref{1Nsect} the Feynman rules for the $1\over N$-expansion of the simplest model with target space $U(N)\over U(1)\times U(1)\times U(N-2)$ are derived. We pass on to the discussion of the Wilson loop of the one-parametric family of flat connections in Section~\ref{Wloopsect} and propose a novel prescription for the regularized non-local charge in Subsection~\ref{regchargesect}. Then we study the limit when the regularization parameter $\epsilon$ vanishes. This limit depends on the operator product expansions of the Noether currents, which are calculated in Section~\ref{OPEsect1}. In particular, the OPE of two holomorphic components $[K_z (z+\epsilon), K_z(z)]$ is necessary to show that the limit exists, and the OPE of the holomorphic/anti-holomorphic components $[K_z (z+\epsilon), K_{\bar{z}}(z)]$ characterizes the anomaly two-form of the non-local charge. The latter OPE is what replaces the zero-curvature condition for the current $K$ in the quantum theory. In the Appendix we provide some details of the calculations (in which case the relevant appendices are referred to in the body of the paper), as well as discuss certain auxiliary, but nevertheless interesting aspects of the theory of flag manifolds.

\tableofcontents

\section{The geometry of flag manifolds}\label{geomsec}

\subsection{The flag manifold as a complex manifold}

\begin{figure}[h]
    \centering 
    \includegraphics[width=0.5\textwidth]{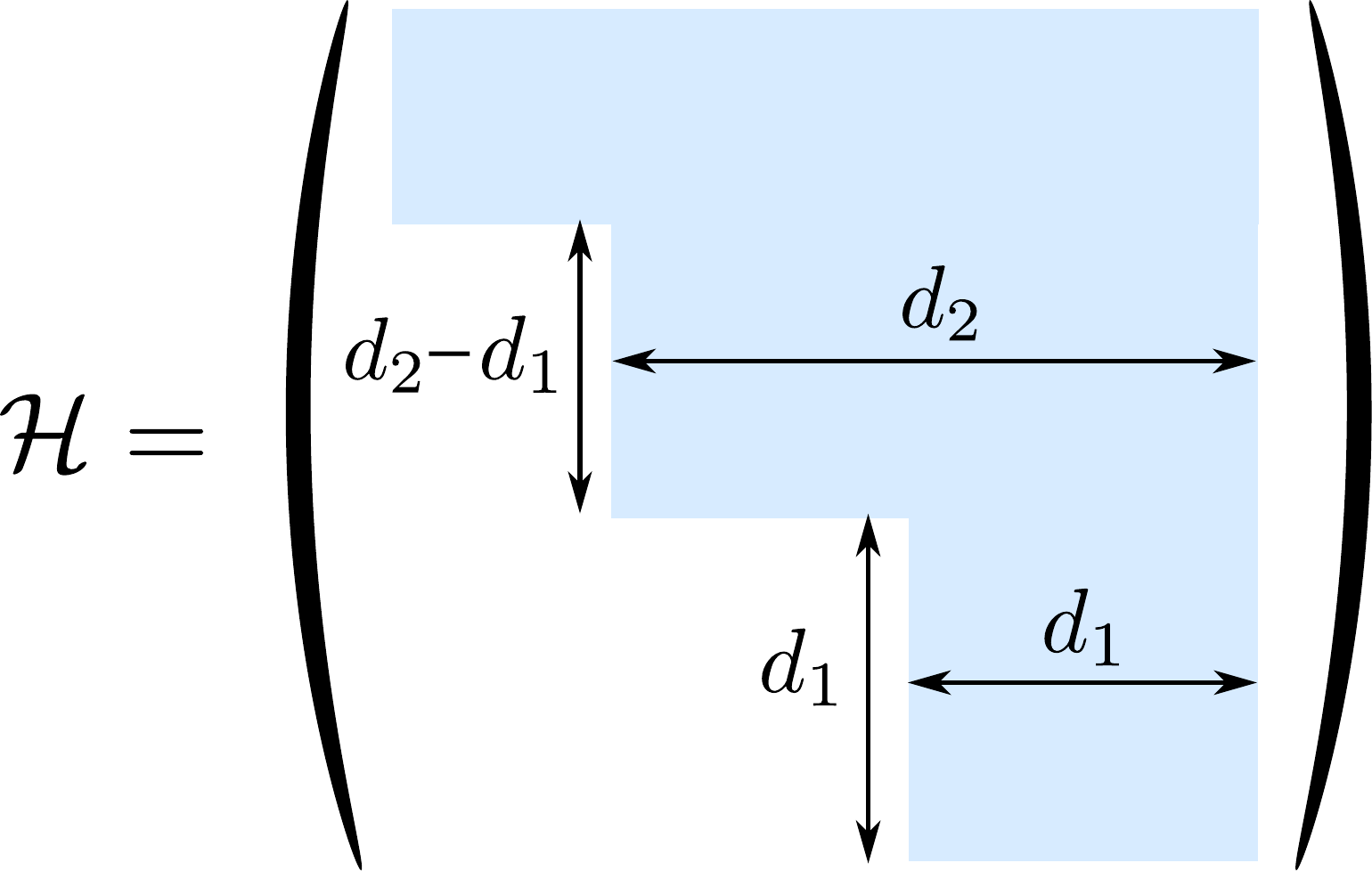}
  \caption{The maximal parabolic subgroup, stabilizing a flag.}
  \label{staircase}
\end{figure}

The flag manifold in $\CC^N$ may be defined as the manifold of linear complex subspaces, embedded to each other:
\bea\label{embeddedspaces}
\mathcal{F}(d_1, \ldots, d_m)=\{0\subset V_1 \subset \ldots \subset V_{m-1} \subset V_m=\CC^N\}\,,
\eea
where $\mathrm{dim}\,V_i=d_i$. The group $GL(N, \mathbb{C})$ acts transitively on this manifold, and the stabilizer of any given point (i.e. of a given sequence of embedded linear spaces) is a maximal parabolic subgroup (`a staircase'), consisting of matrices, depicted in Fig.~\ref{staircase}.

Therefore we can view the flag manifold as a homogeneous space
\bea\label{flagdefhol}
\mathcal{F}(d_1, \ldots, d_m)=GL(N, \mathbb{C})/\mathcal{H}\,.
\eea
It follows that the flag manifold is a complex manifold of complex dimension
\bea
\mathrm{dim}_\CC\,\mathcal{F}(d_1, \ldots, d_m)=N^2-\sum\limits_{i=1}^m\,d_i\,(d_i-d_{i-1})
\eea
Let us denote $\mathrm{dim}\,V_i/V_{i-1}=n_i$. Then one can also present the flag manifold as a quotient space of the unitary group:
\bea\label{flagunitary}
\mathcal{F}_{n_1, \ldots, n_m}=\frac{U(N)}{U(n_1)\times \ldots \times U(n_m)},\quad \sum\limits_{i=1}^m\,n_i=N\,.
\eea
This parametrization is naturally related to a choice of Euclidian metric in $\CC^N$, which allows performing orthogonal decompositions of the form ${V_i\simeq V_{i-1}\oplus V_i/V_{i-1}}$. Let us compute the real dimension and make sure it is twice greater than the complex dimension computed earlier:
\bea
\mathrm{dim}_\mathbb{R}\,\mathcal{F}_{n_1, \ldots, n_m}=N^2-\sum\limits_{i=1}^m\,n_i^2=2\times \mathrm{dim}_\CC\,\mathcal{F}_{d_1, \ldots, d_m}\,.
\eea
Note that sometimes we will denote by $\fl_N$ the complete flag manifold, i.e. the manifold~(\ref{flagunitary}), where all $n_i=1$.

\vspace{0.3cm}\noindent
\textbf{Example.} The projective space $\CP^{N-1}$ may be viewed as the quotient space  $(u_1, \ldots, u_N)\sim \lambda (u_1, \ldots, u_N)$ of $\CC^N/\{0\}$. By analogy the Grassmannian $G(k, N)$ is the quotient of the space of  $k\times N$-matrices of rank  $k$ w.r.t. the group $GL(k, \mathbb{C})$.

\vspace{0.3cm}
We note that the `holomorphic' definition (\ref{flagdefhol}) already implies a choice of complex structure on the flag manifold. The definition (\ref{flagunitary}), in contrast, treats the flag manifold as a real manifold (a complex structure may be defined on it separately -- we will do it below).

\vspace{-0.2cm}
\subsection{Symplectic structures}\label{symplman}

Let us show that $\mathcal{F}_{n_1, \ldots, n_m}$ is a symplectic manifold, i.e. there exists on it a closed 2-form $\omega$, $d\omega=0$, such that $\omega^{\mathrm{dim}_\CC\,\mathcal{F}}\neq 0$ (i.e. it is non-degenerate, $\mathrm{det} \,\omega\neq 0$). In fact, we will be only considering homogeneous (i.e. $U(N)$-invariant) symplectic forms.

\vspace{0.3cm}
More generally, let us consider the homogeneous space $G/H$ and the corresponding Lie algebra decomposition $\mathfrak{g}=\mathfrak{h}\oplus \mathfrak{m}$. A homogeneous space is called reductive if the following relations are fulfilled:
\bea
[\mathfrak{h}, \mathfrak{h}]\subset \mathfrak{h},\quad\quad [\mathfrak{h}, \mathfrak{m}]\subset \mathfrak{m}\,.
\eea
The second requirement is automatically satisfied, if there is an ad-invariant metric on $\mathfrak{g}$. Indeed, let there be elements $h_0, h_1\in \mathfrak{h}$ and $a, b \in \mathfrak{m}$, such that $[h_0, a]=b+h_1$. Let us compute the scalar product with an arbitrary element $\gamma\in \mathfrak{h}$, then we will get $\langle \gamma, [h_0, a]\rangle=\langle \gamma, h_1\rangle$. The l.h.s. is zero (which can be verified by swapping the commutator), therefore $\langle \gamma, h_1\rangle=0$ for all $\gamma$. It follows from the non-degeneracy of the metric that {$h_1=0$.}

\vspace{0.3cm}
There is the following general statement:

\noindent $G$-invariant tensors on a reductive homogeneous space $G/H$ are in correspondence with $H$-invariant tensors on  $\mathfrak{m}$ (the latter is simply a linear space $\mathbb{R}^{\mathrm{dim}\, \mathfrak{m}}$).

\vspace{0.3cm}
In what follows we will be mainly using the unitary representation (\ref{flagunitary}) and the corresponding Lie algebra decomposition
\bea
\mathfrak{g}:=\mathfrak{u}_N=\oplus_{i=1}^m\,\mathfrak{u}_{n_i}\oplus\mathfrak{m},
\eea
where $\mathfrak{m}$ is the orthogonal complement to $\mathfrak{h}$ w.r.t. an ad-invariant metric on $\mathfrak{u}_N=\mathfrak{u}_1\oplus \mathfrak{su}_N$. Since $[\mathfrak{h}, \mathfrak{m}]\subset \mathfrak{m}$, the subgroup $H$ is represented in the space $\mathfrak{m}$, and this representation may be decomposed into irreducibles:
\bea\label{mirreps}
\mathfrak{m}\otimes \CC= \oplus_{i\neq j} \,V_{ij},\quad\quad\textrm{where}\quad\quad V_{ij}=\CC^{n_i n_j}
\eea
Moreover, $V_{ji}=\bar{V}_{ij}$. $V_{ij}$ is the vector space of the bi-fundamental representation of the group $U(n_i)\times U(n_j)$.

\vspace{0.3cm}
We introduce the Maurer-Cartan current $
j=-g^{-1}\,dg,\, g\in U(N),
$
and decompose it as follows:
\bea\label{jdecomp}
j=[j]_\mathfrak{h}+\sum\limits_{i\neq k}\,j_{ik},\quad\quad j_{ik}\in V_{ik}\,.
\eea
According to the general statement mentioned above, the most general invariant two-form is
\bea
\Omega=\sum\limits_{i<j}\,a_{ij}\,\mathrm{Tr}(j_{ij}\wedge j_{ji})\,.
\eea
To check, in which case it is closed, we will take advantage of the flatness of the Maurer-Cartan current, $
dj-j\wedge j=0\,.
$
It follows that
\bea
\mathscr{D}j_{ij}=\sum\limits_{m\neq (i, j)}\,j_{im}\wedge j_{mj}\,
\eea
where $\mathscr{D}$ is the $H$-covariant derivative, defined as follows:
$
\mathscr{D}j_{ij}:=dj_{ij}-\{j_{\mathfrak{h}}, j_{ij}\}\,.
$
From the condition that $\Omega$ is closed it follows that
\bea\label{closed1}
a_{ij}+a_{jk}+a_{ki}=0\quad\quad \textrm{for all pairwise different}\quad\quad (i, j, k)\,.
\eea
The general solution to this equation is
\bea\label{closed2}
a_{ij}=z_i-z_j\,.
\eea
Therefore we have a family of homogeneous symplectic forms with $m-1$ real parameters. These forms may be compactly written as follows:
\bea
\Omega =\mathrm{Tr}(z\,j\wedge j)\,,\quad\quad \textrm{where}\quad\quad z=\mathrm{Diag}(z_1\,\mathds{1}_{n_1}, \ldots, z_m \mathds{1}_{n_m})\,.
\eea
The element $z$  may be normalized to be traceless: $\mathrm{Tr}(z)=0$. The stabilizer $H$ may now be thought of as the stabilizer of the matrix $z\in \mathfrak{u}_N$, and the flag manifold itself -- as an adjoint orbit:
\bea
\mathcal{F}_{d_1, \ldots, d_m}=\{g\,z\,g^{-1},\quad g\in SU(N)\}\,.
\eea

The above formula gives an embedding of the flag manifold into the Lie algebra $\mathfrak{su}_N$. Moreover, this embedding may be identified with the image of the moment map
\bea\label{mommap1}
\mu=g\,z\,g^{-1}\,.
\eea
The action of the group $U(N)$ on the flag manifold, $x\to \tilde{x}$, is as follows:
\bea
g_0 \cdot g(x)= g(\tilde{x})\cdot h_0,\quad g_0\in U(N),\quad h_0 \in U(n_1)\times\cdots \times U(n_m)\, .
\eea
The moment map (\ref{mommap1}) transforms correctly under the action of the group: $\mu(g_0\cdot g)=g_0 \,\mu(g)\,g_0^{-1}$. To check that (\ref{mommap1}) indeed generates Hamiltonian functions for the action of the group $G$ on the flag manifold, let us assume that $v_a$ is a vector field on $\mathcal{F}$, corresponding to the Lie algebra element $T_a$. Then the following holds:
\bea
j(v_a)\equiv i_{v_a}\, j =-g^{-1}\,\nabla_{v_a}g=-g^{-1}\frac{d}{dt}\left(e^{T_a\, t}\,g\right)\big|_{t=0}=- g^{-1} T_a g
\eea
Using this equality, we can check that (\ref{mommap1}) indeed gives Hamiltonian functions for the action of the group:
\bea
d(\mathrm{tr} (\mu T_a))=d(\mathrm{tr} (g\,z\,g^{-1} \,T_a))=\mathrm{tr}(z\,[j,g^{-1}T_a g])=i_{v_a}\,\mathrm{tr}(z j\wedge j)=i_{v_a}\,\Omega
\eea

\subsection{Complex structures}\label{compstructsec0}

A very detailed treatment of complex structures on homogeneous spaces was given as early as in the classic work~\cite{Borel}, so here we mostly present an adaptation of some of these statements to our needs. To start with, on the manifold $U(N)\over U(1)^N$ of complete flags in $\CC^N$ there are $2^{\frac{N(N-1)}{2}}$ invariant almost complex structures, $N!\leq 2^{\frac{N(N-1)}{2}}$ of them being integrable (we note that for large $N$, according to Stirling's formula,  $e^{N\,\log(N)}<e^{\log(2)\,N^2\over 2}$). 

\vspace{0.3cm}
Since the complex structure $\mathscr{J}$ is a certain invariant tensor on $G/H$, according to the  logic mentioned in the previous section, one needs to define an  $H$-invariant action of the operator $\mathscr{J}$ on the linear space $\mathfrak{m}$. Since $\mathscr{J}^2=-1$, the complex structure may be diagonalized over the complex numbers, in which case we need to define a decomposition
\bear\label{mdecomp}
\mathfrak{m}\otimes \CC=\mathfrak{m}_+\oplus \mathfrak{m}_-,\\
\textrm{where}\quad\quad [\mathfrak{h}, \mathfrak{m}_\pm]\subset \mathfrak{m}_\pm\,.
\eear
Here $\mathfrak{m}_\pm$ play the role of holomorphic tangent spaces to $G/H$, i.e. $\mathscr{J}\circ a=\pm i\,a$ for $a\in \mathfrak{m}_\pm$. In (\ref{mirreps}) we decomposed $\mathfrak{m}\otimes \CC$ into irreducible components, therefore, in order to define an almost complex structure on $\mathcal{F}$, one can define the action of  $\mathscr{J}$ as follows:
\bea
\mathscr{J}\circ V_{pq}=\pm\,i\,V_{pq}\quad\quad \textrm{for}\quad\quad 1\leq p<q\leq N\,.
\eea
As a result, one has exactly $2^{\frac{N(N-1)}{2}}$ possibilities. There are several equivalent definitions of integrability of a complex structure:
\begin{itemize}
\item Vanishing of the Nijenhuis tensor: \bea
[\mathscr{J}\circ X, \mathscr{J}\circ Y]-\mathscr{J}\circ([\mathscr{J}\circ X, Y]+[X, \mathscr{J}\circ Y])-[X, Y]=0
\eea
for arbitrary vector fields $X, Y$.
\item Using vector fields: the commutator of two holomorphic vector fields should be holomorphic, i.e.
\bea \label{integrdist1}
(1-i\,\mathscr{J})\,[(1+i\,\mathscr{J})X, (1+i\,\mathscr{J})Y]=0\,.
\eea
(The property (\ref{integrdist1}) may also be stated as the condition that the distribution of holomorphic vector fields is integrable.)
\item Using forms: the holomorphic forms should form a differential ideal, i.e. the following condition should be satisfied: $d(J_-)_a\sim \sum\limits_b \,R_{ab}\wedge (J_-)_b$ for some one-forms $R_{ab}$. 
\end{itemize}
Let us demonstrate, that the last definition implies
\bea\label{algcond0}
[\mathfrak{m}_+, \mathfrak{m}_+]\subset \mathfrak{m}_+,\quad\quad [\mathfrak{m}_-, \mathfrak{m}_-]\subset \mathfrak{m}_-\,,
\eea
if the restriction to $\mathfrak{m}$ of the adjoint-invariant metric $\mathbb{G}$ on $\mathfrak{u}_N$  is Hermitian w.r.t. the chosen almost complex structure $\mathscr{J}$. In general the integrability of an almost complex structure means that $[\mathfrak{m}_+, \mathfrak{m}_+]\subset \mathfrak{m}_+\oplus\, \mathfrak{h}$. To see this, note that an almost complex structure $\mathscr{J}$ is defined by the conditions $\mathscr{J}\circ J_\pm=\pm i \, J_\pm$, where $J_\pm$ are the components of a Maurer-Cartan current:
\bea\label{Jcurrdecomp}
J=-g^{-1}dg=J_\mathfrak{h}+J_++J_-,\quad\quad J_\pm\in \mathfrak{m}_\pm\,.
\eea
Since $dJ-J\wedge J=0$, we get $$dJ_-=\big[-J_0\wedge J_0+(\textrm{terms with}\,J_-) - J_+\wedge J_+\big]_{\mathfrak{m}_-}\,.$$ Therefore for the integrability of $\mathscr{J}$ one should have $[J_+\wedge J_+]_{\mathfrak{m_-}}=0$, i.e. $[\mathfrak{m}_+, \mathfrak{m}_+]\subset \mathfrak{m}_+\oplus \mathfrak{h}$. We see that the conditions  (\ref{algcond0}) therefore define an integrable complex structure. Conversely suppose we have an integrable complex structure on $G/H$, and $\mathfrak{m}_\pm$ are its respective holomorphic/anti-holomorphic subspaces. Then $[a, b]=c+\gamma$, where $a, b, c\in \mathfrak{m}_+$ and $\gamma \in \mathfrak{h}$. Since $\boldlangle \mathfrak{m}_+, \mathfrak{h}\boldrangle_{\mathbb{G}}=0$, computing the scalar product with a generic element $\gamma'\in \mathfrak{h}$, we obtain $\boldlangle \gamma', [a,b] \boldrangle_{\mathbb{G}}=\boldlangle \gamma', \gamma \boldrangle_{\mathbb{G}}$. Using the identity $\boldlangle [a, \gamma'], b \boldrangle_{\mathbb{G}}+\boldlangle \gamma', [a, b] \boldrangle_{\mathbb{G}}=0$, we get $\boldlangle \gamma', \gamma \boldrangle_{\mathbb{G}}=-\boldlangle [a, \gamma'], b \boldrangle_{\mathbb{G}}=\boldlangle a', b \boldrangle_{\mathbb{G}}$, and $a'=[\gamma', a]\in \mathfrak{m}_+$. As discussed earlier, the subspace $\mathfrak{m}_+$ is isotropic, if the metric $\mathbb{G}$ is Hermitian, therefore $\boldlangle \gamma', \gamma \boldrangle_{\mathbb{G}}=0$ for all $\gamma'\in \mathfrak{h}$, which implies $\gamma=0$ due to the non-degeneracy of $\mathbb{G}$. The result $[\mathfrak{m}_+, \mathfrak{m}_+]\subset \mathfrak{m}_+$ follows.
  
\begin{figure}[h]
    \centering
    \includegraphics[width=0.4\textwidth]{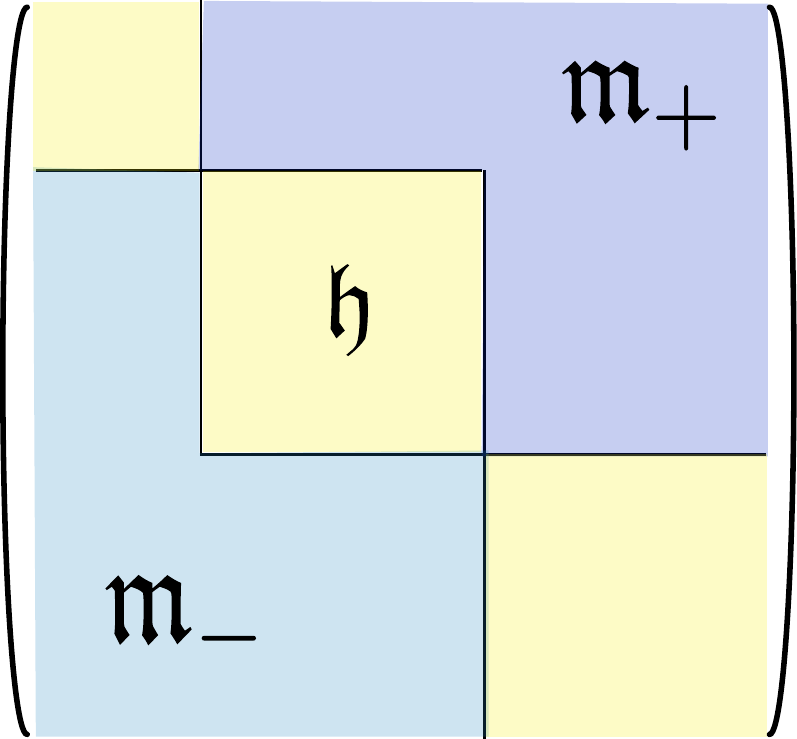}
    \caption{The decomposition (\ref{mdecomp}) of the Lie algebra.}
    \label{figcompstr}
\end{figure}

 \vspace{0.3cm} A typical integrable complex structure on the flag manifold defines the holomorphic/anti-holomorphic subspaces $\mathfrak{m}_\pm$ shown in Fig.~\ref{figcompstr}.

 \vspace{0.3cm}
As discussed above, one can define an \emph{almost complex structure} on $\mathcal{F}_N$ by choosing $\frac{N(N-1)}{2}$ mutually non-conjugate forms $J_{p_1 q_1}, \ldots, J_{m_{\frac{N(N-1)}{2}} n_{\frac{N(N-1)}{2}}}$ and declaring them holomorphic. The remaining $\frac{N(N-1)}{2}$ forms will be therefore anti-holomorphic. To determine, which of those complex structures are integrable, it is useful to use a diagrammatic representation. We draw $N$ vertices, as well as arrows from the node $p_1$ to the node $q_1$, from $p_2$ to $q_2$ and so on (such diagrams are called~`tournaments', see~\cite{SalamonTourn}). The integrability of the almost complex structure, defined in this way, is equivalent to the acyclicity of the graph (the condition that it should not contain closed cycles). For a proof see~\cite{BykovFlag2}, for example.

\vspace{0.3cm}\noindent
One can now establish the following fact:

\vspace{0.3cm}\noindent
\noindent
\textbf{Lemma.} There are exactly $N!$ acyclic diagrams.

\begin{figure}[h]
    \centering 
    \includegraphics[width=0.8\textwidth]{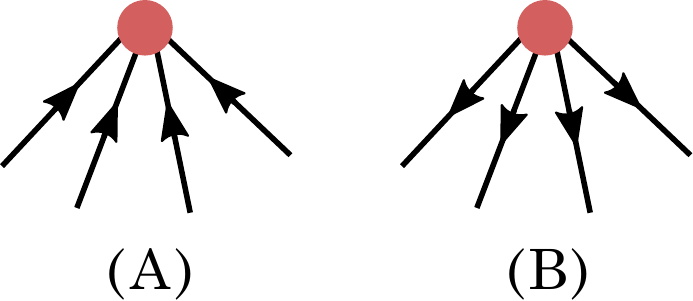}
  \caption{(A) The `sink' vertex,\quad (B) The `source' vertex\,.}
  \label{sourcesinkvertex}
\end{figure}

\vspace{0.3cm}\noindent
\textbf{Proof.} The statement of the lemma implies that there is only one combinatorial type of diagrams, and all acyclic diagrams may be obtained from any one of them by the action of the permutation group  $S_N$. Let us describe this combinatorial type. Every acyclic diagram has a `source'-vertex, in which all the lines are outgoing, and a `sink'-vertex, in which all lines are incoming  (see Fig.~\ref{sourcesinkvertex}). Indeed, if that were not so, every vertex would contain at least, say, one outgoing line. Then one can start at any vertex and follow outgoing lines, until a loop is formed. Let us consider the `source'-vertex. The diagram formed by the remaining $N-1$ vertices together with the edges joining them can be an arbitrary acyclic diagram (as the chosen vertex is a `source', there cannot be cycles containing it). Therefore we have performed the first step of the induction. The subsequent steps consist in finding the `source' vertex in the reduced diagram. It is therefore clear that there always exists a vertex with $i$ outgoing lines for all $i=0, \ldots, N-1$. This statement completely describes the combinatorial structure of the diagram. Equivalently, there is a total ordering on the set of vertices. Different diagrams differ just by a relabeling of the vertices.

We have established that there are $N!$ complex structures on a complete flag manifold $U(N)\over U(1)^N$. Analogously there are $m!$ complex structures on a partial flag manifold $U(N)\over U(n_1)\times \cdots \times U(n_m)$. The number of complex structures may be easily interpreted as follows. Choosing a complex structure is equivalent to choosing a complex quotient space representation~(\ref{flagdefhol}). In order to construct such representation one should choose a sequence of embedded linear spaces of the type~(\ref{embeddedspaces}), and the dimensions of these spaces are given by the partial sums of the integers $n_i$. These dimensions are therefore determined by an ordering of the set $\{n_i\}$, and there are $m!$ such orderings.

\subsection{Cohomology}\label{cohsec}

The second cohomology group of the complete flag manifold is
\bea
\mathrm{H}^2(\fl_N, \Zz)=\Zz^{N-1},
\eea
hence there exist  $N-1$ linearly independent 2-forms, which are the generators of $\Hh^2(\fl_N)$. As a model for $\Hh^2(\fl)$  let us consider the following construction. On $\fl_N$ there are $N$ standard line bundles $L_1, \cdots , L_N$, moreover their sum is a trivial bundle:
\bea\label{triv}
\overset{N}{\underset{i=1}{\oplus}}\;L_i = \fl_N\, \times \CC^N\; .
\eea
The first Chern classes of these bundles give $N$ closed 2-forms: $\Omega_i = c_1(L_i), \; i=1\cdots N$. Due to the condition (\ref{triv}) and the additivity of the first Chern classes $c_1(E\oplus F)=c_1(E)+c_1(F)$ it is clear that the forms $\Omega_i$ are not independent but rather satisfy the relation
\bea\label{triv2}
\sum\limits_{i=1}^N\, \Omega_i = 0
\eea
The two-forms $\Omega_i,\;i=1 \,\cdots\, N$, satisfying the relation (\ref{triv2}), generate $\Hh^2(\fl_N, \Zz)$.

\vspace{0.5cm}
One can also obtain an even more explicit description of the cohomology, which will shed some light on the relation (\ref{triv2}). To this end we note that there exists an embedding
\bea\label{embed1}
i:\;\;\fl_N \,\hookrightarrow \,\underbracket[0.6pt][0.6ex]{\CP^{N-1}\,\times\,\cdots\,\times \, \CP^{N-1}}_{N\;\textrm{factors}} .
\eea
A point $m\in (\CP^{N-1})^{\times N}$ is a collection of $N$ lines in $\CC^N$ through the origin. Those points that correspond to $N$ \emph{pairwise orthogonal} lines are the points lying on the flag manifold $\fl_N$: here one should recall that $\fl_N$ may be thought of as the manifold of sets of $N$ ordered mutually orthogonal lines in $\CC^N$. Let us now consider $(\CP^{N-1})^{\times N}$ as a symplectic manifold with a product symplectic form
\bea\label{sfprod}
\omega=\sum\limits_{i=1}^N\;\omega_i
\eea
One can check (this is proven in \cite{BykovAF}) that (\ref{embed1}) is a Lagrangian embedding w.r.t. this symplectic form, i.e.
\bea\label{lagr1}
\omega|_{\fl_N}=0 .
\eea
Identifying $\Omega_i=[i^\ast(\omega_i)]$ and taking into account (\ref{lagr1}), we obtain the relation (\ref{triv2}). One can analogously prove that a general flag manifold, embedded into a product of Grassmannians $\mathcal{F}_{n_1,\ldots, n_m}\hookrightarrow G_{n_1, N}\times \cdots \times G_{n_m, N}$, is also a Lagrangian submanifold.

\vspace{0.3cm}
The above observations are tightly related to the theory of geometric quantization, i.e. to the construction of the representation theory of a Lie group $G$ using line bundles over its flag manifolds. See our paper~\cite{BykovAF} for a review.

\section{The models}\label{flagmanintrosec}

In this paper we will continue our study of $\sigma$-models proposed in \cite{BykovZeroCurv}. The target spaces of these models are homogeneous \emph{complex} target spaces $\mathcal{M}={G\over H}$, endowed with the `Killing metric' $\mathbb{G}$ and an integrable $G$-invariant complex structure $\mathscr{J}$. The action schematically takes the form:
\bea\label{action}
\mathcal{S}[\mathbb{G}, \mathscr{J}]:=\int_\Sigma\,d^2 z\,\|\dd X\|^2_\mathbb{G}+\int_\Sigma\,X^\ast \omega,
\eea
where $X$ is the map $X: \Sigma \to \mathcal{M}$ and $\omega$ is the K\"ahler form corresponding to the pair $(\mathbb{G}, \mathscr{J})$, defined as $\omega=\mathbb{G}\circ \mathscr{J}$. The peculiarity of such models lies in the fact that one can construct zero-curvature representations for their equations of motion. More precisely, a one-parametric family $\mathcal{A}_u, u\in\CC,$ of flat connections can be constructed from the Noether current $K$ using the formula of Pohlmeyer~\cite{Pohlmeyer}:
\bea\label{connection}
\mathscr{A}_u={1-u \over 2}\,K_z dz+{1-u^{-1}\over 2} \,K_{\bar{z}}d\bar{z}\;.
\eea
In the present paper we define the complex coordinates $z, \bar{z}$ as $\{z, \bar{z}\}={x^1\pm i x^2\over \sqrt{2}}$, so that the line element is $ds^2=2\,dz\,d\bar{z}$, and the metric is $g=\left( \begin{array}{cc}
0 & 1  \\
1 & 0  \end{array} \right)$.

\vspace{0.3cm}
The connections defined in (\ref{connection}) are flat, since the Noether current $K$ itself is flat for the models above. This construction is well-known for symmetric spaces~\cite{Forger}, which e.g. for the group $G=SU(N)$ are the Grassmannians $G_{k, N}$. Our models allow for a richer class of target spaces, in particular including the so-called flag manifolds.

\subsubsection{Dependence on the metric.}
Let us clarify that here and in what follows when referring to the Noether current, we mean a one-form $K$. In fact, the canonical Noether procedure produces a vector field $K^\mu$, satisfying the conservation equation $\dd_\mu (\sqrt{g} K^\mu)=0$. The one-form is obtained from this vector field by lowering the index in the standard way: $K=g_{\mu\nu} K^\nu\,dx^\mu$. The Hodge dual to this one-form is $\ast K={1\over \sqrt{g}}g_{\alpha\beta}\epsilon^{\beta\gamma} K_\gamma\,dx^\alpha$ and
\bear
d\ast K=\dd_\lambda({1\over \sqrt{g}}g_{\alpha\beta}\epsilon^{\beta\gamma} K_\gamma)\,dx^\lambda\wedge dx^\alpha=
\dd_\lambda({1\over \sqrt{g}}\epsilon^{\lambda\alpha}g_{\alpha\beta}\epsilon^{\beta\gamma} K_\gamma)\,dx^1\wedge dx^2=\\
=-{1\over \sqrt{g}}\dd_\lambda(\sqrt{g}\,g^{\lambda\gamma} K_\gamma)\,\mathrm{vol}_{\Sigma_2}\,.
\eear
Let us consider a simple example of a free complex massless boson, the action being $S=\int\,dx^1 dx^2\,\sqrt{g}\,g^{\alpha\beta}\,\dd_\alpha\bar{\psi}\,\dd_\beta\psi$. The corresponding one-form is then $K=i(\bar{\psi}d\psi-d{\bar{\psi}} \psi)$ and, in particular, it is independent of the worldsheet metric $g$. Suppose we now include a certain invariant $B$-field, or even a topological term, replacing the action by $\tilde{S}=S+\kappa\,\int\,i\,d\bar{\psi}\wedge d\psi$. The Noether current should now be replaced by $\tilde{K}=K+\kappa \,\ast d(\bar{\psi} \psi)$, the additional component being a topological current. Nevertheless, the full current $\tilde{K}$ now depends on the \emph{conformal class} of the metric $g$, through the dependence of the Hodge star $\ast$. We come to the conclusion that, in general, both one-forms $K$ and $\ast K$ depend on the conformal class $[g]$ of the worldsheet metric. \hfill$\blacksquare$

\vspace{0.3cm}
In the present paper we will concentrate on the simplest non-symmetric (i.e. non-Grassmannian) example, when 
\bea\label{flagquot}
\mathcal{F}:=\frac{SU(N)}{S(U(n)\times U(m)  \times U(N-m-n))}\,.
\eea
This manifold may be thought of as the space of mutually orthogonal (w.r.t. some metric on $\CC^N$) linear subspaces $E_n, E_m, E_{N-m-n}$ of $\CC^N$, passing through the origin, of dimensions $n, m$ and $N-m-n$. Clearly, these subspaces span all of $\CC^N$.

\section{Relation to models with $\mathbb{Z}_m$-graded target spaces}\label{gradedsec}

Although the manifold~(\ref{flagquot}) is not a symmetric space, i.e. not a space with a $\mathbb{Z}_2$-grading, it is a space with a $\mathbb{Z}_3$-grading. Before explaining the concept, we note that natural examples of $\mathbb{Z}_3$-graded spaces are provided by twistor spaces of symmetric spaces~\cite{Salamon} and nearly K\"ahler homogeneous spaces~\cite{Butruille}. In this context the manifold~(\ref{flagquot}) may be seen as the twistor space of $G_{n,N}$, $G_{m,N}$ or $G_{N-m-n,N}$.

\vspace{0.3cm}
A homogeneous space $G\over H$ is called $\mathbb{Z}_m$-graded (or $m$-symmetric), if the Lie algebra $\mathfrak{g}$ of its isometry group admits the following decomposition:
\bea\label{Zm}
\mathfrak{g}=\oplus_{i=0}^{m-1}\,\mathfrak{g}_i,\quad\quad [\mathfrak{g}_i, \mathfrak{g}_j]\subset \mathfrak{g}_{i+j\;\mathrm{mod}\;m},\quad\quad \mathfrak{g}_0=\mathfrak{h}\,.
\eea
In this language the ordinary symmetric spaces are $2$-symmetric spaces. Then, one can show, that, similarly to what happens for symmetric spaces, the e.o.m. of certain $\sigma$-models with $\mathbb{Z}_m$-graded target-spaces may be rewritten as flatness conditions of a one-parametric family of connections.

\vspace{0.3cm}
A related class of models was considered in \cite{Young} and subsequently studied in~\cite{Beisert}. The action studied in \cite{Young} has the form
\bea\label{actionY}
\widetilde{\mathcal{S}}:=\int_\Sigma\,d^2 x\,\|\dd X\|^2_G+\int_\Sigma\,X^\ast \widetilde{\omega},
\eea
where $\widetilde{\omega}$ is a 2-form constructed using the $\mathbb{Z}_m$-decomposition of the Lie algebra (\ref{Zm}). Note that, if \;$\widetilde{\omega}$\; were the K\"ahler form, one would obtain precisely the action  (\ref{action}). Now we come to the precise definition of $\widetilde{\omega}$. Decompose the current $J=-g^{-1}dg$ according to (\ref{Zm}):
\bea\label{curr}
J=-g^{-1}dg=\sum\limits_{i=0}^{m-1}\;J^{(i)},\quad\quad\textrm{where}\quad J^{(i)}\in \mathfrak{g}_i\;.
\eea
The form $\widetilde{\omega}$ is defined as follows:
\bea\label{omegaY}
\widetilde{\omega}={1\over 2}\sum\limits_{k=1}^{m-1}{(m-k)-k\over m}\,\tr(J^{(k)}\wedge J^{(m-k)})
\eea
This formula poses an important question. According to (\ref{omegaY}), the form $\widetilde{\omega}$ depends on the $\mathbb{Z}_m$-grading on the Lie algebra. On the other hand, generally a given Lie algebra $\mathfrak{g}$ may have many different gradings (with different, or same, values of $m$). The question is: are the models defined by (\ref{actionY})-(\ref{omegaY}), corresponding to different gradings of $\mathfrak{g}$, different?

\vspace{0.3cm}
Before answering this question, first we review the construction of cyclic gradings on semi-simple Lie algebras, which was completed long ago \cite{Kac}. Let us consider, for simplicity, the case of $\mathfrak{g}=\mathfrak{su}(N+1)$. A cyclic grading may be constructed as follows\footnote{Here we restrict ourselves to the grading of type $A_N^{(1)}$.}: one picks a system of $N$ simple positive roots $\alpha_1, \ldots \alpha_N$, as well as the maximal negative root $\alpha_{N+1}=-\alpha_1-\ldots-\alpha_N$\footnote{In the paper of Kac \cite{Kac} the roots $\alpha_1, \ldots \alpha_{N+1}$ are seen as the positive simple roots of the corresponding affine Lie algebra $\widehat{A}_N$. Consider the case $N=2$. The simple positive roots of the loop algebra $\mathfrak{su}_3(t, t^{-1})$ may be chosen as follows: $$\alpha_1=\left( \begin{array}{ccc}
0 & 1 & 0 \\
0 & 0 & 0 \\
0 & 0 & 0  
  \end{array} \right), \;\;\;\alpha_2=\left( \begin{array}{ccc}
0 & 0 & 0 \\
0 & 0 & 1 \\
0 & 0 & 0  
  \end{array} \right), \;\;\;\alpha_0=t\,\left( \begin{array}{ccc}
0 & 0 & 0 \\
0 & 0 & 0 \\
1 & 0 & 0  
  \end{array} \right)\,.$$ In this context the latter root $\alpha_0$ -- the analog of $\alpha_{N+1}$ -- is customarily called `imaginary'. In fact, the whole theory of cyclic Lie algebra gradings is formulated by Kac naturally in terms of affine Lie algebras and their Dynkin diagrams.}. Then one assigns to these $N+1$ roots arbitrary (non-negative integer) gradings $m_1, \ldots m_N, m_{N+1}$. The gradings of all other roots are determined by the Lie algebra structure, and the value of $m$ is calculated as
\bea\label{msum}
m=m_1+\ldots+m_{N+1}\,.
\eea
In usual matrix form, this grading looks as follows:
\bea\label{gradingM}\left( \begin{array}{C{0.8cm}C{0.8cm}C{0.8cm}C{0.8cm}C{0.8cm}}
0 & $\mathbf{m_1}$ & & &\\
 & 0 & $\mathbf{m_2}$ & &\\
 & & 0 & $\mathbf{\ddots}$ & \\
  & &   & 0 & $\mathbf{m_{N}}$  \\
$\mathbf{m_{\tiny{N+1}}}$ & & & & 0 \end{array} \right)\eea
The subalgebra $\mathfrak{g}_0$, which determines the denominator $H$ of the quotient space $G/H$, is determined by those $m_i$'s, which are zero. For example, if all $m_i>0$, the resulting space is the manifold of complete flags $SU(N+1)\over S(U(1)^{N+1})$.

\vspace{0.3cm}
In general, for a choice of grading determined by the set $m_1, \ldots, m_{N+1}$ some of the subspaces $\mathfrak{g}_i$ will be identically zero. Therefore a natural restriction to adopt is to require that $\mathfrak{g}_i\neq 0$ for all $i \;(\mathrm{mod}\; m)$. We will call such a grading \emph{admissible}. This still leaves a wide range of possibilities. For example, in the case of $SU(3)$ the following is a complete list of admissible gradings (up to the action of the Weyl \mbox{group $S_3$):}
\bear
&&\mathbb{Z}_2:\;\;
\left( \begin{array}{ccc}
0 & \mathbf{0} & 1 \\
0 & 0 & \mathbf{1} \\
\mathbf{1} & 1 & 0  
  \end{array} \right), \quad \mathbb{Z}_3:\;\;
\left( \begin{array}{ccc}
0 & \mathbf{1} & 2 \\
2 & 0 & \mathbf{1} \\
\mathbf{1} & 2 & 0  
  \end{array} \right), \quad \left( \begin{array}{ccc}
0 & \mathbf{0} & 1 \\
0 & 0 & \mathbf{1} \\
\mathbf{2} & 2 & 0  
  \end{array} \right),\\ && 
    \mathbb{Z}_4:\;\; \left( \begin{array}{ccc}
0 & \mathbf{1} & 2 \\
3 & 0 & \mathbf{1} \\
\mathbf{2} & 3 & 0  
  \end{array} \right), \quad 
 \mathbb{Z}_5:\;\;\left( \begin{array}{ccc}
0 & \mathbf{1} & 3 \\
4 & 0 & \mathbf{2} \\
\mathbf{2} & 3 & 0  
  \end{array} \right),\quad \left( \begin{array}{ccc}
0 & \mathbf{1} & 2 \\
4 & 0 & \mathbf{1} \\
\mathbf{3} & 4 & 0  
  \end{array} \right),\\
&& \mathbb{Z}_6:\;\; \left( \begin{array}{ccc}
0 & \mathbf{1} & 3 \\
5 & 0 & \mathbf{2} \\
\mathbf{3} & 4 & 0  
  \end{array} \right),\quad   \mathbb{Z}_7:\;\; \left( \begin{array}{ccc}
0 & \mathbf{1} & 3 \\
6 & 0 & \mathbf{2} \\
\mathbf{4} & 5 & 0  
  \end{array} \right)
  \eear
The $\mathbb{Z}_2$-grading and the second $\mathbb{Z}_3$-grading correspond to the homogeneous space $SU(3)/S(U(2)\times U(1))=\CP^2$, and all other gradings correspond to the flag manifold $\mathcal{F}_{1,1,1}$.

\vspace{0.3cm}
We will now give an answer to the question posed above: what is the relation between the $\sigma$-models with the action (\ref{actionY}), taken for different gradings on the corresponding Lie algebra? Our statement is~\cite{BykovCyclic}:

\vspace{0.3cm}\noindent
\textbf{Proposition.}
For homogeneous spaces of the unitary group, the models defined by (\ref{actionY})-(\ref{omegaY}) with different $A_N^{(1)}$-type gradings on $\mathfrak{g}$ are classically equivalent
 to the model defined by the action (\ref{action}) with some choice of complex structure on the target-space

\vspace{0.3cm}
In fact, one has a precise statement about the relation of the $B$-fields in the two models. To formulate it, we `solve' the constraint (\ref{msum}) as follows\footnote{Formula (\ref{inneraut}) implies that the cyclic automorphism $\widehat{\Omega}$ of the Lie algebra, which defines the $\mathbb{Z}_m$ grading, can be represented as follows: $\widehat{\Omega}(a)=\Omega a \Omega^{-1}$, where $\Omega=\mathrm{diag}(e^{2\pi i \,\frac{n_1}{m}}, \ldots, e^{2\pi i \,\frac{n_{N+1}}{m}})$.}:
\bea\label{inneraut}
m_k=n_k-n_{k+1}
\eea
where $n_k$ are integers and $n_{N+2}\equiv n_1-m$. We then have (see~\cite{BykovCyclic} for a proof):
\bea
\tilde{\omega}=\omega-2\, \sum\limits_{i=1}^{N} \,\frac{n_i}{m}\,dJ_{ii}
\eea
We see that, irrespective of the choice of the grading (which is now encoded in the integers $n_i$), the form $\tilde{\omega}$ differs from the K\"ahler form by a topological term. This topological term, clearly, depends on the chosen grading, but does not contribute to the equations of motion.

\vspace{0.3cm}\noindent
\textbf{Comment.} Although the flag manifolds (\ref{flagunitary}) are $\mathbb{Z}_m$-graded spaces, the two classes of target spaces -- $\mathbb{Z}_m$-graded and complex homogeneous spaces -- do not coincide. For example, one has the space $\frac{G_2}{SU(3)}\simeq S^6$. The stability subgroup $SU(3)$ acts on the tangent space $\mathfrak{m}=\mathbb{R}^6$ via $V\oplus \bar{V}$, where $V\simeq \CC^3$ is the standard representation. Therefore it has a unique almost complex structure, which is \emph{not} integrable (see the review~\cite{AgricolaS6}). On the other hand, it is a nearly K\"ahler manifold and is $\mathbb{Z}_3$-graded~\cite{Butruille}. On the other side of the story, one has the complex manifold $S^1\times S^3\simeq U(2)$ (see~\cite{BykovCyclic} for a discussion), which may be viewed as a $\mathbb{T}^2$-bundle over $\CP^1$ (the simplest flag manifold). This manifold is \emph{not} a $\mathbb{Z}_m$-graded homogeneous space of the group ${G=U(2)}$.

\vspace{0.3cm}
We also note that the construction of Lax connections for models with $\mathbb{Z}_m$-graded spaces was explored in~\cite{Beisert}. The fact that the corresponding integrals of motion are in involution was proven, for instance, in~\cite{Lacroix}.

\section{The GLSM representation}\label{GLSMsect}

In this section we will construct a gauged linear sigma-model (GLSM) representation for the models~(\ref{action}) introduced earlier (for flag manifold target spaces~$\mathcal{F}$). We recall that these models depend explicitly on the complex structure $\mathscr{J}$. Now, picking a complex structure $\mathscr{J}$ on $\mathcal{F}$ is equivalent to picking a total ordering of the subspaces $E_n, E_m, E_{N-m-n}$. For the purposes of constructing a $1\over N$-expansion any of the two orderings, where $E_{N-m-n}$ is maximal, will suffice (the reason will become clear shortly). Let us prove, first of all, that it is always possible to achieve this, without loss of generality. The point here is that the action~(\ref{action}), albeit depending on the complex structure, might produce the same equations of motion even for different choices of complex structure. This is due to the fact, that for certain complex structures, which we denote by $\mathscr{J}_1$ and $\mathscr{J}_2$, the difference in the two actions may just be a topological term:
\bea\label{toptermdiff}
\mathcal{S}[\mathscr{J}_1]-\mathcal{S}[\mathscr{J}_2]=\int\limits_\Sigma\,\mathscr{O}_{12},\quad\quad d\mathscr{O}_{12}=0\,.
\eea
Let us describe precisely the situation when this happens. To this end we recall that, as was established at the end of section~\ref{compstructsec0}, the complex structures are in a one-to-one correspondence with an ordering of the mutually orthogonal spaces $\CC^{n_1}, \ldots \CC^{n_m}$, that are a point in a flag manifold $U(N)\over U(n_1)\times \cdots U(n_m)$. 

\vspace{0.3cm}\noindent
\textbf{Proposition.} The actions $\mathcal{S}[\mathscr{J}_1]$ and $\mathcal{S}[\mathscr{J}_2]$ differ by a topological term, as in~(\ref{toptermdiff}), if and only if the corresponding sequences of spaces $\{\CC^{n_1}, \ldots \CC^{n_m}\}$ differ by a cyclic permutation.

\vspace{0.3cm}\noindent
\textbf{Proof.} Let us call $\mathscr{J}$ the standard complex structure, whose holomorphic subspace $\mathfrak{m}_+$ is given by upper-block-triangular matrices. Then $\mathscr{J}_1=\sigma_1(\mathscr{J})$ and $\mathscr{J}_2=\sigma_2(\mathscr{J})$ for some permutations $\sigma_{1}, \sigma_2\in S_m$. We recall the notation $J_{ij}$ from (\ref{jdecomp}). The corresponding K\"ahler forms are
\bear
\omega_1=i\,\sum\limits_{i<j}\,\Tr(J_{\sigma_1(i)\sigma_1(j)}\wedge J_{\sigma_1(j)\sigma_1(i)})\\
\omega_2=i\,\sum\limits_{i<j}\,\Tr(J_{\sigma_2(i)\sigma_2(j)}\wedge J_{\sigma_2(j)\sigma_2(i)})
\eear
Upon introducing the notation $\sigma(J_{ij}):=J_{\sigma(i)\sigma(j)}$, we may write the difference of the two forms as
\bea\label{omegadiff1}
\omega_1-\omega_2=i\,\sigma_1\left(\sum\limits_{i<j}\,\Tr(J_{ij}\wedge J_{ji})-\sum\limits_{i<j}\,\Tr(J_{\tau^{-1}(i)\tau^{-1}(j)}\wedge J_{\tau^{-1}(j)\tau^{-1}(i)})\right),\; 
\eea
where $\tau^{-1}=\sigma_1^{-1}\sigma_2\,.$ This reduces the problem to that of $\mathscr{J}_1=\mathscr{J}$ and $\mathscr{J}_2=\tau^{-1}(\mathscr{J})$. Note that the exterior derivative commutes with the permutation $\sigma$, due to the following simple fact following from the Maurer-Cartan equation:
$\sigma(dJ_{ij})=\sigma(\sum\limits_k\,J_{ik}\wedge J_{kj})=\sum\limits_k\,\sigma(J_{ik})\wedge \sigma(J_{kj})=d\sigma(J_{ij})$, where to arrive at the last equality one has to make a change of the dummy summation index $k\to \sigma(k)$.

\vspace{0.3cm}
We wish to show that $d(\omega_1-\omega_2)=0$ implies that $\tau$ is a cyclic permutation.
To this end we rewrite the above difference as follows:
\bear
&& \omega_1-\omega_2=i\,\sigma_1\left( \sum\limits_{i, j}\,\alpha_{ij}\,\Tr(J_{ij}\wedge J_{ji})\right),\\ && \textrm{where}\quad \alpha_{ij}={1\over 2}\big(\mathrm{sgn}(j-i)-\mathrm{sgn}(\tau(j)-\tau(i))\big)\in\{-1, 0, 1\}\,.
\eear
(We have made a change of dummy variables $i\to\tau(i)$ and $j\to\tau(j)$ in the second sum in (\ref{omegadiff1})).
Closedness of this form requires that (see~(\ref{closed1})-(\ref{closed2}))
\bea
\alpha_{ij}={1\over 2}\big(\mathrm{sgn}(j-i)-\mathrm{sgn}(\tau(j)-\tau(i))\big)=z_i-z_j\,.
\eea
Let us consider the case $j>i$. Then $z_i=z_j$ if $\tau(j)>\tau(i)$ and $z_i=z_j+1$ if $\tau(j)<\tau(i)$. This means that $\{z_i\}_{i=1\ldots m}$ form a non-increasing sequence, and moreover the difference between any two elements is either zero or $1$. This is only possible if the set has the form $(\underbracket[0.6pt][0.6ex]{z, \ldots, z}_{K}, \underbracket[0.6pt][0.6ex]{z-1, \ldots z-1}_{m-K})$. Accordingly the original sequence of $m$ consecutive numbers can be split into two consecutive sets:
\bea
1\ldots m=(I_1, I_2)\,.
\eea
Since $\tau(j)<\tau(i)$ for ($i\leq K$, $j>K$), the permutation acts as follows:
\bea
\tau(I_1, I_2)=(\tau(I_2), \tau(I_1))\,.
\eea
Moreover, since $\tau(i)<\tau(j)$ for $i<j\leq K$ and the image $\tau(I_1)$ is $(m-K+1,\ldots m)$, a moment's thought shows that $\tau(i)=m-K+i$ for $i=1\ldots K$. Analogously $\tau(j)=j-K$ for $j=K+1\ldots m$. Therefore $\tau$ is nothing but a $K$-fold cyclic permutation `to the left' (or $m-K$-fold to the right).

\vspace{0.3cm}
Since for $i<j$ the non-zero $\alpha_{ij}$ are the ones, for which $\tau(j)<\tau(i)$, this implies $j=K+1\ldots m$ and $i=1\ldots K$. These $\alpha_{ij}$ are equal to $1$, therefore
\bea
\omega_1-\omega_2=i\,\sigma_1 \left(\mathop{\sum_{ i=1\,\ldots\, K, }}_{ j= K+1\,\ldots\, n } \,\Tr(J_{ij}\wedge J_{ji})\right)\,,
\eea
which is easily seen to be proportional to the (generalized) Fubini-Study form on the Grassmannian $G_{L, N}$, where $L=\sum\limits_{i=1}^K \,n_i$. Conversely, one shows that for a cyclic permutation the difference between $\omega_1$ and $\omega_2$ is the closed form written above. $\blacksquare$

\vspace{0.3cm}
Returning to the manifold~(\ref{flagquot}), we may assume that we have picked an ordering of $E_n, E_m, E_{N-m-n}$, in which $E_{N-m-n}$ is maximal (by making a cyclic permutation if necessary). The manifold $\mathcal{F}$ with the complex structure defined by this ordering may then be thought of as the space of embedded linear complex spaces:
\bea
\mathcal{F}=\left\{ \;0\subset V_1 \subset V_2 \subset V_3=\CC^N\; \right\}\,,
\eea
where $V_1 \simeq \CC^n$ and $V_2\simeq \CC^{m+n}$. As shown in our paper \cite{BykovGLSM}, the model (\ref{action}) can be formulated in this case as a gauged linear $\sigma$-model, in terms of a `matter field' $\varphi \in \mathrm{Hom}(V_2, \CC^N)$ and an auxiliary field $\mathcal{A}=\mathcal{A}_z dz+\mathcal{A}_{\bar{z}} d\bar{z}\in \mathfrak{u}_{m+n}\otimes T^\ast \Sigma$:
\bea\label{GLSMlagr}
\mathcal{L}=\Tr\left((\mathscr{D}_z \varphi)^\dagger (\mathscr{D}_z \varphi)\right)+i\,\Tr\left(\lambda\,\left(\varphi^\dagger \varphi-{N\over g^2}\mathds{1}_{m+n}\right)\right)
\eea
The field $\varphi$ is subject to a normalization constraint $\varphi^\dagger \varphi={N\over g^2} \mathds{1}_{m+n}$, which is imposed by the Lagrange multiplier $\lambda$. Here $g$ is the coupling constant of the model and the dependence on $N$ is chosen to suit the $1\over N$-expansion, constructed in the next section. To simplify the situation even further, we will restrict to the case $m=n=1$, i.e. we will consider the target space of the form
\bea\label{simpletarget}
\mathcal{F}=\frac{U(N)}{U(1)\times U(1)\times U(N-2)}\,.
\eea
In this case we can express the matrix $\varphi$ in terms of its two column vectors $\varphi=(u, v)$, then the normalization constraint says that $\|u\|=\|v\|={\sqrt{N}\over g}$ and $\bar{u}\circ v=0$. The covariant derivates in the above Lagrangian are defined in the standard way:
\bea
\mathscr{D}_z \varphi=\dd_z \varphi-i\,\varphi\,\mathcal{A}_z,\quad\quad (\mathscr{D}_z \varphi)^\dagger=\dd_{\bar{z}} \varphi^\dagger+i\,\mathcal{A}_{\bar{z}}\,\varphi^\dagger\,.
\eea
Note that we assume the connection $\mathcal{A}$ to be Hermitian, meaning that $\mathcal{A}^\dagger=\mathcal{A}$, hence $(\mathcal{A}_z)^\dagger=\mathcal{A}_{\bar{z}}$. An important point is that the `connection' $\mathcal{A}$ in the model above has to be taken in a non-conventional triangular form:
\bea\label{Atriang}
\mathcal{A}_z=\left( \begin{array}{cc}
a_z & 0  \\
c_z & b_z  \end{array} \right),\quad\quad \mathcal{A}_{\bar{z}}=(\mathcal{A}_z)^\dagger=\left( \begin{array}{cc}
a_{\bar{z}} & c_{\bar{z}}  \\
0 & b_{\bar{z}}  \end{array} \right)
\eea
The form of $\mathcal{A}$ is the main difference of our models from the models with symmetric target spaces. In particular, this triangular form implies that there is only the $U(1)^2$ gauge symmetry, for which $a$ and $b$ are the gauge fields, whereas $c$ is another auxiliary field transforming in the bi-fundamental representation of $U(1)^2$. Had we not imposed the constraints (\ref{Atriang}), we would have the gauge group $U(2)$, and the target space of the model would be the Grassmannian $G_{2, N}$.

\vspace{0.3cm}
Integrating out the auxiliary fields $a, b, c$ we get the following form of the Lagrangian (for details see the Appendix\,\ref{app}):
\bea\label{LagrAfterElim}
\mathcal{L}=\dd_{\bar{z}}\bar{u}\circ \left(\mathds{1}_N-{g^2\over N}\left(u\otimes\bar{u}+v\otimes \bar{v}\right)\right)\circ \dd_z u+\dd_{\bar{z}}\bar{v}\circ \left(\mathds{1}_N-{g^2\over N} v\otimes \bar{v}\right)\circ \dd_z v\,.
\eea
This expression may be simplified, if we introduce an additional matrix-valued field $w: \CC^{N-2}\to \CC^N$, defined by the completeness relation in $\CC^N$:
\bea
u\otimes u^\dagger+v\otimes v^\dagger+\Tr_{\CC^{N-2}}(w\otimes w^\dagger)={N\over g^2}\mathds{1}_N\,.
\eea
This relation simply means that the matrix $w$ contains $N-2$ vectors, which are orthogonal to $u$ and $v$ and mutually orthonormal. We then obtain
\bea
{N\over g^2} \mathcal{L}=\Tr\left((\dd_{\bar{z}}\bar{u}\circ w)(w^\dagger \circ \dd_z u)\right)+\Tr\left((\dd_{\bar{z}}\bar{v}\circ w) (w^\dagger\circ \dd_z v)\right)+(\dd_{\bar{z}}\bar{v}\circ u) (\bar{u}\circ \dd_z v).
\eea
This is precisely the type of Lagrangian considered in \cite{BykovZeroCurv}, \cite{BykovFlag1}, \cite{BykovFlag2}. In this way we have established the equivalence of the model (\ref{action}) with its gauged linear representation (\ref{GLSMlagr}).

\vspace{0.3cm}
Before proceeding further, let us make one more observation regarding the Lagrangian (\ref{GLSMlagr}). A direct calculation shows that the difference
\bear\label{lagrdifftop}
&&\mathcal{L}-\widetilde{\mathcal{L}}=\Tr\left((\mathscr{D}_z \varphi)^\dagger (\mathscr{D}_z \varphi)\right)-\Tr\left((\mathscr{D}_{\bar{z}} \varphi)^\dagger (\mathscr{D}_{\bar{z}} \varphi)\right)=\\&&=\Tr\left(\dd_{\bar{z}}\varphi^\dagger \dd_z \varphi-\dd_z \varphi^\dagger \dd_{\bar{z}}\varphi\right)\sim \textrm{the pull-back of}\;i\,\Tr\left(d\varphi\wedge d\varphi^\dagger\right)
\eear
The last line is clearly a total derivative, being the pull-back to the worldsheet of a Fubini-Study form on the Grassmannian $G_{2, N}$ (With the normalization $\varphi^\dagger \varphi={N\over g^2} \mathds{1}_2$ the Fubini-Study form is ${N\over g^2} \Omega_{FS}\sim i\,\Tr\left(d\varphi\wedge d\varphi^\dagger\right)$). Taking advantage of this fact, we will replace the Lagrangian (\ref{GLSMlagr}) by a more symmetric one
\bea\label{Lagr10}
\mathscr{L}=\Tr(\D_\mu \varphi^\dagger \,\D_\mu \varphi )+i\,\Tr\left(\lambda\,\left(\varphi^\dagger \varphi-{N\over g^2} \mathds{1}_2\right)\right)
\eea
Above it was claimed that the Noether current $K$ derived from the action (\ref{action}) is flat, i.e. $dK-K\wedge K=0$. Let us check this explicitly. To start with, the Noether current has the form
\bea\label{Noether1}
K=\frac{2 g^2}{N}\,\left(\varphi\,(\mathscr{D}_z \varphi)^\dagger\,d\bar{z}-\mathscr{D}_z \varphi\,\varphi^\dagger\,dz\right)\,.
\eea
Note that $K\in \mathfrak{u}_N$ and therefore $K^\dagger = -K$. Now,
\bear\label{flatexplicit}
&&dK-K\wedge K=
\frac{2 g^2}{N}\,\left(2\,\D_z\varphi\, (\D_z\varphi)^\dagger+\varphi \,\D_z \D_{\bar{z}}\varphi^\dagger +\D_{\bar{z}} \D_z\varphi \,\varphi^\dagger-\right.\\ \nonumber &&\left.-\frac{2 g^2}{N}\,\D_z\varphi\,(\varphi^\dagger \varphi)\,\D_{\bar{z}}\varphi^\dagger+\frac{2 g^2}{N}\,\varphi \,(\D_{\bar{z}}\varphi^\dagger \D_z\varphi)\,\varphi^\dagger \right)\,dz\wedge d\bar{z}
\eear
Let us now simplify this expression, using the normalization condition $\varphi^\dagger \varphi={N\over g^2}\mathds{1}_2$, as well as the equations of motion following from the Lagrangian (\ref{Lagr10}):
\bear\label{eqs1}
&&\D_{\bar{z}}\D_z \varphi={1\over 2}\varphi\,(i\lambda+i F_{z\bar{z}}),\quad\quad \D_z\D_{\bar{z}}\varphi^\dagger={1\over 2}(i\lambda+i F_{z\bar{z}})\,\varphi^\dagger
\\ \nonumber
&&\varphi^\dagger\D_z\varphi=- \D_z\varphi^\dagger \varphi=\left( \begin{array}{cc}
0 & \bullet  \\
0 & 0  \end{array} \right),\quad\quad  \D_{\bar{z}}\varphi^\dagger \varphi=-\varphi^\dagger \D_{\bar{z}}\varphi=\left( \begin{array}{cc}
0 & 0  \\
 \bullet & 0  \end{array} \right)
\eear
(We have used the identities $[D_z, D_{\bar{z}}]\varphi=-i\varphi F_{z\bar{z}}$, $[D_z, D_{\bar{z}}]\varphi^\dagger=i F_{z\bar{z}}\varphi^\dagger $).\\
Note an important difference with the standard case of symmetric target spaces (Grassmannians), where one has $\varphi^\dagger\D_z\varphi=\D_{\bar{z}}\varphi^\dagger \varphi=0$.

\vspace{0.3cm}
It follows from the two equations in (\ref{eqs1}) that $i\lambda$ is Hermitian: $(i\lambda)^\dagger=i\lambda$. On the other hand, multiplying the first equation by $\varphi^\dagger$ from the left, we get
\bea
{1\over 2}i\lambda={g^2\over N}\varphi^\dagger \D_{\bar{z}}\D_z \varphi-{1\over 2}i F_{z\bar{z}}={g^2\over N} \D_{\bar{z}}(\varphi^\dagger \D_z \varphi)-{g^2\over N} \D_{\bar{z}}\varphi^\dagger \,\D_z \varphi-{1\over 2}i F_{z\bar{z}}
\eea
The last two terms in the r.h.s. are manifestly Hermitian, whereas the first term turns out to be upper-triangular:
\bea
\D_{\bar{z}}(\varphi^\dagger \D_z \varphi)=\left( \begin{array}{cc}
0 & \widetilde{\bullet}  \\
0 & 0  \end{array} \right),\quad\quad \widetilde{\bullet}=(\dd_{\bar{z}}+i\,(a_{\bar{z}}-b_{\bar{z}}))\,\bar{u}\circ \dd_z v\,.
\eea
Hermiticity of $i\lambda$ then implies that $\D_{\bar{z}}(\varphi^\dagger \D_z \varphi)=0$ and therefore
\bea\label{lambda}
{1\over 2}i\lambda=-{g^2\over N}\D_{\bar{z}}\varphi^\dagger \,\D_z \varphi-{1\over 2}i F_{z\bar{z}}\,.
\eea
Using all this information in (\ref{flatexplicit}), we find that $dK-K\wedge K=0$, as claimed earlier.

\subsection{General flag manifolds}

Although the calculations in the following sections will concern the case of the target space~(\ref{simpletarget}), for completeness let us introduce the GLSM-formulation in the general case of the flag manifolds $\frac{U(N)}{U(n_1)\times \cdots \times U(n_m)}$, $\sum\limits_{i=1}^m\,n_i=N$. We assume that an arbitrary complex structure $\mathscr{J}$ has been chosen, which is tantamount to choosing an ordering of the factors
\bea\label{ordering1}
U(n_1)\times \cdots \times U(n_m)\,,
\eea
as was explained in Section~\ref{compstructsec0}. We can always reshuffle $n_1, \ldots, n_m$ in such a way that the ordering is the literal ordering of factors in (\ref{ordering1}). In this case we are dealing with the manifold of complex flags
\bea
0\subset V_1\subset \ldots \subset V_m=\CC^N,\quad\quad \mathrm{dim}_{\CC}\,V_k=\sum\limits_{i=1}^k\,n_i\,.
\eea
We define $M:=\mathrm{dim}\,V_{m-1}$ and introduce the matter field $\varphi\in \mathrm{Hom}(\CC^M, \CC^N)$. Its columns parametrize $M$ vectors that define the flag. They are orthonormal:
\bea
\varphi^\dagger \varphi={N\over g^2}\mathds{1}_M\,.
\eea
This is an analog of the moment map constraint from the theory of K\"ahler quotients. Now we introduce an analogue of the gauge field
\bear
&&\!\!\!\!\!\!\!\!\!\!\!\!\!\!\!\!\!\!\!\!\mathcal{A}=\mathcal{A}_z dz+\mathcal{A}_{\bar{z}} d\bar{z},\\ \nonumber \\ \label{gaugef}
&&\!\!\!\!\!\!\!\!\!\!\!\!\!\!\!\!\!\!\!\!\mathcal{A}_z :=\left(\begin{array}{ccccc}
(A_{11})_z &  0 &  0& \cdots & 0 \\
(A_{21})_z & (A_{22})_z &0& \cdots & 0 \\
\vdots & \ddots & \ddots & \ddots & 0 \\
(A_{m-1\, 1})_z & (A_{m-1\, 2})_z &\cdots & \cdots & (A_{m-1\, m-1})_z  
\end{array}\right),\quad \mathcal{A}_{\bar{z}}=(\mathcal{A}_z)^\dagger\,
\eear
and the covariant derivative
\bea
\mathscr{D}_\mu \varphi:=\dd_\mu \varphi-i\,\varphi\,\mathcal{A}_\mu\,.
\eea
In (\ref{gaugef}) the diagonal blocks $(A_{jj})_z$ are of size $n_j \times n_j$ and represent gauge fields for the respective gauge groups $U(n_j)$. The off-diagonal blocks $(A_{ij})_z, i\neq j$ are additional `matter fields' transforming in bi-fundamental representations of the corresponding gauge groups $U(n_i)\times U(n_j)$. Note that the gauge group $U(n_m)$ does not appear in the above formulas -- this is analogous to the situation familiar from the $\CP^{N-1}$-model, where the standard GLSM-representation involves only the $U(1)$ gauge field, despite the fact that the quotient space form $\CP^{N-1}\simeq\frac{U(N)}{U(1)\times U(N-1)}$ involves a product of gauge groups $U(1)\times U(N-1)$.

\vspace{0.3cm}
The Lagrangian of the model is a direct generalization of~(\ref{Lagr10}):
\bea\label{genGLSMlagr}
\mathscr{L}=\mathrm{Tr}((\mathscr{D}_\mu \varphi)^\dagger \,\mathscr{D}_\mu \varphi)+\mathrm{Tr}\left(\lambda\,\left(\varphi^\dagger \varphi-{N\over g^2}\mathds{1}_M\right)\right)\,.
\eea
As a result, we have obtained a theory that is very similar to the Grassmannian $G_{M, N}$ sigma-model, but with a `reduced' gauge field.

\section{Feynman rules for the $1\over N$-expansion}\label{1Nsect}

We now return to the case of the target space (\ref{simpletarget}). In the previous section we introduced the Lagrangian (\ref{Lagr10}) of our model. The Lagrangian in the form~(\ref{GLSMlagr}) is not the most convenient one for carrying out the $1\over N$-expansion, as it would lead to mixed $\langle\mathcal{A}\,\lambda \rangle$ propagators. For this reason we will be using the Lagrangian in the form~(\ref{Lagr10}):
\bea\label{Lagr1}
\mathscr{L}=\Tr(\D_\mu \varphi^\dagger \,\D_\mu \varphi )+i\,\Tr(\lambda\,(\varphi^\dagger \varphi-{N\over g^2}\,\mathds{1}_2))
\eea
One should keep in mind the normalization of the metric that leads to $D_\mu\otimes D_\mu=D_{\bar{z}}\otimes D_z+D_z\otimes D_{\bar{z}}$. Note that we have also normalized the fields differently now:
\bea\label{newnorm}
\varphi^\dagger \varphi={1\over \hat{g}^2}\,\mathds{1}_2={N\over g^2}\,\mathds{1}_2\,.
\eea
This is the proper normalization for carrying out the $1\over N$-expansion.
We see that in the large-$N$ limit the coupling constant $\hat{g}\sim {1\over \sqrt{N}}$ vanishes, just like in the large-$N$ limit of gauge theories. The renormalized coupling constant $g$ stays fixed in the large-$N$ limit.

\vspace{0.3cm}
Quantization of the theory (\ref{Lagr1}) as it stands will lead to one-loop divergences. To get rid of them, we will introduce additional fermionic ghost fields $\chi$ -- the Pauli-Villars regulators:
\bear\label{Lagr2}
\widehat{\mathscr{L}}=&&\Tr\left((\mathscr{D}_\mu \varphi)^\dagger (\mathscr{D}_\mu\varphi)\right)+\Tr\left((\mathscr{D}_\mu \chi)^\dagger (\mathscr{D}_\mu \chi)\right)+M^2\,\Tr(\chi^\dagger \chi)+\\ \nonumber &&+i\Tr\left(\Lambda(\varphi^\dagger \varphi+\chi^\dagger\chi-{N\over g^2}\,\mathds{1}_2)\right)\,.
\eear
Integrating over the `matter fields' $\varphi, \chi$ in the path integral, we obtain the effective action:
\bea\label{effact}
\mathcal{S}^{\mathrm{eff}}=-N\,\Tr \,\mathrm{Log}(-\mathscr{D}_{\mu} \mathscr{D}_\mu +i \Lambda)+N\,\Tr \,\mathrm{Log}(-\mathscr{D}_{\mu} \mathscr{D}_\mu +M^2+i \Lambda)-i\,{N\over g^2}\,\int\,d^2z\,\Tr \,\Lambda\,.
\eea
We will now find the saddle point w.r.t. $\Lambda$, assuming that at the saddle point $\Lambda_0=-i\,m^2\,\mathds{1}_2$ and $\langle \mathcal{A}_z\rangle=\langle \mathcal{A}_{\bar{z}}\rangle=0$:
\bea
\int\,d^2p\,\left(\frac{1}{p^2+m^2}-\frac{1}{p^2+M^2+m^2}\right)={1\over g^2}\,.
\eea
This has the solution
\bea
m^2=\frac{M^2\,e^{-{1\over g^2}}}{1-e^{-{1\over g^2}}}\,.
\eea
At the end of the day we wish to get rid of the ghost fields $\chi$, so we will take the limit $M\to\infty$, in such a way that the mass $m$ of the field $\varphi$ stays finite, therefore in this limit we would need to take $g\to0_+$ accordingly.

\begin{figure}[h]
    \centering 
    \includegraphics[width=\textwidth]{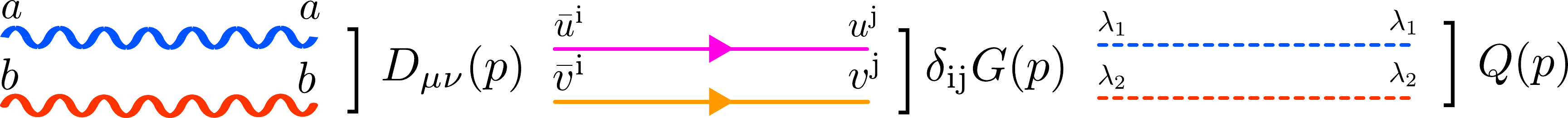}
  \caption{The propagators of the (doubled) $\CP^{N-1}$ model.}
  \label{figprop}
\end{figure}
\begin{figure}[h]
    \centering 
    \includegraphics[width=0.65\textwidth]{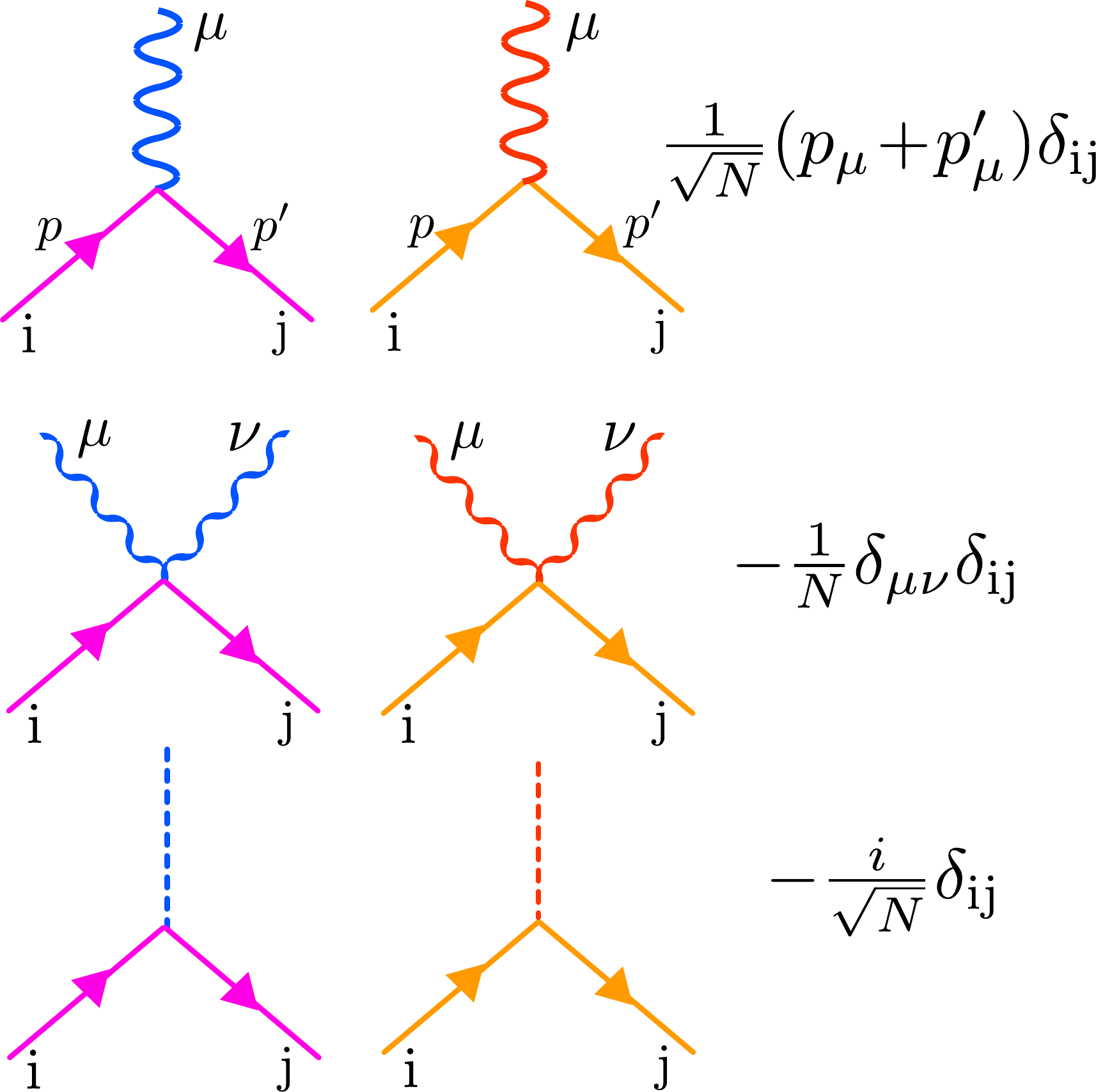}
  \caption{The vertices of the (doubled) $\CP^{N-1}$ model.}
  \label{figvert}
\end{figure}
\begin{figure}[h]
    \centering 
    \includegraphics[width=0.8\textwidth]{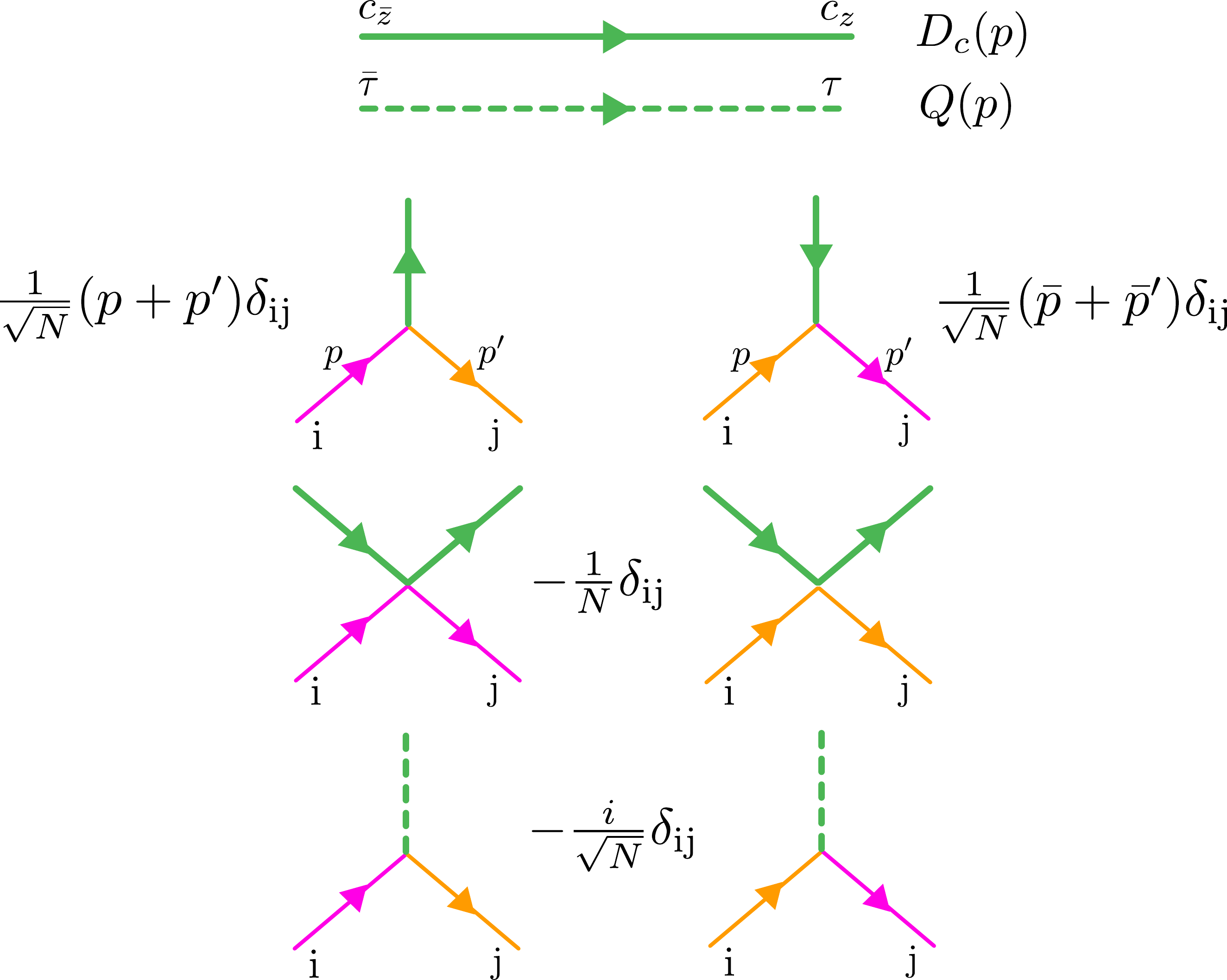}
  \caption{Propagators and vertices of the new fields $\tau, c_z$. Here $p=p_z$ and $\bar{p}=p_{\bar{z}}$ are the holomorphic/anti-holomorphic components of the momentum.}
  \label{fignewfields}
\end{figure}

The Lagrangian (\ref{Lagr2}) may be rewritten in the form of two coupled $\CP^{N-1}$-model Lagrangians:
\bear\label{LagrCPN2}
&&\!\!\!\!\!\!\!\!\!\!\widehat{\mathscr{L}}=|D_\mu^{(a)} u|^2+|D_\mu^{(b)} v|^2+\\ \nonumber
&&\!\!\!\!\!\!\!\!\!\!+i\,(c_{\bar{z}}\,\bar{v}\circ \dd_z u-c_z\,\dd_{\bar{z}}\bar{u}\circ v+c_z\,\bar{u}\circ \dd_{\bar{z}}v-c_{\bar{z}}\dd_z\bar{v}\circ u)+\,c_z\,c_{\bar{z}}\,(|u|^2+|v|^2)+\\ \nonumber
&&\!\!\!\!\!\!\!\!\!\! +i\lambda_1\, (\|u\|^2-N)+i\lambda_2\, (\|v\|^2-N)+i\tau\, \bar{u}\circ v+i\bar{\tau}\,\bar{v}\circ u
 + \,\textrm{terms with ghost fields}\,.
\eear
The two $\CP^{N-1}$ models, parametrized by vectors $u$ and $v$, interact via the matter field $c$, as well as through the orthogonality constraint $\bar{u}\circ v=0$, implemented with the help of the Lagrange multiplier $\tau$. The first line in (\ref{LagrCPN2}) encodes the propagators and interaction vertices of two $\CP^{N-1}$ models -- these are shown in Fig.~\ref{figprop} and Fig.~\ref{figvert} respectively. Apart from these, our model has the additional complex fields $\tau$ and $c_z$ transforming in the bi-fundamental representation of $U(1)\times U(1)$. Their respective propagators and interactions are shown in Fig.~\ref{fignewfields}.

\vspace{0.3cm}
In implementing the steepest descent method for the large-$N$ limit, we will parametrize $\Lambda=\Lambda_0+\frac{1}{\sqrt{N}}\widetilde{\Lambda}$, $\mathcal{A}=\frac{1}{\sqrt{N}}\widetilde{\mathcal{A}}$. The expansion of the effective action $\mathcal{S}^{\mathrm{eff}}$ starts at the quadratic order in $\widetilde{\lambda}, \widetilde{\mathcal{A}}$, and all subsequent orders are suppressed by powers of $\sqrt{N}$. The propagators shown in the figures are obtained by expanding the effective action in (\ref{effact}) to quadratic order in the fields, which yields in momentum~space:
\bear\label{effact}
&&\mathcal{S}^{\mathrm{eff}}=\int\,d^2p\,\left(\Gamma_{\mu\nu}(\vec{p})\times\left[a_\mu(\vec{p})\,a_\nu(-\vec{p})+b_\mu(\vec{p})\,b_\nu(-\vec{p})\right]+\right.\\ \nonumber&&\left.+\widetilde{\Gamma}(p)\,c_{\bar{z}}(\vec{p})\,c_z(-\vec{p}) +\Gamma_\lambda(\vec{p})\, \mathrm{Tr}(\lambda(p)\lambda(-p))\right)
\eear
The functions $\Gamma_{\mu\nu}(\vec{p})$ and $\Gamma_\lambda(\vec{p})$ were computed already in \cite{DAdda1}. The function~$\widetilde{\Gamma}(p)$ is found in a similar way, and altogether we have:
\bear
&\Gamma_\lambda(\vec{p})={1\over 2}A(p)={1\over 2}\int\,\frac{d^2k}{(k^2+m^2)((k+p)^2+m^2)}&\\ \label{GammaMuNu}
&\Gamma_{\mu\nu}(\vec{p})={1\over 2}(\delta_{\mu\nu}-\frac{p_\mu p_\nu}{p^2})\,\left((p^2+4 m^2) \,A(p)-{1\over \pi}\right)&\\
&\widetilde{\Gamma}(p)=2\,\Gamma_{z\bar{z}}(\vec{p})={1\over 2}\left((p^2+4 m^2) \,A(p)-{1\over \pi}\right)&
\eear

In the effective action there are no mixed terms of the form $\int\,d^2p\, \Gamma_\mu(p)\,a_\mu(p)\,\lambda(p)$ due to the fact that the corresponding one-loop diagram vanishes:
\bea\label{gammamu}
\Gamma_\mu(p) = \int\,\frac{d^2k}{(2\pi)^2}\,\frac{2k_\mu+p_\mu}{(k^2+m^2)((k+p)^2+m^2)}=\{k\to -p-k\}=-\Gamma_\mu(p)=0\,.
\eea
In the Lorenz gauge the propagators featuring in the figures have the following~expressions:
\bear
&&G(p)=\frac{1}{p^2+m^2},\quad\quad Q(p)=A(p)^{-1}\\
&&D_{\mu\nu}(\vec{p})=(\delta_{\mu\nu}-\frac{p_\mu p_\nu}{p^2})\,\left[(p^2+4 m^2) \,A(p)-{1\over \pi}\right]^{-1}\\
&&D_c(p)=2\,\left[(p^2+4 m^2) \,A(p)-{1\over \pi}\right]^{-1}
\eear

\section{The Wilson loop and conserved charges}\label{Wloopsect}

The main object of interest that we wish to be able to calculate is the following correlation function, with a Wilson loop of the flat connection inserted:
\bea
Q(\Gamma):=\langle \prod\limits_{i\in I}\, u_{i}(z_i,\bar{z}_i)\,\prod\limits_{j\in J}\, v_{j}(z_j, \bar{z}_j)\,\prod\limits_{k\in K}\,\mathcal{A}_k(z_k,\bar{z}_k)\; \mathrm{P}_0\,e^{-\int_\Gamma\,\mathscr{A}_u}\rangle\,,
\eea
where $\mathscr{A}_u$ is given by formula (\ref{connection}). The notation $\mathrm{P}_0$ is meant to emphasize that the Wilson loop depends on the choice of a base point on the contour, which we call $p_0\in \Gamma$. On the worldsheet $\Sigma \simeq \mathbb{R}^2$, when the contour $\Gamma$ is infinite, it is often convenient to choose infinity as the base point. The Noether current $K$ entering this formula is the one corresponding to the $U(N)$-symmetry acting on the $\varphi$ fields: $\varphi \to g\circ \varphi, \,g\in U(N)$.
Therefore the Noether current one-form is (see~(\ref{Noether1}))
\bea
K=\frac{2g^2}{N}\,(\varphi\,(\mathscr{D}_z \varphi)^\dagger\,d\bar{z}-\mathscr{D}_z \varphi\,\varphi^\dagger\,dz)\,.
\eea
The notation $Q(\Gamma)$ is meant to emphasize that we will be interested in the dependence on the contour $\Gamma$. Note that classically the connection $\mathscr{A}_u$ is flat, hence $\mathrm{P}_0\,e^{-\int_\Gamma\,\mathscr{A}_u}$ is independent of $\Gamma$, provided one fixes the homology class of $\Gamma$ (and assuming the base point $p_0$ is left intact). Nevertheless, the quantity $Q(\Gamma)$ will in general depend on $\Gamma$ due to the UV-divergences that will need to be regularized, and this gives rise to the quantum anomalies in the higher conservation laws of the model, as first calculated in \cite{AbdallaAnomaly}.

\subsubsection{Example.}
It is useful to consider a correlation function with an insertion of a conserved charge $\mathcal{Q}$ in a simpler model. In fact, the model will be as simple as possible, namely a two-dimensional theory with the Lagrangian
\bea
\mathcal{L}=\dd_z \phi\,\dd_{\bar{z}}\phi\,.
\eea
This Lagrangian exhibits a shift symmetry of the form $\phi\to\phi+a$, where $a$ is a constant. The corresponding Noether current~is
$
J=\dd_z\phi\,dz+\dd_{\bar{z}}\phi\,d\bar{z}\,
$
and the conserved charge~is
\bea
\mathcal{Q}=\int\limits_\Gamma\,\ast J=\int\limits_\Gamma\,i\,(\dd_z\phi\,dz-\dd_{\bar{z}}\phi\,d\bar{z})\,.
\eea
\begin{figure}[h]
    \centering 
    \includegraphics[width=0.35\textwidth]{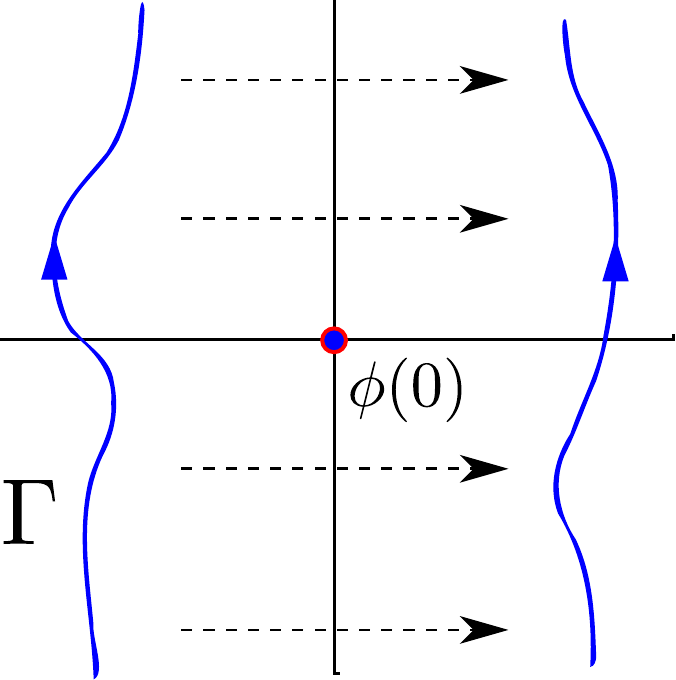}
  \caption{Moving the contour $\Gamma$ past an operator inserted at the origin.}
  \label{movecont}
\end{figure}

The only connected correlation function to consider is $\langle\mathcal{Q}\,\phi(0) \rangle$, where we have chosen to place the second operator at the origin. The contour $\Gamma$ will run along the $y$-axis at a fixed value of $x$, though one could as well consider a curved contour, as shown in Fig.~\ref{movecont}. Using the fact that the two-dimensional propagator is $\langle \phi(z, \bar{z}) \,\phi(0)\rangle=-{1\over 4\pi}\,\log{|z|^2}$, we get for the correlation function
\bea
\langle\mathcal{Q}\,\phi(0) \rangle={1\over 2\pi}\,\int\limits_{-\infty}^{\infty}\,\frac{x\,dy}{x^2+y^2}={1\over 2}\,\mathrm{sgn}(x).
\eea
The value is independent of the position of the contour $\Gamma$, as long as it does not cross the point of insertion of the local operator (in this case the origin). The discontinuity of the correlation function that arises when the contour is moved past a local operator is compatible with the Ward identity for the symmetry in question. \hfill $\blacksquare$

\vspace{0.3cm}
The first non-local charge (as well as the usual local charge) may be calculated by expanding the path-ordered exponent $\mathscr{W}_0(u):=\mathrm{P}_0\,e^{-\int_\Gamma\,\mathscr{A}_u}$ around the point $u=1$ to second order. In order to obtain charges that are manifestly (anti)-Hermitian, it will be useful to parametrize $u=e^{-i\theta}$ and expand to second order in $\theta$. First we note that, to this order,
\bear
&&\mathscr{A}_u={i \theta \over 2}\,(K_z dz-K_{\bar{z}}d\bar{z})+{\theta^2 \over 4}\,(K_z dz+K_{\bar{z}}\,d\bar{z})+\ldots=\\
&&=\frac{\theta}{2} \,\ast K+{\theta^2 \over 4} \,K+\ldots
\eear
Here $\ast$ is the Hodge star, defined by $\ast dz=i \,dz$. We are now in a position to expand the path-ordered exponent:
\bea
\mathscr{W}_0(u)=\mathds{1}-\frac{\theta}{2} \,\int\limits_\Gamma\,\ast K-{\theta^2 \over 4}\,\left[\int\limits_\Gamma\,K-\mathrm{P}\,\int\limits_\Gamma \,\ast K\,\int\limits_\Gamma\,\ast K\right]+\ldots
\eea
The two first conserved charges are therefore
\bear
&&\mathcal{Q}_1=\int\limits_\Gamma\,\ast K \\
&&\widehat{\mathcal{Q}}_2=\int\limits_\Gamma\,K -\mathrm{P}\,\int\limits_\Gamma \,\ast K\,\int\limits_\Gamma\,\ast K\,.
\eear
$\mathcal{Q}_1$ is the usual Noether charge corresponding to the $SU(N)$-symmetry. The `conservation' of the charges means in this context that they are independent of smooth deformations of the contour $\Gamma$. For the charge $\mathcal{Q}_1$ this follows from the Stokes theorem and the conservation of the current, i.e. $d\ast K=0$.
We will follow the conventions of~\cite{AbdallaAnomaly} and further simplify the charge $\widehat{\mathcal{Q}}_2$, using the fact that we are free to add to it an arbitrary function of the charge $\mathcal{Q}_1$. The square of this charge is
\bea\nonumber
\mathcal{Q}_1^2=\int\,dt\,ds\,(\ast K)_t \,(\ast K)_s=\int\limits_{t<s}+\int\limits_{t>s}=\int\limits_{t<s}\,dt\,ds\,[(\ast K)_t \,(\ast K)_s+(\ast K)_s \,(\ast K)_t]
\eea
Therefore we introduce the charge
\bea
\mathcal{Q}_2=\widehat{\mathcal{Q}}_2+{1\over 2}\mathcal{Q}_1^2= \int\limits_\Gamma\,K -{1\over 2}\, \int\limits_{t<s}\,dt\,ds\,[(\ast K)_t, (\ast K)_s]
\eea
The charge so defined also has the nice property that it lies in the Lie algebra of $G$, $\mathcal{Q}_2\in \mathfrak{g}$. In order to automatically obtain charges with values in the Lie algebra, one should consider the `Berry connection' $\mathscr{B}=(\mathscr{W}_0(u))^{-1}\frac{d \mathscr{W}_0(u)}{d\theta}$ and expand it around $\theta=0$.

\vspace{0.3cm}
Let us check that, classically, the charge $\mathcal{Q}_2$ is independent of $\Gamma$. To this end we introduce the one-form (the lower limit of integration -- point $p_0$ -- is again the base point, which enters the definition of the Wilson loop)
\bea\label{Sform}
S(p):=[\left(\int\limits_{p_0}^p\,\ast K\right), \ast K(p)].
\eea
Here $\int\limits_{p_0}^p\,\ast K$ is the primitive of the one-form $\ast K$, which is well-defined due to the fact that $\ast K$ is closed (conservation of the current). In fact, if the point $p$ was always restricted to lie on $\Gamma$, the integral  $\int\limits_{p_0}^p\,\ast K$ could be understood as an integral along $\Gamma$, even without the closedness of $\ast K$. However we will now wish to appeal again to Stokes' theorem in the interior of $D_{\delta \Gamma}$ (this is the space between the two contours, see Fig.~\ref{contdefcorr}), and therefore we will need a one-form well-defined inside~$D_{\delta \Gamma}$.

\vspace{0.3cm}
With this said, we may now write the charge $\mathcal{Q}_2$ as 
\bea
\mathcal{Q}_2=\int\limits_\Gamma\,(K -{1\over 2}S)\,.
\eea
The variation again may be computed using Stokes' theorem. Taking into account that $dS=2\,\ast K \wedge \ast K=2 \,K\wedge K$ and that the current $K$ is flat, $dK-K\wedge K=0$,we find that the variation is zero:
\bea
\delta_\Gamma \mathcal{Q}_2(\Gamma)=0\,.
\eea

\subsection{The regularized charge}\label{regchargesect}

In the quantum theory, due to the short-distance singularities in the product of two currents, one needs to define a regularized version of the non-local charge $Q_2$. This will involve the splitting of coincident points in the definition of the one-form $S$ in~(\ref{Sform}). One possibility is to define the following $\epsilon$-regularized form:
\bea\label{Seps}
S_\epsilon(p):=[\left(\int\limits_{p_0}^{p+\epsilon}\,\ast K\right), \ast K(p)].
\eea
\begin{figure}[h]
    \centering 
    \includegraphics[width=0.6\textwidth]{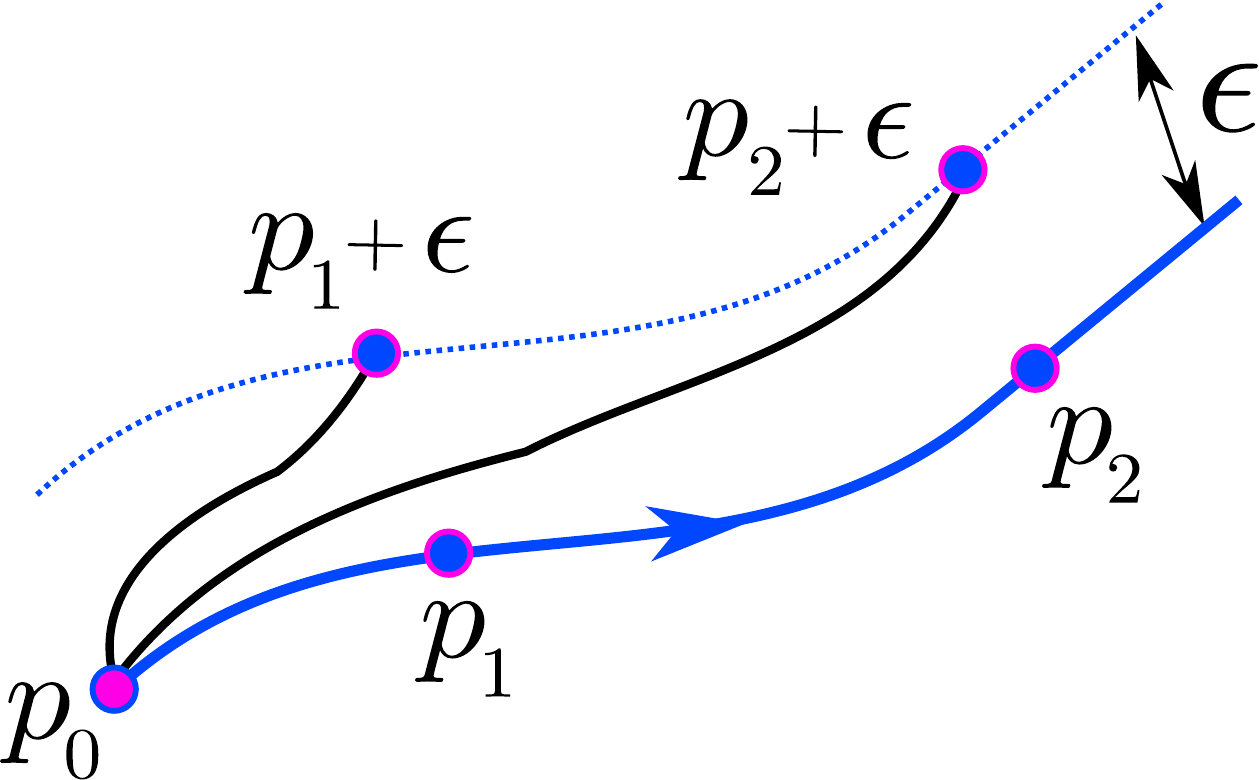}
  \caption{The paths of integration from the base point $p_0$ to a given point $p+\epsilon$ in the definition of the operator $S_\epsilon$. The points $p_0, p_1, p_2$ lay on the contour $\Gamma$.}
  \label{figcontour}
\end{figure}

Here it is implied that $\epsilon$ is a \emph{fixed} two-vector, and we are considering the case when $\Sigma_2\simeq \mathds{R}^2$, equipped with a flat metric, when the notion of addition `$p+\epsilon$' makes sense. The one-form $S_\epsilon$  depends on the contour of integration in~(\ref{Seps}) in the topological sense, i.e. it depends on whether the contour winds around the point $p$. Therefore we will fix a certain value of $\epsilon$ (such as the one shown in Fig.~\ref{figcontour}) and define the auxiliary contours in~(\ref{Seps}) for this value of $\epsilon$. For instance, for contours $\Gamma$ of the type shown in  Fig.~\ref{figcontour} one can define the auxiliary contour by requiring that it lay on one side of~$\Gamma$. Then we can gradually change the value of $\epsilon$ so that it circles around the origin, $\epsilon\to e^{2\pi i} \epsilon$. As a result, $S_\epsilon$ will gain a monodromy of the form
\bea
S_{e^{2\pi i} \epsilon}-S_\epsilon=[\left(\oint\,\ast K\right), \ast K(p)]\,,
\eea
where the integral is around a loop centered at $p$. Since the current $K$ is conserved, $d\ast K=0$, the value of the integral does not depend on the size of the loop. Yet it is not zero, due to the presence of the operator insertion of $\ast K(p)$ at the center of the loop. However, the loop may be shrunk to be arbitrarily small, and the value of the commutator may be calculated from the most singular term in the OPE of the two currents. Such analysis below will bring us to the following answer:
\bea\label{ambigcomm}
[\left(\oint\,\ast K\right), \ast K(p)]=2g^2\,\ast K(p)
\eea
In order to cancel this monodromy, we will consider the operator
\bea\label{regQcharge}
\mathcal{Q}_\epsilon(\Gamma):=\int\limits_\Gamma\,\left(\left[\mathbf{a}+{1\over 2\pi}\log{(\epsilon)}\right]\,K_z dz+\left[\bar{\mathbf{a}}+{1\over 2\pi}\log{(\bar{\epsilon})}\right]\,K_{\bar{z}}d\bar{z}-{1\over 2 g^2}\,S_\epsilon\right)\,,
\eea
where $\mathbf{a}$ is a constant, independent of $\epsilon$, to be chosen later. It is clear that the  operator  defined above is invariant as the regularization parameter $\epsilon$ rotates around the origin: $\epsilon \to e^{2\pi i}\,\epsilon$. This operator $\mathcal{Q}_\epsilon$ is our candidate for the regularized version of the charge~$\mathcal{Q}_2$.

\vspace{0.3cm}
We wish to prove the following:
\begin{itemize}
\item There exists a limit $\underset{\epsilon\to 0}{\mathrm{lim}} \,\mathcal{Q}_\epsilon$
\item The limit depends on the curve $\Gamma$ through an anomaly 2-form $\Omega_A$, namely
\bea\label{anomdef0}
\delta_\Gamma \left(\underset{\epsilon\to 0}{\mathrm{lim}} \,\mathcal{Q}_\epsilon\right)=\int\limits_{D_{\delta\Gamma}}\,\Omega_A\,,
\eea
where $D_{\delta\Gamma}$ is a two-dimensional domain bounded by the original and final curves $\Gamma_1, \Gamma_2$ (see Fig.~\ref{contdefcorr}).
\end{itemize}
Let us start from the second point by calculating the difference $\mathcal{Q}_\epsilon(\Gamma_1)-\mathcal{Q}_\epsilon(\Gamma_2)$ of the values of the charge on two different contours. We use Stokes' theorem and the conservation of the current to obtain:
\bear\nonumber
\mathcal{Q}_\epsilon(\Gamma_1)-\mathcal{Q}_\epsilon(\Gamma_2)
=&& \!\!\!\!\!\!\!\!\!\!\int\limits_{D_{\delta \Gamma}}\,dz\wedge d\bar{z}\,\left[\left(\mathrm{Re}(\mathbf{a})+{1\over 2\pi} \log{|\epsilon|}\right)\,(\dd_z K_{\bar{z}}-\dd_{\bar{z}}K_z)\,-\right. \\ \label{anomdef}&& \!\!\!\!\!\!\!\!\!\!\left.-\,{1\over 2 g^2}\left([K_z(p+\epsilon), K_{\bar{z}}(p)]+[K_z(p), K_{\bar{z}}(p+\epsilon)]\right)\right]\,.
\eear
Therefore one of the goals  will be  the calculation  of the OPE of the two currents $[K_z(p+\epsilon), K_{\bar{z}}(p)]$ as $\epsilon\to 0$.

\subsection{The operator product expansion}\label{OPEsect1}

We start by recalling the definition of the current~(\ref{Noether1}):
\bea\label{Noethernew}
K=\frac{2g^2}{N}\,\left(\varphi\,(\mathscr{D}_z \varphi)^\dagger\,d\bar{z}-\mathscr{D}_z \varphi\,\varphi^\dagger\,dz\right)\,.
\eea

First we will calculate the OPE between two holomorphic components of the current, and as a result we will be able to prove the first point above -- that there exists a limit $\underset{\epsilon\to 0}{\mathrm{lim}} \,\mathcal{Q}_\epsilon$.

\vspace{0.3cm}
In what follows we will encounter the function $D(z)$ -- the massive scalar propagator in two dimensions (recalling the definition $\{z, \bar{z}\}={x^1\pm i x^2\over \sqrt{2}}$ of the complex coordinates):
\bear\label{Gpropexp}
D(z)&=&\int\,\frac{d^2p}{(2\pi)^2}\,\frac{e^{i\,(p_1 x^1+p_2 x^2)}}{p_1^2+p_2^2+m^2}=\frac{1}{2\pi}\,K_0(\sqrt{2} m|z|)=\\ \nonumber
&=& -{1\over 4\pi} \left(1+{m^2|z|^2\over 2}\right)\,\log{\left(m^2 |z|^2\over 2\right)}-{\gamma\over 2\pi}+{1-\gamma\over 2\pi}{m^2 |z|^2\over 2}+\ldots
\eear
The final line captures the first few terms in the expansion of $D(z)$ for $z\to 0$.

\vspace{0.3cm}
Let us also make a reservation regarding the technique that we use in the present section. Most of the diagrams that we will be calculating (see~Figs.~\ref{OPEfig} and \ref{OPEquadr}) really correspond to the theory of a scalar field $\varphi$ in a background gauge field $\mathcal{A}$. Therefore these diagrams are the same as the ones that would arise in the corresponding Grassmannian sigma-model with target space $G_{M, N}$. For this reason in these calculations we will simply treat $\mathcal{A}$ as a non-abelian external gauge field (which can later be restricted to the form~(\ref{gaugef})). To simplify the figures, we will still be drawing the same diagrams as for a single copy of the $\CP^{N-1}$-model, but with the understanding that the fields carry an additional index. The sole role of this index is to make sure the ordering of the fields is taken into account. The only diagrams where the gauge field enters in the internal lines are the ones of Fig.~\ref{OPEquadr2}, and for these diagrams we carry out a more thorough analysis in Section~\ref{respartsec}.

\subsubsection{The commutator $[(\ast K)_z(z_1), (\ast K)_z(z_2)]$}

We now pass to the calculation of the OPE of the commutator of two $(\ast K)_z$-components. To this end we consider the product
\bea
(\ast K)_z(z_1) \cdot (\ast K)_z(z_2)=-{4g^4\over N^2}\,\mathscr{D}_z \varphi(z_1)\left(\varphi^\dagger(z_1) \mathscr{D}_z \varphi(z_2)\right) \varphi^\dagger(z_2)
\eea
We write
\bea
\varphi^\dagger(z_1) \mathscr{D}_z \varphi(z_2)=-N\cdot \dd_{z_1}D(z_1-z_2) \,\mathds{1}_2+O(\log{(|z_1-z_2|)})
\eea
Note that in the commutator $[(\ast K)_z(z_1), (\ast K)_z(z_2)]$ the logarithmic (as well as finite) terms will disappear as they are symmetric under $z_1\leftrightarrow z_2$, therefore we have:
\bea
[(\ast K)_z(z_1), (\ast K)_z(z_2)]={8g^4\over N}\,\dd_{z_1}D(z_1-z_2)\,\colon\mathscr{D}_z \varphi(z_1) \varphi^\dagger(z_2)\colon+\ldots
\eea
We have inserted a normal ordered product in the r.h.s., since the self-contractions of the field $\varphi$ would give a contribution proportional to the unit operator $\mathds{1}_N$, which clearly cannot enter in the expansion of a commutator.

\vspace{0.3cm}
From (\ref{Gpropexp}) it follows that $\dd_{z_1}D(z_1-z_2)=-{1\over 4\pi}\frac{1}{z_1-z_2}+O(z_1-z_2)$, where $O(z_1-z_2)$ denotes terms that vanish in the limit $z_1\to z_2$. Therefore
\bea
[(\ast K)_z(z_1), (\ast K)_z(z_2)]={g^2\over \pi}\frac{1}{z_1-z_2}\,K_z(z_2)+O(|z_1-z_2|^0)
\eea
Analogously for the $\bar{z}$-components we have:
\bea\label{antiKOPE}
[(\ast K)_{\bar{z}}(z_1), (\ast K)_{\bar{z}}(z_2)]={g^2\over \pi}\frac{1}{\bar{z}_1-\bar{z}_2}\,K_{\bar{z}}(z_2)+O(|z_1-z_2|^0)
\eea
The above two equalities are already sufficient to prove (\ref{ambigcomm}). Indeed, in the limit of an infinitesimally small contour around $w$ we have
\bear\label{ambigcomm2} \nonumber
&&[\left(\oint\,(\ast K)_z\,dz+\oint\,(\ast K)_{\bar{z}}\,d{\bar{z}}\right), \ast K(w)]=\\ 
&&=\oint\,{g^2\over \pi}\frac{dz}{z-w}\,K_w(w) dw+\oint\,{g^2\over \pi}\frac{d\bar{z}}{\bar{z}-\bar{w}} \,K_{\bar{w}}(w) d\bar{w}=\\ \nonumber
&&=2g^2(i\, K_w(w)\,dw-i \,K_{\bar{w}}(w)\,d\bar{w})=2g^2\ast K(w)
\eear

A similar derivation is used to find the behavior of $S_\epsilon$ for $\epsilon\to 0$:
\bear\nonumber
&&S_\epsilon(w)=  \left[\left(\int\limits_0^{w+\epsilon}\,(\ast K)_z\,dz+\int\limits_0^{\bar{w}+\bar{\epsilon}}\,(\ast K)_{\bar{z}}\,d{\bar{z}}\right), \ast K(w)\right]\sim\\ 
&&\underset{\epsilon\to 0}{\sim}\;\;{g^2\over \pi}\,\log{(\epsilon)}\,K_w(w) dw+{g^2\over \pi}\,\log{(\bar{\epsilon})}\,K_{\bar{w}}(w) d\bar{w}+\ldots\,,
\eear
where $\ldots$ denotes finite terms. It is now obvious that the integrand in (\ref{regQcharge}) has a finite limit for $\epsilon\to 0$. This limit is, in particular, independent of the angle at which $\epsilon$ approaches zero.

\subsubsection{The commutator $[(\ast K)_z(z_1), (\ast K)_{\bar{z}}(z_2)]$}\label{holantiholcomm}

\begin{figure}[h]
    \centering 
    \includegraphics[width=0.5\textwidth]{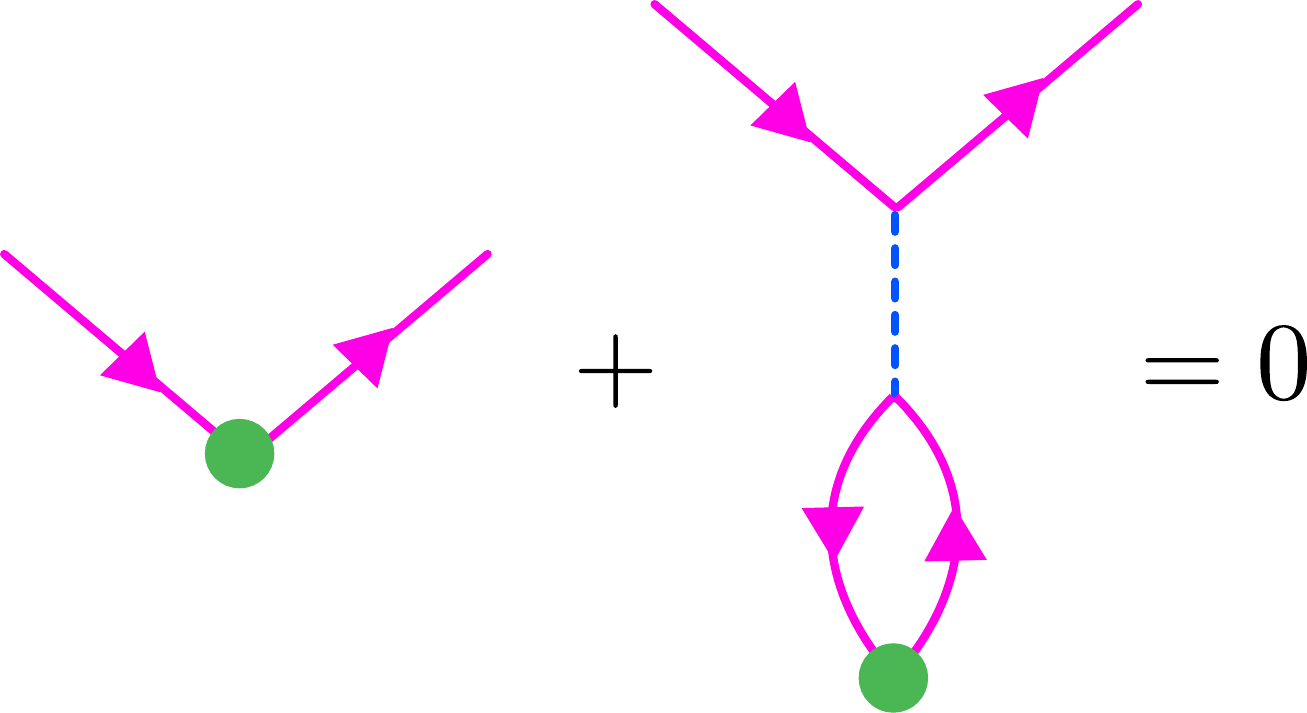}
  \caption{A proof that $:\varphi^\dagger(z)  \varphi(z):=0$.}
  \label{lambdacancel}
\end{figure}

First of all, using the above definition, we calculate
\bea
\ast K=-\frac{2ig^2}{N}\,\left(\mathscr{D}_z \varphi\,\varphi^\dagger\,dz+\varphi\,(\mathscr{D}_z \varphi)^\dagger\,d\bar{z}\right)\,,
\eea
therefore
\bea
(\ast K)_z(z_1) \cdot (\ast K)_{\bar{z}}(z_2)=-\frac{4 g^4}{N^2}\,\mathscr{D}_z \varphi(z_1)\left(\varphi^\dagger(z_1)  \varphi(z_2)\right) (\mathscr{D}_z \varphi)^\dagger(z_2)
\eea
Classically $\varphi^\dagger(z_1)  \varphi(z_1)={N\over g^2}\,\mathds{1}_2$. Here however we need to write down the OPE of the following form
\bea
\varphi^\dagger(z_1)  \varphi(z_2)=N\cdot D(z_1-z_2)\,\mathds{1}_2+:\varphi^\dagger(z_1)  \varphi(z_1):+\ldots
\eea
where $\ldots$ denotes terms that vanish as $z_2\to z_1$. We will now show that $:\varphi^\dagger(z_1)  \varphi(z_1):=0$. Indeed, this follows from the cancellation of two diagrams shown in Fig.~\ref{lambdacancel}.

\begin{figure}[h]
    \centering 
    \includegraphics[width=0.3\textwidth]{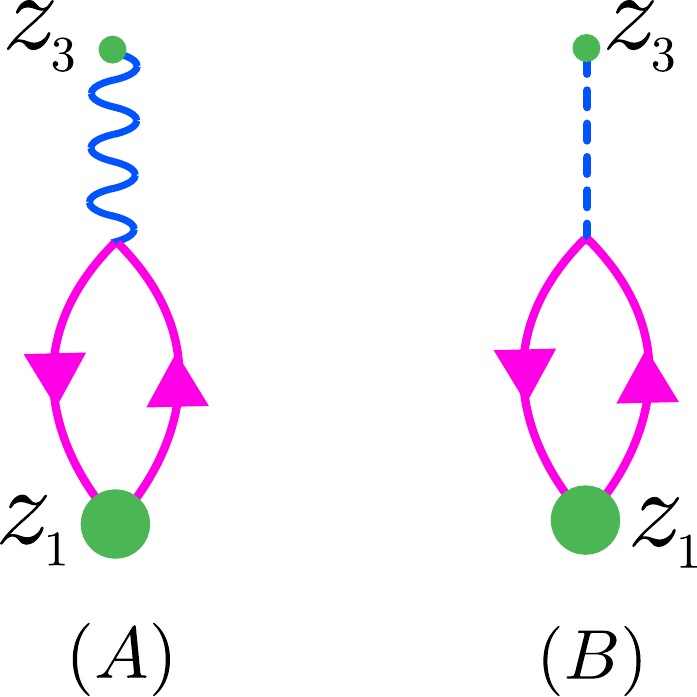}
  \caption{Diagrams that contribute to the correlation functions $(A)$ $\langle \varphi^\dagger(z_1)  \varphi(z_2) \mathcal{A}(z_3)\rangle$ and $(B)$ $\langle \varphi^\dagger(z_1)  \varphi(z_2) \lambda(z_3)\rangle$ in the limit $z_2\to z_1$.}
  \label{phiphiexppic}
\end{figure}

\vspace{0.3cm}
Note that, in principle, one should also consider the correlation functions $\langle \varphi^\dagger(z_1)  \varphi(z_2) \mathcal{A}(z_3)\rangle$ and $\langle \varphi^\dagger(z_1)  \varphi(z_2) \lambda(z_3)\rangle$ (see Fig.~\ref{phiphiexppic}). One can show, however, that
\bear
&&\underset{z_2\to z_1}{\mathrm{lim}}\langle \varphi^\dagger(z_1)  \varphi(z_2) \mathcal{A}(z_3)\rangle=0\,,\\ \label{deltacorrfunc1}
&&\underset{z_2\to z_1}{\mathrm{lim}}\langle \varphi^\dagger(z_1)  \varphi(z_2) \lambda(z_3)\rangle\sim\,\delta(z_1-z_3)\,.
\eear

\begin{figure}[h]
    \centering 
    \includegraphics[width=0.8\textwidth]{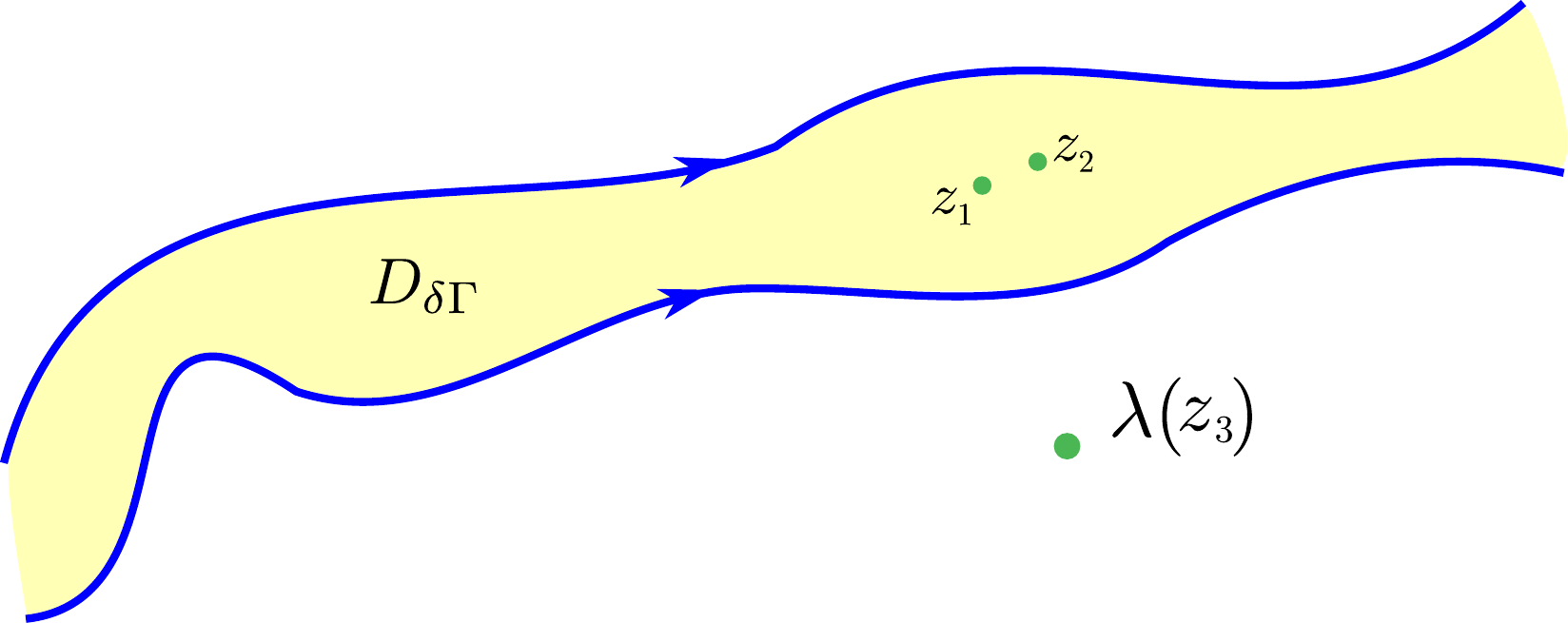}
  \caption{The point $z_3$, where the operator $\lambda$ is inserted, is outside of the relevant integration domain $D_{\delta\Gamma}$. Since according to~(\ref{deltacorrfunc1}) the corresponding correlation function~(\ref{deltacorrfunc2}) is proportional to $\delta(z_1-z_3)$, it is zero for the depicted configuration.}
  \label{contdefcorr}
\end{figure}

The diagram $(A)$ vanishes for symmetry reasons, as its value is equal to $\Gamma_\mu$ from~(\ref{gammamu}).
As for the second diagram, the loop integral is precisely equal to the inverse of the $\lambda$-field propagator $Q(p)$, and therefore in coordinate space the diagram $(B)$ is proportional to $\delta(z_1-z_3)$. Such a term does not contribute to the anomaly, as at the level of correlation functions we have the following result~(see~Fig.~\ref{contdefcorr}):
\bea\label{deltacorrfunc2}
\langle \delta_\Gamma \mathcal{Q}\cdot \lambda(z_3)\rangle =\int\limits_{D_{\delta\Gamma}}\,\langle \Omega_A \,\cdot\lambda(z_3)\rangle=0\quad\quad\textrm{for} \quad z_3 \quad\textrm{outside of}\quad D_{\delta\Gamma}\,.
\eea

Let us also recall that $\varphi: \CC^2\to \CC^N$ is a $(N\times 2)$-matrix. Let us denote the $i$-th row by $\varphi^i$ and, following \cite{AbdallaAnomaly}, consider the off-diagonal entries of $\left((\ast K)_z(z_1) \cdot (\ast K)_{\bar{z}}(z_2)\right)^{ij}$ ($i\neq j$) in the OPE. This will slightly simplify things, as in this case the contractions between the outer fields $\varphi^i$ and $(\varphi^\dagger)^j$ are prohibited. Then we get
\bea
\left((\ast K)_z(z_1) \cdot (\ast K)_{\bar{z}}(z_2)\right)^{ij}\underset{z_2\to z_1}{\to}-\frac{4g^4}{N}\,\left[\mathscr{D}_z \varphi \,\mathscr{D}_{\bar{z}} \varphi^\dagger\right]_{z_1}^{ij}\,D(z_1-z_2)\,,
\eea
where it is understood that one should only keep the non-vanishing terms in the expansion of $D(z_1-z_2)$ as $z_2\to z_1$. Using the definition of the current $K$, we find that
\bear\nonumber
\dd_z K_{\bar{z}}-\dd_{\bar{z}} K_z&=&{4g^2\over N}\,\mathscr{D}_z \varphi \,\mathscr{D}_{\bar{z}} \varphi^\dagger+{2g^2\over N} \,(\varphi\,\D_{z}\D_{\bar{z}}\varphi^\dagger+\D_{\bar{z}}\D_{z}\varphi\,\varphi^\dagger)=\\
&=& {4g^2\over N}\,\mathscr{D}_z \varphi \,\mathscr{D}_{\bar{z}} \varphi^\dagger+{2g^2\over N} \,\varphi\,(i\lambda+i F_{z\bar{z}})\,\varphi^\dagger\,,
\eear
where in the last line we have used the e.o.m. (\ref{eqs1}). Substituting this in the above OPE, we get

\bear\label{leadordOPE11}
&&\!\!\!\!\!\!\!\!\!\!\left((\ast K)_z(z_1) \cdot (\ast K)_{\bar{z}}(z_2)\right)^{ij}\underset{z_2\to z_1}{\to}\\ \nonumber
&&\quad\quad\quad\quad\quad\quad\to g^2\left(\dd_{\bar{z}} K_z-\dd_z K_{\bar{z}}+{2g^2\over N} \left[\varphi(i\lambda+i F_{z\bar{z}})\varphi^\dagger\right]\right)^{ij}_{z_1}\!D(z_1-z_2)
\eear

\begin{figure}[h]
    \centering 
    \includegraphics[width=\textwidth]{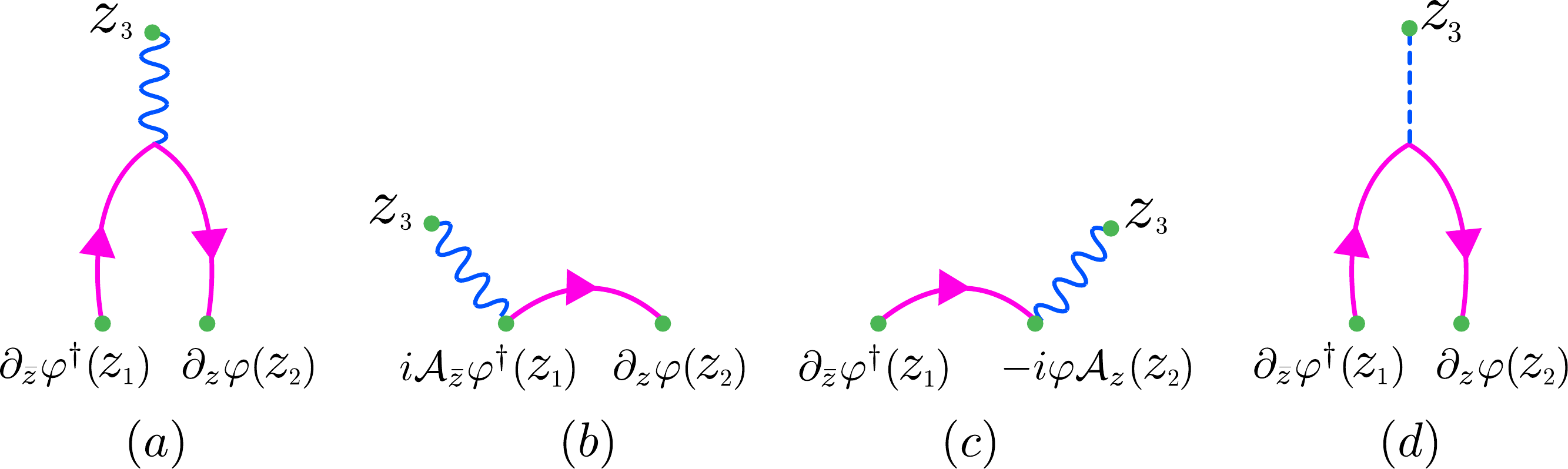}
  \caption{The diagrams contributing at order $1\over \sqrt{N}$ to the OPE (\ref{OPEexp}).}
  \label{OPEfig}
\end{figure}

Multiplying the currents in the opposite order, we obtain
\bea
(\ast K)_{\bar{z}}(z_1)\cdot (\ast K)_z(z_2) =-\frac{4g^4}{N^2}\,\varphi(z_1)\left(\mathscr{D}_{\bar{z}} \varphi^\dagger(z_1)\cdot \mathscr{D}_z \varphi(z_2)   \right) \varphi^\dagger(z_2)
\eea
Classically we see from (\ref{lambda}) that we could replace $\underset{z_2\to z_1}{\mathrm{lim}} {g^2\over N} \mathscr{D}_{\bar{z}} \varphi^\dagger(z_1)\cdot \mathscr{D}_z \varphi(z_2)=-{1\over 2}i\lambda(z_1)-{1\over 2}i\,F_{z\bar{z}}(z_1)$. At the quantum level the corresponding statement is that the following OPE holds:
\bea\label{OPEexp}
{1\over N}\,\mathscr{D}_{\bar{z}} \varphi^\dagger(z_1)\cdot \mathscr{D}_z \varphi(z_2)\underset{z_2\to z_1}{\to}\,\kappa_\lambda(z_1-z_2)\,i\lambda(z_1)+\kappa_{F}(z_1-z_2)\,i\,F_{z\bar{z}}(z_1)\,,
\eea
where $\kappa_\lambda(z)$ and $\kappa_F(z)$ are two coefficient functions to be found, at leading order~$1\over \sqrt{N}$, from the diagrams shown in Fig.~\ref{OPEfig}.

The sum of diagrams $(a)$, $(b)$ and $(c)$ leads to the following integral\;\footnote{Note that here, and in what follows, we use the non-rescaled fields $\mathcal{A}, \lambda$ in the external lines of various diagrams. As explained in Section~\ref{1Nsect}, to justify the result within the $1\over N$-expansion, one should keep in mind that $\lambda=\frac{1}{\sqrt{N}}\widetilde{\lambda}$, $\mathcal{A}=\frac{1}{\sqrt{N}}\widetilde{\mathcal{A}}$. It is in this sense that we count the orders of the $1\over N$-expansion in Figs.~\ref{OPEfig},~\ref{OPEquadr}.}\;\footnote{In writing out explicit expressions for the Feynman integrals we will suppress factors of~$N$, coming from loops of matter fields, and reinstate them only in the final expressions for the OPE's.}:
\bear\nonumber
&&\!\!\!\!\!\!\!\!\!\!I\!=\!\int\!\!\frac{d^2k}{(2\pi)^2}\!\left(\frac{k_{\bar{z}}(k_z+p_z)}{(k^2+m^2)((k+p)^2+m^2)}\,\left((2k_{\bar{z}}+p_{\bar{z}})\mathcal{A}_z(p)+(2k_{z}+p_{z})\mathcal{A}_{\bar{z}}(p)\right)\,-\right. \\  \label{anomint1}  &&\left. \quad\quad\quad\quad\quad-\frac{k_{\bar{z}}}{k^2+m^2}\,\mathcal{A}_z(p)-\frac{k_z+p_z}{(p+k)^2+m^2}\,\mathcal{A}_{\bar{z}}(p) \right)\,e^{ik\epsilon}
\eear
By a direct calculation one can show (see Appendix~\ref{intsimpl}) that, up to terms which vanish in the limit $\epsilon \to 0$, it is equal to
\bea\label{anomint2}
I=(p_{\bar{z}}\mathcal{A}_z-p_z\mathcal{A}_{\bar{z}})\,(\Gamma_{z\bar{z}}+\frac{1}{4\pi})-{1\over 2}(p_{\bar{z}}\mathcal{A}_z-p_z\mathcal{A}_{\bar{z}})\,D(\epsilon)+\ldots\,,
\eea
where $\Gamma_{z\bar{z}}$ is a component of the $\Gamma_{\mu\nu}$-tensor defined in the effective action~(\ref{effact}). The contribution of $\Gamma_{z\bar{z}}(p)$ may be discarded in the OPE, as again it leads to contact terms in the correlation functions (analogously to the situation described in Fig.~\ref{contdefcorr}).

\vspace{0.3cm}
Let us now evaluate the diagram $(d)$. It corresponds to the integral
\bea
I_d=-i\lambda(p)\,\int\,\frac{d^2k}{(2\pi)^2}\,\frac{k_{\bar{z}}(k_z+p_z)}{(k^2+m^2)((k+p)^2+m^2)}\times e^{ik\epsilon}
\eea
Using manipulations similar to the ones described in Appendix~\ref{intsimpl}, one can show that up to terms vanishing in the limit $\epsilon\to 0$ this integral is equal to
\bea\label{anomint4}
I_d=(m^2+p_z p_{\bar{z}})\,\Gamma_{\lambda}(p)\,i\lambda(p)\,-{1\over 2} i\lambda(p)\,D(\epsilon)+\ldots
\eea
Once again, the terms proportional to $\Gamma_{\lambda}(p)\,i\lambda(p)$ are contact terms and may be dropped in the OPE.

\vspace{0.3cm}
As a result we obtain the following OPE:
\bear\label{leadordOPE}
&&\!\!\!\!\!\!\!\!\!\!\left((\ast K)_{\bar{z}}(z_1)\cdot (\ast K)_z(z_2)\right)^{ij}\!\underset{z_2\to z_1}{\to} \\ \nonumber
&&\quad\quad\quad\quad\quad\quad\to {2g^4\over N}\left(\left[\varphi\,(i\lambda+i F_{z\bar{z}})\,\varphi^\dagger\right] D(\epsilon)-{1\over 2\pi }\,\varphi \, i F_{z\bar{z}}\,\varphi^\dagger\right)^{ij}_{z_1}+\ldots
\eear
Using the expansions (\ref{leadordOPE11}) and (\ref{leadordOPE}), we find the following OPE for the commutator of two currents (for $i\neq j$):
\bea\label{OPEleadfull}
\left[(\ast K)_z(z_1),  (\ast K)_{\bar{z}}(z_2)\right]^{ij}\underset{z_2\to z_1}{\to}g^2\left(\left[\dd_{\bar{z}} K_z-\dd_z K_{\bar{z}}\right]\,D(\epsilon)+{g^2\over \pi N}\,\varphi \, i F_{z\bar{z}}\,\varphi^\dagger\right)^{ij}_{z_1}+\ldots
\eea
In the last two formulas by $F_{z\bar{z}}$ we mean the linear part of the full non-abelian field strength. In the next subsection we will see that the calculation of the OPE at the next order in the $1\over N$-expansion amounts to completing the field strength to the non-abelian form.

\subsubsection{Order $1\over N$}\label{ord1Nsect}

\begin{figure}[h]
    \centering 
    \includegraphics[width=0.8\textwidth]{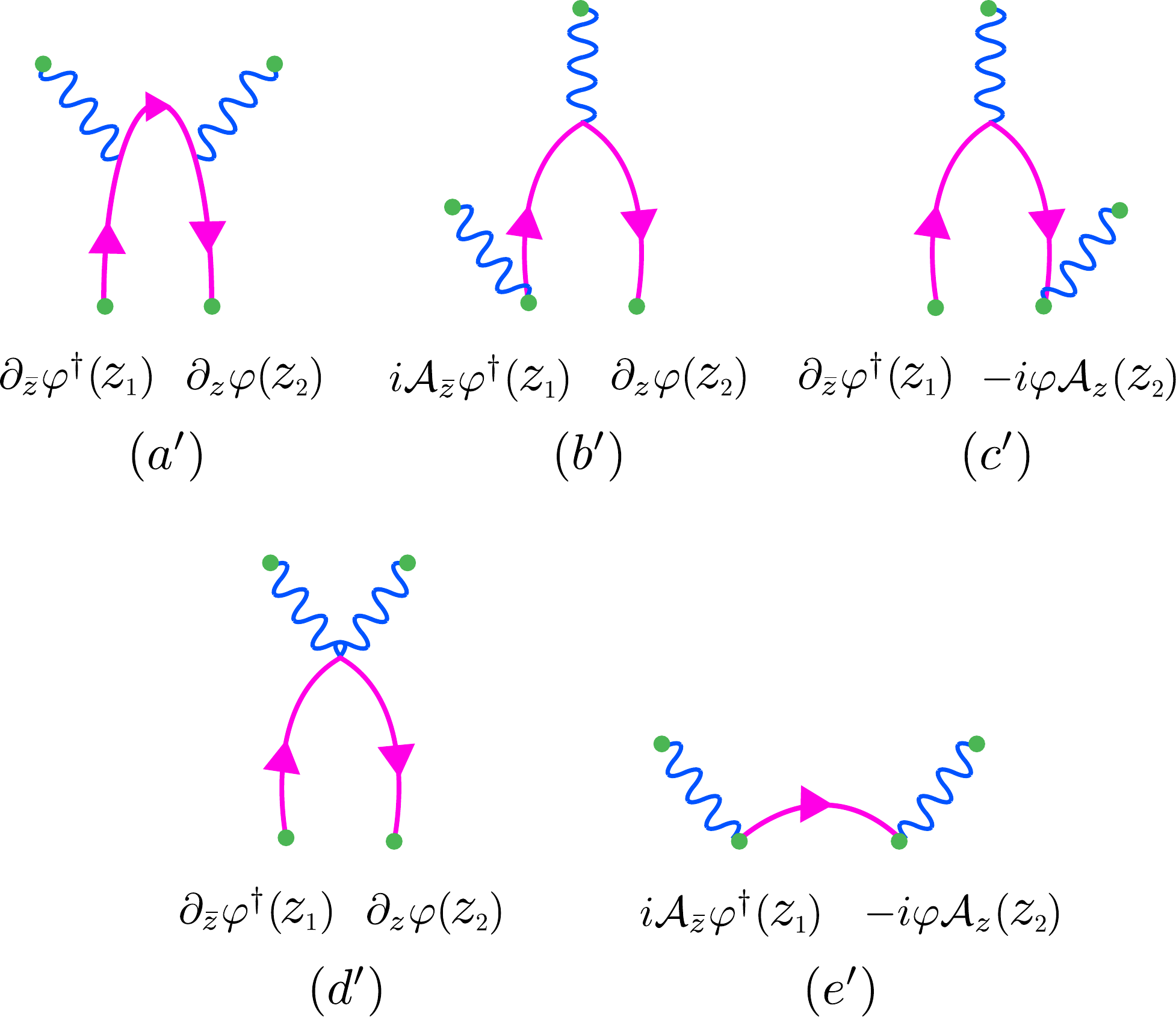}
  \caption{The diagrams at order $1\over N$ in the OPE (\ref{OPEexp}).}
  \label{OPEquadr}
\end{figure}

\begin{figure}[h]
    \centering 
    \includegraphics[width=0.9\textwidth]{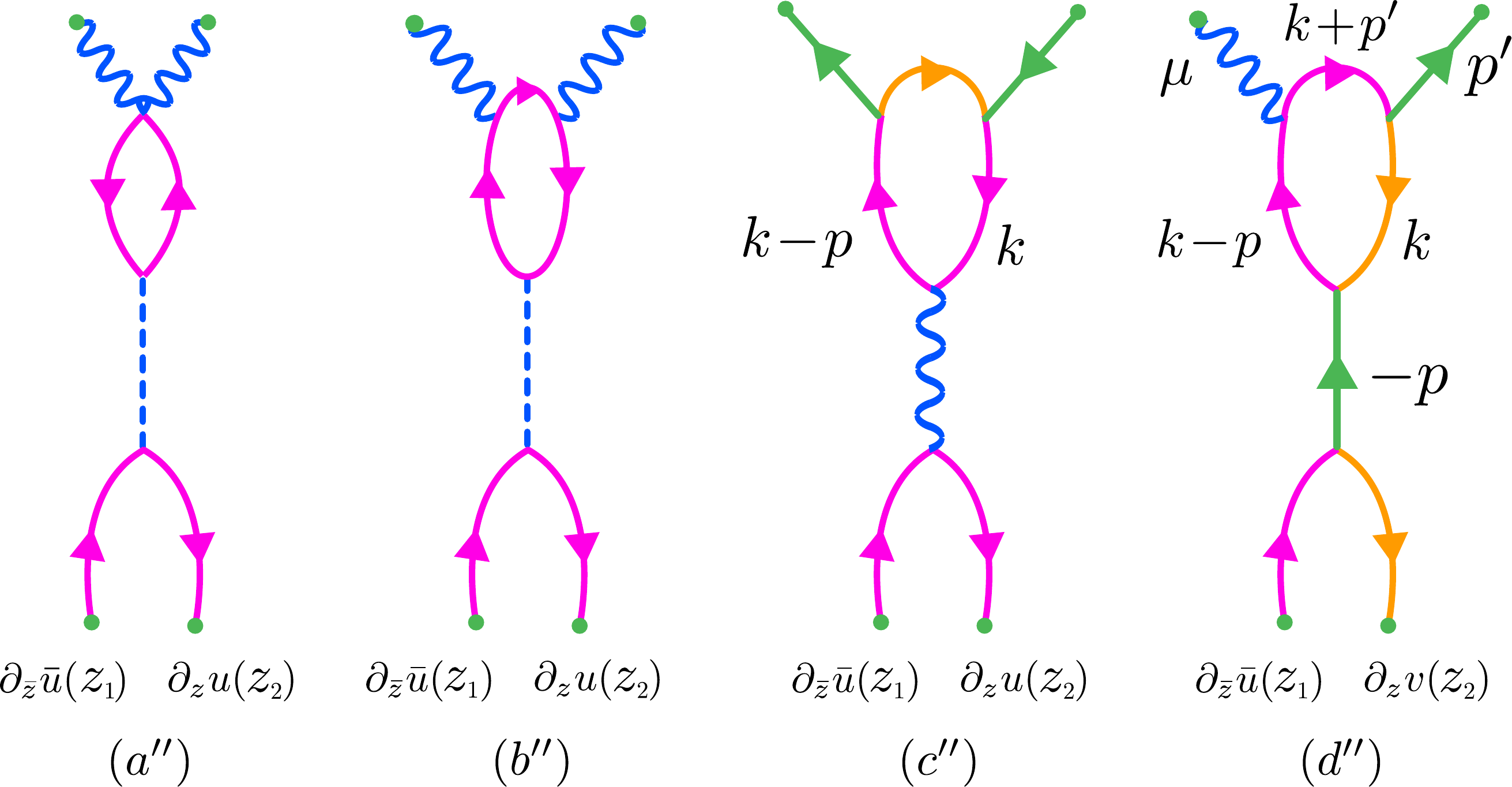}
  \caption{Some of the additional diagrams at order $1\over N$ in the OPE (\ref{OPEexp}), which are obtained by attaching a one-loop (upper) diagram with three external legs to one of the external lines in the diagrams of Fig.~\ref{OPEfig}. The flow of the momentum~$p$ is chosen to comply with the convention in Fig.~\ref{OPEfig} and in the formula~(\ref{anomint1}).}
  \label{OPEquadr2}
\end{figure}

So far we have calculated the OPE to leading order, $1\over \sqrt{N}$. At the next order, $1\over N$, we have the diagrams depicted in Figs.~\ref{OPEquadr} and~\ref{OPEquadr2}.
First of all, in the integrand corresponding to Fig.~$(a')$ we will do the following~rewriting:
\bea\label{k2decomp}
k_{\bar{z}}(k_z+p_z)= {1\over 4}\left[\left((k+p)^2+m^2\right)+\left(k^2+m^2\right)\right]+\underbracket[0.6pt][0.6ex]{{1\over 2}\left(k_{\bar{z}}p_z-k_zp_{\bar{z}}\right)-{1\over 2} \left(p_z p_{\bar{z}}+m^2\right)}_{\textrm{Residual part}}
\eea
Let us start by considering the two terms in square brackets (the contribution of the residual part will be analyzed in the next subsection). They may be used to cancel the leftmost or the rightmost propagator in Fig.~$(a')$. Cancelling the propagator $1\over k^2+m^2$, for example, one is left with the following integral (after the change of variable $k\to k-p_1$):
\bea
I_{a'}^{(1)}\simeq {1\over 4}\,\int\,\frac{d^2 k}{(2\pi)^2} \,e^{ik\epsilon}\,\frac{\mathcal{A}_{\mu}(p_1)(2k-p_1)_\mu\,(2k+p_2)_\nu \mathcal{A}_{\nu}(p_2)}{((k+p_2)^2+m^2)(k^2+m^2)}+(p_1\leftrightarrow p_2)
\eea
Here $p_1$ and $p_2$ are the momenta carried by the external gauge fields, and $\simeq$ means `up to terms that vanish in the limit $\epsilon \to 0$'. It is easily seen that in the bracket $(2k-p_1)_\mu$ one can drop the term, proportional to $p_1$, as the integral multiplying this term vanishes by symmetry in the limit $\epsilon\to 0$. In this limit the integral is indistinguishable from
\bear\nonumber
\!\!\!\!\!\!\!\!\!\! I_{a'}^{(1)}\simeq \!\!\!\!\!\!\!\!\!\!&&{1\over 4}\,\int\,\frac{d^2 k}{(2\pi)^2} \,e^{ik\epsilon}\,\frac{\mathcal{A}_{\mu}(p_1)(2k+p_2)_\mu\,(2k+p_2)_\nu \mathcal{A}_{\nu}(p_2)}{((k+p_2)^2+m^2)(k^2+m^2)}+(p_1\leftrightarrow p_2)=\\ \label{Id1}
&&={1\over 4}\,\tau_{z\bar{z}}(p_2|\epsilon)\,(\mathcal{A}_{z}(p_1)\mathcal{A}_{\bar{z}}(p_2)+\mathcal{A}_{\bar{z}}(p_1)\mathcal{A}_{z}(p_2))+\\ \nonumber &&+ {1\over 4}\,\tau_{\bar{z}\bar{z}}(p_2|\epsilon)\,\mathcal{A}_{z}(p_1)\mathcal{A}_{z}(p_2)+
{1\over 4}\,\tau_{zz}(p_2|\epsilon)\,\mathcal{A}_{\bar{z}}(p_1)\mathcal{A}_{\bar{z}}(p_2)+(p_1\leftrightarrow p_2)\,,
\eear
where we have introduced the notation
\bea
\tau_{\mu\nu}(p|\epsilon):=\int\,\frac{d^2 k}{(2\pi)^2} \,e^{ik\epsilon}\,\frac{(2k+p)_\mu\,(2k+p)_\nu }{((k+p)^2+m^2)(k^2+m^2)}\,.
\eea
Apart from the contribution $I_{a'}^{(1)}$, there will be a contribution $I_{a'}^{(2)}$, where instead of cancelling the propagator $1\over k^2+m^2$ one cancels the propagator $1\over (k+p)^2+m^2$. The only difference in the resulting expression will be that in the argument of the function $\tau$ one will have to replace $p_1\to p_2$.

\vspace{0.3cm}
We now pass to the calculation of diagrams $(b')$ and $(c')$. Their sum is
\bear\label{bcdiags}
&&I_{b'}+I_{c'}\simeq\\ \nonumber&&\simeq
-\int\,\frac{d^2 k}{(2\pi)^2} \,e^{ik\epsilon}\,\frac{\mathcal{A}_{\bar{z}}(p_1) k_z\,(2k+p_2)_\nu \mathcal{A}_{\nu}(p_2)+(2k+p_2)_\nu \mathcal{A}_{\nu}(p_2) \mathcal{A}_{z}(p_1) k_{\bar{z}}}{((k+p_2)^2+m^2)(k^2+m^2)}+\\ \nonumber&&+\;(p_1\leftrightarrow p_2)\simeq
-{1\over 2}\tau_{z\bar{z}}(p_2|\epsilon)\left(\mathcal{A}_{\bar{z}}(p_1) \mathcal{A}_{z}(p_2)+\mathcal{A}_{\bar{z}}(p_2) \mathcal{A}_{z}(p_1) \right)-\\ \nonumber&&-{1\over 2}\tau_{\bar{z}\bar{z}}(p_2|\epsilon) \mathcal{A}_{z}(p_2)\mathcal{A}_{z}(p_1)-{1\over 2}\tau_{zz}(p_2|\epsilon) \mathcal{A}_{\bar{z}}(p_1)\mathcal{A}_{\bar{z}}(p_2)+(p_1\leftrightarrow p_2)
\eear
Altogether (\ref{Id1}), a similar expression with the replacement $\tau(p_2)\to \tau(p_1)$ (which we call $I_{a'}^{(2)}$) and~(\ref{bcdiags}) give the following result:
\bear\label{abcsum}
&& I_{a'}^{(1)}+I_{a'}^{(2)}+I_{b'}+I_{c'}=\\ \nonumber&&={1\over 4}\,\tau_{z\bar{z}}(p_2|\epsilon)\,([\mathcal{A}_{z}(p_1),\mathcal{A}_{\bar{z}}(p_2)]+[\mathcal{A}_{z}(p_2),\mathcal{A}_{\bar{z}}(p_1)])+\\ \nonumber &&+ {1\over 4}\,\tau_{\bar{z}\bar{z}}(p_2|\epsilon)\,[\mathcal{A}_{z}(p_1),\mathcal{A}_{z}(p_2)]+
{1\over 4}\,\tau_{zz}(p_2|\epsilon)\,[\mathcal{A}_{\bar{z}}(p_2),\mathcal{A}_{\bar{z}}(p_1)]+(p_1\leftrightarrow p_2)=\\ \nonumber
&& ={1\over 4}\,(\tau_{z\bar{z}}(p_2|\epsilon)+\tau_{z\bar{z}}(p_1|\epsilon))\,([\mathcal{A}_{z}(p_1),\mathcal{A}_{\bar{z}}(p_2)]+[\mathcal{A}_{z}(p_2),\mathcal{A}_{\bar{z}}(p_1)])+\\ \nonumber
&&+ {1\over 4}\,(\tau_{\bar{z}\bar{z}}(p_2|\epsilon)-\tau_{\bar{z}\bar{z}}(p_1|\epsilon))\,[\mathcal{A}_{z}(p_1),\mathcal{A}_{z}(p_2)]+\\ \nonumber
&&+{1\over 4}\,(\tau_{zz}(p_2|\epsilon)-\tau_{zz}(p_1|\epsilon))\,[\mathcal{A}_{\bar{z}}(p_2),\mathcal{A}_{\bar{z}}(p_1)]
\eear

We will now relate the functions $\tau_{\mu\nu}$ to the effective action $\Gamma_{\mu\nu}$, which is defined as the limit $M\to \infty$ of the following integral:
\bear\nonumber
&&\Gamma_{\mu\nu}^{(M^2)}=-\frac{1}{2}
 \int\frac{d^2k}{(2\pi)^2}\biggl(\frac{(2k+p)_\mu(2k+p)_\nu}{(k^2+m^2)((k+p)^2+m^2)}-(m^2\mapsto m^2+M^2)\biggr)+{}
\\ \label{gammamunu}
 &&\quad\quad\quad\quad+\,\delta_{\mu\nu}\int\frac{d^2k}{(2\pi)^2}\biggl(\frac{1}{k^2+m^2}-(m^2\mapsto m^2+M^2)\biggr).
\eear
Its $(z\bar{z})$-component may be equivalently rewritten as
\bea\label{GammaMM}
\Gamma_{z\bar{z}}^{(M^2)}\!=\!-\frac{1}{2}
 \int\frac{d^2k}{(2\pi)^2}\biggl(\frac{(2k+p)_z(2k+p)_{\bar{z}}}{(k^2+m^2)((k+p)^2+m^2)}-2\, \frac{1}{k^2+m^2}\biggr)-(m^2\to m^2+M^2).
\eea
Note that the last term, in the limit $M\to\infty$, is equal to
$
{\int\frac{d^2k}{(2\pi)^2}\,\frac{1}{(k^2+1)^2}={1\over 4\pi}}
$. 
Introducing momentarily a factor $e^{ik \epsilon}$ in the (convergent) integral in~(\ref{GammaMM}), recalling the definition of $\tau_{z\bar{z}}$ and taking the limit $M\to \infty$, we may rewrite (\ref{GammaMM})  in the following way:
\bea
\Gamma_{z\bar{z}}+{1\over 4\pi}=-{1\over 2}\,\underset{\epsilon\to 0}{\mathrm{lim}}\,(\tau_{z\bar{z}}(p|\epsilon)-2 D(\epsilon))
\eea
In other words, as $\epsilon$ goes to 0, we have the asymptotic behavior
\bea\label{tauzbarz}
\tau_{z\bar{z}}(p|\epsilon)=2 D(\epsilon)-2\left(\Gamma_{z\bar{z}}+{1\over 4\pi}\right)+\ldots
\eea
Analogously one can show that
\bea\label{tauzz}
\tau_{zz}(p|\epsilon)=-2\Gamma_{zz}-{1\over 2\pi}\,\frac{\bar{\epsilon}}{\epsilon}+\ldots
\eea
(For completeness we prove this relation in the Appendix~\ref{onemore}.)

\vspace{0.3cm}
Substituting the above relations in~(\ref{abcsum}), we arrive at the expression
\bear\label{abcsum}
&&\!\!\!\!\!\!\!\!\!\! I_{a'}^{(1)}+I_{a'}^{(2)}+I_{b'}+I_{c'}=\\ \nonumber
&&\!\!\!\!\!\!\!\!\!\! =\left(D(\epsilon)-{1\over 4\pi}\right)\,([\mathcal{A}_{z}(p_1),\mathcal{A}_{\bar{z}}(p_2)]+[\mathcal{A}_{z}(p_2),\mathcal{A}_{\bar{z}}(p_1)])-\\ \nonumber
&&\!\!\!\!\!\!\!\!\!\!-{1\over 2}\,(\Gamma_{z\bar{z}}(p_2)+\Gamma_{z\bar{z}}(p_1))\,([\mathcal{A}_{z}(p_1),\mathcal{A}_{\bar{z}}(p_2)]+[\mathcal{A}_{z}(p_2),\mathcal{A}_{\bar{z}}(p_1)])+\\ \nonumber
&&\!\!\!\!\!\!\!\!\!\!+ {1\over 2}\,(\Gamma_{\bar{z}\bar{z}}(p_1)-\Gamma_{\bar{z}\bar{z}}(p_2))\,[\mathcal{A}_{z}(p_1),\mathcal{A}_{z}(p_2)]+{1\over 2}\,(\Gamma_{zz}(p_1)-\Gamma_{zz}(p_2))\,[\mathcal{A}_{\bar{z}}(p_2),\mathcal{A}_{\bar{z}}(p_1)]
\eear
We will now argue that, just as we did earlier, the parts proportional to the effective action (i.e. the last two lines) may in fact be dropped.
First we rewrite them as
\bear\label{gammmup1part}
-{1\over 2}\,\Gamma_{z\mu}(p_1) [\mathcal{A}_{\mu}(p_1),\mathcal{A}_{\bar{z}}(p_2)]-{1\over 2}\,\Gamma_{\mu\bar{z}}(p_1)\,[\mathcal{A}_{z}(p_2),\mathcal{A}_{\mu}(p_1)]+(p_1\leftrightarrow p_2)
\eear
If one takes for $\mathcal{A}_{\mu}(p_1)$ the diagonal (gauge) part, say $a_\mu(p_1)$, and computes a correlation function with the field strength operator $(p_1)_\mu a_\nu(p_1)-(p_1)_\nu a_\mu(p_1)$, the effective action part will fully cancel, leaving contributions which in coordinate space are again local (w.r.t. the point where the field strength operator is inserted). Such contributions can be dropped according to the logic explained earlier. On the other hand, if $\mathcal{A}_{\mu}(p_1)$ is one of the off-diagonal `matter' fields, say $c_{\bar{z}}(p_1)$, and one computes a correlation function with the insertion of an operator $c_z$, there will be an additional non-local piece proportional (in momentum space) to ${\Gamma_{zz}(p)\over \Gamma_{z\bar{z}}(p)}=-{p_z\over p_{\bar{z}}}$. In the next subsection we will see, however, that in this case there is an additional contribution from the `residual part' in~(\ref{k2decomp}) which completely cancels out the $\Gamma_{zz}$-piece in the formula~(\ref{gammmup1part}).

\vspace{0.3cm}
Anyway~(\ref{abcsum}) is not yet the final answer, as we still have to analyze the diagrams~$(d')$ and~$(e')$. The integrand of the diagram~$(d')$ has in the numerator the product $k_{\bar{z}}(k_z+p_z)$, for which we can use the decomposition~(\ref{k2decomp}). The first two terms can again be used to cancel the propagators, which results in the following contribution (the residual part is again left for a later analysis):
\bear
I_{d'}^{(1)}+I_{d'}^{(2)}&&\!\!\!\!\!\!\!\!=-{1\over 2} D(\epsilon)\,A_\mu(p_1)A_\mu(p_2)+(p_1\leftrightarrow p_2)=\\ \nonumber &&\!\!\!\!\!\!\!\!=-{1\over 2} D(\epsilon)\,\left(A_{\bar{z}}(p_1)A_z(p_2)+A_{z}(p_1)A_{\bar{z}}(p_2)\right)+(p_1\leftrightarrow p_2)
\eear
The diagram $(e')$ gives $I_{e'}=D(\epsilon) A_{\bar{z}}(p_1) A_z(p_2)+(p_1\leftrightarrow p_2)$, so that the sum of two diagrams is
\bear\nonumber
&&I_{d'}^{(1)}+I_{d'}^{(2)}+I_{e'}={1\over 2} D(\epsilon)\,\left(A_{\bar{z}}(p_1)A_z(p_2)-A_{z}(p_1)A_{\bar{z}}(p_2)\right)+(p_1\leftrightarrow p_2)=\\ \label{sum2}
&&\quad\quad\quad\quad\quad\quad\quad\,=-{1\over 2} D(\epsilon)\,\left( [ A_z(p_1), A_{\bar{z}}(p_2)]+[ A_z(p_2), A_{\bar{z}}(p_1)]\right)
\eear
Together, (\ref{abcsum}) and (\ref{sum2}) amount to (after dropping the $\Gamma$-terms)
\bear\nonumber
&&I_{a'}^{(1)}+I_{a'}^{(2)}+I_{b'}+I_{c'}+I_{d'}^{(1)}+I_{d'}^{(2)}+I_{e'}\sim\\ \nonumber&&\quad\quad\quad\quad\quad\sim\left({1\over 2}\,D(\epsilon)-{1\over 4\pi}\right) \left([\mathcal{A}_{z}(p_1),\mathcal{A}_{\bar{z}}(p_2)]+[\mathcal{A}_{z}(p_2),\mathcal{A}_{\bar{z}}(p_1)] \right)
\eear
Comparing with the leading order OPE~(\ref{leadordOPE}), we see that the quadratic terms in $\mathcal{A}$ simply amount to completing the linearized field strength to the full non-abelian field strength $F_{z\bar{z}}=\dd_z \mathcal{A}_{\bar{z}}-\dd_{\bar{z}}\mathcal{A}_z+i\,[\mathcal{A}_z, \mathcal{A}_{\bar{z}}]$. 

\subsubsection{The residual part}\label{respartsec}

Finally we need to analyze the contributions that the residual part of the decomposition~(\ref{k2decomp}) makes to the values of the diagrams~($a'$) and ($d'$). Our claim is that these contributions cancel partially against the diagrams obtained by the procedure described in Fig.~\ref{OPEquadr2}, and partially against the $\Gamma$-terms in~(\ref{abcsum}), producing contributions that are already included in the linear terms~(\ref{leadordOPE}) of the OPE. To start with, it is clear that the Feynman integrals corresponding to the diagrams in Fig.~\ref{OPEquadr2} are essentially products
\bea\nonumber
\textrm{Lower diagram}\times \textrm{Connecting propagator}\times \textrm{Upper diagram}\,.
\eea
Since the lower diagrams are the diagrams of Fig.~\ref{OPEfig}, we already know the corresponding expressions: these are given by~(\ref{anomint2}) and~(\ref{anomint4}). If we dropped in the latter expressions the parts with the effective action $\Gamma$, we would obtain exactly the correlation functions of the r.h.s. of~(\ref{leadordOPE}) with two additional $\mathcal{A}$-operators. Therefore, up to the $\Gamma$-terms, the values of the diagrams in Fig.~\ref{OPEquadr2} are already included in the linear terms of the OPE~(\ref{leadordOPE}). As for the $\Gamma$-parts of the diagrams on Fig.~\ref{OPEquadr2}, they cancel with the `residual parts' of the diagrams $(a')$ and $(d')$, as well as with certain $\Gamma$-terms in~(\ref{abcsum}). The mechanism is as follows: the functions $\Gamma_\lambda(p)$ and $\Gamma_{z\bar{z}}(p)$ cancel against the corresponding intermediate propagators of Fig.~\ref{OPEquadr2}, leaving truncated diagrams whose associated Feynman integrals  cancel against the `residual parts' of the diagrams $(a')$ and $(d')$, possibly up to contact terms.

\vspace{0.3cm}
Let us elaborate this phenomenon at the example of the rightmost two diagrams in Fig.~\ref{OPEquadr2}. We will be using the Feynman rules derived in Section~\ref{1Nsect}.

\vspace{0.3cm}
\textbf{Diagram $(c'')$.} The $\Gamma$-dependent part of the short-distance expansion of the lower part of the diagram is given by~(\ref{anomint2}) -- it is equal to $(p_{\bar{z}}a_z-p_z a_{\bar{z}})\,\Gamma_{z\bar{z}}$, where $a$ is the $U(1)$ gauge field corresponding to the blue line. Multiplying by the propagators of the gauge fields, we see that the coupling to the upper part of the diagram is given by $(p_{\bar{z}}D_{z\bar{z}} (2k-p)_z+p_{\bar{z}}D_{zz} (2k-p)_{\bar{z}} -p_z D_{z\bar{z}} (2k-p)_{\bar{z}}-p_z D_{\bar{z}\bar{z}} (2k-p)_{z})\,\Gamma_{z\bar{z}}$. Recalling that $D_{z\bar{z}}={1\over 8} \Gamma_{z\bar{z}}^{-1}$ and $D_{zz}=-{1\over 8}\frac{p_z}{p_{\bar{z}}}\Gamma_{z\bar{z}}^{-1}$, we get 
${1\over 2}(p_{\bar{z}} k_z-p_z k_{\bar{z}})$. This cancels exactly against the analogous contribution in~(\ref{k2decomp}).

\vspace{0.3cm}
\textbf{Diagram $(d'')$.} Just like above, we read off the $\Gamma$-dependent part of the short-distance expansion of the lower part of the diagram from~(\ref{anomint2}) -- in this case it is equal to $-p_{z}c_{\bar{z}}\,\Gamma_{z\bar{z}}$, where $c_{\bar{z}}$ is the off-diagonal auxiliary field in~(\ref{Atriang}). We now supply the propagator of the $c$-field, which is $D_c={1\over 2}\Gamma_{z\bar{z}}^{-1}$ and the coupling to the upper vertex, which is $(2k-p)_{\bar{z}}$. Altogether this gives $-{1\over 2}p_z(2k-p)_{\bar{z}}$. This does not completely cancel the contribution of~(\ref{k2decomp}), the sum of the two terms being instead  ${1\over 4}(p^2-2kp)={1\over 4}((k-p)^2+m^2)-{1\over 4}(k^2+m^2)$. This allows canceling the propagators in the upper diagram, leading to the following expression:
\bear\nonumber
&&{1\over 4}\int\,\frac{d^2k}{(2\pi)^2}\,\frac{(2k-p+p')_\mu\,(2k+p')_z}{(k+p')^2+m^2}\left(\frac{1}{k^2+m^2}-\frac{1}{(k-p)^2+m^2}\right)=\\ \nonumber
&&={1\over 2}\,(\Gamma_{\mu z}(p+p')-\Gamma_{\mu z}(p'))
\eear
To obtain the latter expression, one needs to subtract the analogous integral with the replacement $m^2\to m^2+M^2$ (which in itself vanishes in the limit $M\to\infty $),  split the resulting integral into two parts, use the definition~(\ref{gammamunu}) of $\Gamma_{\mu\nu}^{(M^2)}$ and then take the limit $M\to \infty$. Now, the above expression is to be multiplied by $a_\mu(p+p') c_{\bar{z}}(p')$ and it has to be added to the terms~(\ref{gammmup1part}) obtained from the diagram $(a')$ of Fig.~\ref{OPEquadr}. In~(\ref{gammmup1part}) we will assume that $\mathcal{A}_\mu(p_1)$ corresponds to the field $c_{\bar{z}}(p_1)$ (hence $p_1=p'$) and $\mathcal{A}_\mu(p_2)$ corresponds to $a_\mu(p+p')$ (hence $p_2=p+p'$). Adding the above expression to~(\ref{gammmup1part}) and restricting the fields as described, we get 
\bear\nonumber
&&{1\over 2}\,(\Gamma_{\mu z}(p_2)-\Gamma_{\mu z}(p_1))a_\mu(p_2) c_{\bar{z}}(p_1)+\\ \nonumber &&+{1\over 2}\,\Gamma_{zz}(p_1) c_{\bar{z}}(p_1)a _{\bar{z}}(p_2)-{1\over 2}\,\Gamma_{z\bar{z}}(p_1)\,a_{z}(p_2)c_{\bar{z}}(p_1)-{1\over 2}\,\Gamma_{z\mu}(p_2) a_{\mu}(p_2) c_{\bar{z}}(p_1)=\\ \nonumber
&& =-\Gamma_{z\bar{z}}(p_1)a_z(p_2) c_{\bar{z}}(p_1)
\eear
This is again a contact term: computing a correlation function with an external $c_z$-operator produces a local contribution, which may be dropped.

\subsection{The anomaly}\label{anomalysection}

We are now ready to collect the results that concern the OPE's of the Noether currents taken at two nearby points obtained in the previous section and calculate the anomaly two-form. To this end, we recall the formulas~(\ref{anomdef}) (the definition of the anomaly) and (\ref{OPEleadfull}) (the OPE):
\bear\nonumber
\mathcal{Q}_\epsilon(\Gamma_1)-\mathcal{Q}_\epsilon(\Gamma_2)
=&&\!\!\!\!\!\!\!\!\!\! \int\limits_{D_{\delta \Gamma}}\,dz\wedge d\bar{z}\,\left[\left(\mathrm{Re}(\mathbf{a})+{1\over 2\pi} \log{|\epsilon|}\right)\,(\dd_z K_{\bar{z}}-\dd_{\bar{z}}K_z)\,-\right. \\ \nonumber && \!\!\!\!\!\!\!\!\!\!\left.-\,{1\over 2 g^2}\left([K_z(p+\epsilon), K_{\bar{z}}(p)]+[K_z(p), K_{\bar{z}}(p+\epsilon)]\right)\right]\,.
\eear
\bea \nonumber
\left[K_z(z_1),  K_{\bar{z}}(z_2)\right]^{ij}\underset{z_2\to z_1}{\to}g^2\left(\left[\dd_{\bar{z}} K_z-\dd_z K_{\bar{z}}\right]\,D(\epsilon)+{g^2\over \pi N}\,\varphi \, i F_{z\bar{z}}\,\varphi^\dagger\right)^{ij}_{z_1}+\ldots
\eea
Since, according to (\ref{Gpropexp}), $D(\epsilon)= -{1\over 4\pi} \log{\left(m^2 |\epsilon|^2\over 2\right)}-{\gamma\over 2\pi}+\ldots$, we will set $\mathrm{Re}(\mathbf{a})={1\over 4\pi} \log{\left(m^2 \over 2\right)}+{\gamma\over 2\pi}$. Then in the limit $\epsilon \to 0$ we obtain the following result:
\bear\label{anomanswer}
&&\mathcal{Q}_\epsilon(\Gamma_1)-\mathcal{Q}_\epsilon(\Gamma_2)
=\int\limits_{D_{\delta \Gamma}}\,dz\wedge d\bar{z}\;\;\colon\!\!-{g^4\over \pi N}\,\varphi \, i F_{z\bar{z}}\,\varphi^\dagger\colon,
\\ \nonumber &&
\textrm{where}\quad\quad F_{z\bar{z}}=\dd_z \mathcal{A}_{\bar{z}}-\dd_{\bar{z}}\mathcal{A}_z+i\,[\mathcal{A}_z, \mathcal{A}_{\bar{z}}]\,,
\eear
and the `gauge field' $\mathcal{A}$ is in general of the restricted form (\ref{gaugef}) (for the particular case of the flag manifold~(\ref{simpletarget}) it is of the form~(\ref{Atriang})). Note that formally all our calculations were performed for the off-diagonal matrix elements ($i\neq j$), therefore since $\mathcal{Q}_\epsilon\in \mathfrak{su}_N$, one would need to subtract the trace-part of the respective expressions (which is proportional to the identity matrix $\mathds{1}_N$). Instead, we have inserted the normal ordering in the r.h.s. of~(\ref{anomanswer}), as in this case the trace-part vanishes due to the property $:\varphi^\dagger(z)  \varphi(z):=0$ discussed earlier in Section~\ref{holantiholcomm} (see Fig.~\ref{lambdacancel}).

\section{Conclusion and outlook}

In the present paper we investigated the classical and quantum properties of the flag manifold sigma-models, introduced in our work~\cite{BykovZeroCurv, BykovFlag1}. In fact, the models of~\cite{BykovZeroCurv} constitute a slightly broader class, allowing for complex homogeneous target spaces, however the subclass of flag manifold models is representative and interesting enough for an elaborate study. As we showed in Section~\ref{gradedsec}, models of this class may also be understood in terms of the theory of sigma-models with $\mathbb{Z}_m$-symmetric target spaces considered in~\cite{Young}.  Throughout the paper for simplicity we restricted to the case when the isometry group is $SU(N)$, although we expect that similar analysis might be carried out for other compact semi-simple groups. With the aim of constructing the $1\over N$-expansion for the flag manifold sigma-models, we first reformulated these models as gauged linear sigma-models in Section~\ref{GLSMsect}. The construction is non-trivial, as the target spaces in question are not K\"ahler. Nevertheless, formally the result is akin to the GLSM-formulation for Grassmannian (K\"ahler) target spaces, albeit with a `gauge field' of a special (restricted) form. In the next section, using this formulation, we derived the Feynman rules for the simplest non-symmetric-space model. Subsequently we introduced a regularized version of the non-local charge, which is different from the original definition of L\"uscher~\cite{Luscher}. As opposed to the definition in~\cite{Luscher}, our version of the regularized charge depends on the integration contour $\Gamma$ through an anomaly two-form and does not depend on the parametrization of the contour. We compute this anomaly two-form explicitly in Sections~\ref{OPEsect1}, \ref{anomalysection}. It formally coincides with the anomaly for the Grassmannian models~\cite{AbdallaGeneral}, but the gauge fields entering the anomaly should be taken in restricted form, of the type mentioned earlier. It is an important and interesting question, how the anomaly may be cancelled. Most likely this can be done by introducing fermions, i.e. by a mechanism similar to the one of~\cite{AbdallaFermions, AbdallaCancel}. The investigation of the peculiarities of this mechanism in application to the present models will be a subject of future work.

\vspace{0.5cm}
\noindent\textbf{Acknowledgements.}
{\footnotesize
I would like to thank I.~Aref'eva, S.~Frolov, E.~Ivanov, S.~Ketov, O.~Lechtenfeld, A.~Maltsev, M.~Semenov-Tian-Shansky, A.~Tseytlin, K.~Zarembo, P.~Zinn-Justin
 for valuable comments. I am indebted to Prof.~A.A.Slavnov and to my parents for support and encouragement. I would like to thank the Institut des Hautes \'Etudes Scientifiques, where part of this work was done, and in particular V.~Pestun for hospitality.}

\vspace{0.3cm}
\appendix
\begin{center}
{\Large Appendix}
\end{center}

\section{Flag manifolds and elements of representation theory}\label{appBWB}

This Appendix lies somewhat outside the main line of exposition in the present paper. Yet we have included it to demonstrate, how flag manifolds arise in a well-known physical situation. Incidentally this makes a neat connection to the applications of flag manifolds in representation theory, discussed below in Section~\ref{appQuant}.

\vspace{0.3cm}
It is well-known, how one can describe a classical particle, interacting with an external electromagnetic field $A_\mu$. The action has the form
\bea
\mathcal{S}=\int\,dt\,\frac{g_{\mu\nu} \dot{x}^\mu \dot{x}^\nu}{2}-\int\,A=\int\,dt\,\left(\frac{g_{\mu\nu} \dot{x}^\mu \dot{x}^\nu}{2}-A_\mu \dot{x}^\mu\right)\,.
\eea
The question is, how to write an analogous action for the case when the gauge field is non-abelian, or, simply speaking, when it has additional gauge indices $A_\mu^{ab}$. The answer is that the particle should possess additional degrees of freedom, taking values in a certain flag manifold, corresponding to the representation, in which the particle transforms. In other words, one should enlarge the phase space as follows~\cite{Sternberg} (here $\mathcal{M}$ is the configuration space):
\bea\label{Phasespace}
T^\ast\mathcal{M}\to T^\ast\mathcal{M}\times \mathcal{F}\;(\textrm{the flag manifold})\,.
\eea
We start by rewriting the standard action of a particle in first-order form:
\bea
\mathcal{S}=\int\,dt\,\left(p_\mu\dot{x}^\mu-\frac{g^{\mu\nu}p_\mu p_\nu}{2}-A_\mu \dot{x}^\mu\right)
\eea
Upon enlarging the phase space we can analogously write down the non-abelian action as follows ($A$ is assumed Hermitian):
\bea\label{flagaction1}
\mathcal{S}=\int\,p_\mu\,dx^\mu-\int\,dt\,\mathcal{H}(x, p)+\int\,\left(\theta-\tr(A\,\mu) \right)\,,
\eea
where $\theta$ is the canonical one-form, defined by the condition
\bea
d\theta=\Omega\quad\quad (=\textrm{the symplectic form on}\;\;\mathcal{F})\,,
\eea
and $\mu$ is the moment map for the action of the group $G$ on $\mathcal{F}$. We note that the form $\theta$ is only defined up to the addition of a total derivative, $\theta\to\theta+dh$, but the difference only affects the boundary terms in the action. In the case of periodic boundary conditions one may even write
\bea
\int\limits_{\Gamma}\,\theta=\int\limits_D\,\Omega,
\eea
where $D$ is a disc, whose boundary is the curve $\Gamma$: $\dd D=\Gamma$. In fact this term is nothing but the one-dimensional version of the Wess-Zumino-Novikov-Witten term~\cite{WZW, NovCurr, WittenCurr}.

\vspace{0.3cm}
One needs to show that the expression so obtained is gauge-invariant. For simplicity let us consider the standard representation of  $SU(N)$ (when the representation space is $\CC^N$): in this case the relevant flag manifold is the projective space $\CP^{N-1}$. Let us normalize the homogeneous coordinates $(z_1, \ldots, z_N)$ on $\CP^{N-1}$:
\bea\label{zknorm}
\sum\limits_{k=1}^N\,|z_k|^2=1\,.
\eea
One still has the remaining gauge group $U(1)$, which acts by multiplication of all coordinates $z_k$ by a common phase. The Fubini-Study form on $\CP^{N-1}$, when written in homogeneous coordinates, looks as follows:
\bea
\omega_{FS}=\frac{i}{\|z\|^2}\,\left(dz_k\wedge d\bar{z}_k-\frac{\bar{z}_n d z_n\wedge z_m d\bar{z}_m }{\|z\|^2}\right)\,,
\eea
but it may be simplified, if one uses the above normalization:
\bea
\omega_{FS}=i\,dz_k\wedge d\bar{z}_k\,.
\eea
Then we have the following expressions for $\theta$ and $\mu$:
\bea
\theta=i\,z_k\,d\bar{z}_k,\quad\quad \mu=z\otimes \bar{z}-\frac{1}{N}\,\mathds{1}_N\,.
\eea
The part of the action, corresponding to the motion in the `internal' space (in this case the projective space), has the form
\bear
\tilde{\mathcal{S}}=\int\,dt\,\left(i\,z_k\,\dot{\bar{z}}_k-\bar{z}_m\,(A_\mu)_{mn}\,\dot{x}^\mu z_n\right)=\\
=-\int\,dt\,\bar{z}_m (i\,\dot{z}_m+(A_{\mu})_{mn}\,\dot{x}^\mu\,z_n)\,,
\eear
and one should take into account that the normalization condition (\ref{zknorm}) is also implied. From the second form of the action it is evident that it is gauge-invariant w.r.t. the transformations
\bea
z\to g(x(t))\circ z\,\quad\quad A_\mu\to g A_\mu g^{-1}-i\,\dd_\mu g\,g^{-1}\,.
\eea
To make it even more obvious, we note that the exterior derivative of the one-form $\theta-\tr(A\,\mu)$ (viewed as a form on the enlarged phase space~(\ref{Phasespace})) produces a two-form, which is explicitly gauge-invariant:
\bear\label{gaugeinv2form}
&&d(\theta-\tr(A\,\mu))=i\,\mathscr{D}z_k\wedge \mathscr{D}\bar{z}_k-\tr(F\,\mu)\,,\\ \nonumber
&&\mathscr{D}z=dz-i\,A\,z,\quad\quad \mathscr{D}\bar{z}=d\bar{z}+i\,\bar{z}\,A,\quad\quad 
F=dA-i A\wedge A\,.
\eear
Each of the two terms in (\ref{gaugeinv2form}) is separately gauge-invariant, however~(\ref{gaugeinv2form}) is the only linear combination of them, which is closed (and therefore locally is an exterior derivative of a one-form).

\vspace{0.3cm}
Next we wish to write out the equations of motion on the flag manifold, which follow from the action above. Let us concentrate for simplicity on the $SU(2)$-case, where the sphere $S^2$ plays the role of a flag manifold. Instead of using the spinor $(z_1, z_2)$, we can parametrize it in a more standard way, with the help of a unit vector $\vec{n}\in \mathbb{R}^3$. The equations take the form
\bea
\dot{\vec{n}}=\vec{A}\times \vec{n},\quad\quad \textrm{where}\quad\quad \vec{A}=\{A_\mu^a\,\dot{x}^\mu\}_{a=1, 2, 3}
\eea
is a vector of components of the gauge field in the basis of Pauli matrices. We see that the equations are \emph{linear} in $\vec{n}$, and the condition
\bea
\vec{n}^2=\mathrm{const.}
\eea
itself is a consequence of the equations, i.e. the motion takes place on a sphere in $\mathbb{R}^3$. This is a general fact. Indeed, in the case of a general compact simple Lie algebra $\mathfrak{g}$ with basis $\{\tau_i\}$ we can introduce a variable $n=\sum \,x^i\,\tau_i\in \mathfrak{g}$, and the equations will then take the form
\bea\label{adjmotion}
\dot{n}=[A_\mu \dot{x}^\mu, n]\,,
\eea
or, in terms of the variables $x^i$,
\bea
\dot{x}^i=f_{jk}^i\,(A_\mu \dot{x}^\mu)^j\,x^k\,.
\eea
It is in this form that this system of equations was discovered in~\cite{Wong}. The motion defined by these equations in reality takes place on flag manifolds embedded in $\mathfrak{g}$, since the  `Casimirs'
\bea
C_J=\tr(n^J),\quad\quad J=1, 2, \ldots
\eea
are integrals of motion of the system (\ref{adjmotion}). We have thus established a connection with the formulation through flag manifolds used earlier.

\subsection{`Quantization' of the symplectic form on flag manifolds}\label{appQuant}

One of the approaches to quantization is related to considering path integrals of the following form\footnote{Another approach to the quantization of coadjoint orbits, which is also based on the path integral, was developed in~\cite{AFS}.}:
\bea\label{pathint}
\int\,\prod\limits_i\,d\varphi_i\,e^{i\,\mathcal{S}}
\eea
where  the exponent contains the action (\ref{flagaction1}). The connection $\theta$ is not a globally-defined one-form on the flag manifold. Indeed, let us consider the simplest case of $\mathcal{F}=\CP^1=S^2$. The most general invariant symplectic form is as follows:
\bea
\omega=\frac{a}{2}\,\sin{\vartheta}\, d\vartheta\wedge d\phi
\eea
with an arbitrary constant $a$. It can be also written in the form $\omega=-\frac{a}{2}\, dz\wedge d\phi$, where $z=\cos{\vartheta}$ is the  $z$-coordinate of a given point on the sphere. Since the latter form is nothing but the area element of a cylinder, it implies that the projection of a sphere to the cylinder preserves the area. Since the action $\mathcal{S}$ entering the exponent in (\ref{pathint}) involves a term $\int \,\theta$, where $\theta$ is a connection satisfying $d\theta=\omega$, standard arguments familiar from Wess-Zumino-Novikov-Witten theory~\cite{WZW, NovCurr, WittenCurr} lead to the requirement that the coefficient $a$ is quantized according to $\int \omega\in 2\pi n, n\in \mathbb{Z}$. By analogy, in the case when the manifold $\mathcal{F}$ has several non-trivial basic cycles $C_1, \ldots, C_{N-1}$,  in order for the exponent $e^{i\,\int\limits_\Gamma\, \theta}$ to be well-defined for any contour  $\Gamma$, the integral of the symplectic form $\Omega$ over any 2-cycle $C_i$ (which in the case of a flag manifold is always a sphere $S^2$) should be quantized:
\bea
\int\limits_{C_i}\,\Omega \in 2\pi \mathbb{Z}\quad\quad\textrm{for every 2-cycle}\quad C_i\in H_2(\mathcal{F}, \mathbb{Z})\,.
\eea
Let us construct these 2-cycles explicitly for the case when $\mathcal{F}$ is a complete flag manifold
\bea\label{complflagapp}
\mathcal{F}({1,\ldots, N-1, N})=\{0\subset \CC \subset \CC^2 \subset \ldots \subset \CC^N\}=\frac{U(N)}{U(1)^N}\,.
\eea
It can be parametrized using $N$ orthonormal vectors $u_i, i=1\ldots N$, $u_i\circ \bar{u}_j=\delta_{ij}$, defined modulo phase transformations: $u_k\sim e^{i\alpha_k}\,u_k$. As we showed in Section~\ref{symplman}, the most general symplectic form on  $\mathcal{F}({1,\ldots, N-1, N})$ may be written as follows:
\bear\label{flagsymplform2}
&&\Omega=i\,\sum\limits_{i<j}\,a_{ij}\,J_{ij}\wedge J_{ji},\quad\quad\textrm{where}\quad\quad J_{ij}=u_i\circ d\bar{u}_j\\
&& \textrm{and}\quad\quad a_{ij}=z_i-z_j\,.
\eear
If one fixes $N-2$ out of $N$ lines defined by the vectors $u_1, \ldots, u_N$, the remaining free parameters define the configuration space of ordered pairs of mutually orthogonal lines, passing through the origin and laying in a plane, orthogonal to the $N-2$ fixed lines. This configuration space is nothing but the sphere $\CP^1$:
\bear\nonumber
&&\{u_{i_1}, \ldots , u_{i_{N-2}}\quad\textrm{are fixed},\quad\quad u_{i_{N-1}}, u_{i_N}\in (u_{i_1}, \ldots , u_{i_{N-2}})^\perp \quad \\ \nonumber&&\textrm{are mutually orthogonal and otherwise generic}\}\simeq (\CP^1)_{i_{N-1}, i_{N}}\,.
\eear

\vspace{0.3cm} 
Let us now fix the permutation $(i_1, \ldots, i_N)$ in such a way that $z_{i_m}$ would form a non-increasing sequence, i.e. $z_{i_m}\geq z_{i_n}$ for $m<n$ (in the case of the complete flag manifold~(\ref{complflagapp}) the sequence should be strictly decreasing, whereas the `non-increasing' case in general corresponds to partial flag manifolds in a natural way). We can rewrite the symplectic form as follows: $\Omega= i\,\sum\limits_{m<n}\,a_{i_mi_n}\,J_{i_mi_n}\wedge J_{i_ni_m}$, and uniquely fix a complex structure, in which the one-forms $J_{i_mi_n}, m<n,$ are holomorphic. After such a permutation we may choose $(\CP^1)_{{i_m},i_{m+1}}$ as a basis in the homology group $H_2(\mathcal{F}_{1,\ldots, N-1, N}, \mathbb{Z})$, with the orientation of the spheres induced by the complex structure. Then the integrals of the symplectic form over these cycles will be positive:
\bea\label{Omegaint}
\int\limits_{(\CP^1)_{{i_m},i_{m+1}}}\,\Omega=z_{i_m}-z_{i_{m+1}}\in 2\pi \mathbb{Z}^+\,,\quad\quad m=1\ldots N-1\,.
\eea

In order for the value of the integral to be an integer, one should choose $z_i$ in the form
\bea
(z_1, \ldots, z_N)=\lambda (1, \ldots, 1)+(\tilde{z}_1, \ldots, \tilde{z}_N)\,, \quad \lambda \in \mathbb{R},\quad \tilde{z}_i\in \mathbb{Z}\,.
\eea
We may view $(\CP^1)_{{i_m},i_{m+1}}$ as the positive simple  roots, and $\Omega$ is, in this case, a dominant weight. Adding to $\{\tilde{z}_i\}$ a vector, proportional to $\lambda (1, \ldots, 1)$, does not change the values of the integrals.
Using this property, we may normalize the values $\{\tilde{z}_i\}$ in such a way that their sum would be zero. According to the general theory of adjoint orbits, the flag manifold under consideration is the orbit of the element
\bea
z=\left( \begin{array}{ccc}
\tilde{z}_1 & 0 & 0 \\
\vdots & \ddots  & \vdots \\
0 & 0 & \tilde{z}_N \end{array} \right)\in \mathfrak{su}_N\,.
\eea
The fundamental weights are, by definition, the ones, whose value on simple positive roots is equal to zero, apart from a single root, on which the fundamental weight has value one: $\langle \Omega_{F_{[j]}}, \lambda_{[i]}\rangle=\delta_{ij}$. These weights correspond to the highest weights of fundamental representations. According to the   theory described above, they correspond to orbits of the elements
\bea
z_{F_{[i]}}=\mathrm{Diag}(\underbracket[0.6pt][0.6ex]{1, \ldots , 1}_{i}, \underbracket[0.6pt][0.6ex]{0, \ldots , 0}_{N-i})\,.
\eea
According to (\ref{Omegaint}), in this case we have
\bea
\int\limits_{(\CP^1)_{k,k+1}}\,\Omega_{F_{[i]}}=\delta_{ik}\,.
\eea
The adjoint orbit in question is the Grassmannian $G_{i, N}$. The general theory that we have described is nothing but `geometric quantization' for the case of flag manifolds.

\section{Eliminating auxiliary fields}\label{app}
In this Appendix we demonstrate explicitly how the auxiliary fields $a, b, c$ may be eliminated from the Lagrangian (\ref{GLSMlagr}) with $m+n=2$ and the gauge field given by~(\ref{Atriang}). To this end, we write the Lagrangian as follows:
\bear
&&\mathcal{L}=\dd_{\bar{z}}\bar{u}\circ \dd_z u+\dd_{\bar{z}}\bar{v}\circ \dd_z v+\\ \nonumber
&&+ {N\over g^2} a_{\bar{z}}a_z+a_{\bar{z}}(i\,\bar{u}\circ \dd_z u)-a_z (i\,\dd_{\bar{z}}\bar{u}\circ u)+\\ \nonumber
&&+ {N\over g^2} b_{\bar{z}}b_z+b_{\bar{z}}(i\,\bar{v}\circ \dd_z v)-b_z (i\,\dd_{\bar{z}}\bar{v}\circ v)+\\ \nonumber
&&+{N\over g^2} c_{\bar{z}}c_z+c_{\bar{z}}(i\,\bar{v}\circ \dd_z u)-c_z (i\,\dd_{\bar{z}}\bar{u}\circ v)
\eear
Here we have taken advantage of the orthonormality relations $\|u\|=\|v\|={\sqrt{N}\over g}$ and $\bar{u}\circ v=0$. Completing the squares, we can obtain relations of the form
\bea
{N\over g^2} a_{\bar{z}}a_z+a_{\bar{z}}(i\,\bar{u}\circ \dd_z u)-a_z (i\,\dd_{\bar{z}}\bar{u}\circ u)=-{g^2\over N}(\bar{u}\circ \dd_z u)(\dd_{\bar{z}}\bar{u}\circ u)+\textrm{complete square}\,.
\eea
Therefore upon setting the fields $a, b, c$ equal to their stationary values, we get the Lagrangian in the form (\ref{LagrAfterElim}).

\section{The point splitting of L\"uscher}

In this section we recall the point splitting method for the non-local charge, introduced in the original paper~\cite{Luscher}.
In order to measure this splitting one needs to introduce a metric $g$ on the worldsheet $\Sigma_2$ (note that classically everything depended only on the conformal class $[g]$ of the metric). For simplicity we will now consider the situation when $\Sigma_2\simeq \mathds{R}^2$, equipped with a flat metric. As in the body of the paper, we will assume that the contour $\Gamma$ is infinite (a straight line, for example) and we will fix the reparametrization invariance on the worldline by choosing a gauge
\bea
(\dot{x}^\mu)^2=1\,.
\eea
Here $x^\mu: \Gamma\simeq \mathds{R}_t\to \Sigma_2$ is the parametrization of $\Gamma$ by a variable $t$ taking values from minus infinity to plus infinity. If $\Gamma$ was a contour of finite length, one could still parametrize it by a variable $t$ with infinite range, however one would need to choose a gauge $(\dot{x}^\mu)^2= f^2(t)$, so that the length $\int\limits_{-\infty}^{\infty} \,|f(t)|\,dt<\infty$ was finite.

\vspace{0.3cm}
The regularized non-local charge of~\cite{Luscher} has the form:
\bea\label{Q2reg}
Q_2^\epsilon=Z(\epsilon)\,\int\limits_{-\infty}^{\infty}\,dt\,\dot{x}^\mu\,K_\mu-{1\over 2}\,\int\limits_{-\infty}^{\infty}\,ds\,\int\limits_{-\infty}^{s-\epsilon} \,dt\,[(\ast K)_t, (\ast K)_s]\,,
\eea
where $\epsilon$ is now a positive real number. The factor $Z(\epsilon)$ has been inserted to cancel the divergence in the short-distance expansion of the two currents in the second term. Next we compute the variation of this charge under deformations of the contour $\delta x^\mu$ (sometimes we will denote this variation by $\delta_\Gamma$). Upon integrating by parts in the terms involving $\delta\dot{x}^\mu$, we obtain:
\newcommand*{\Scale}[2][4]{\scalebox{#1}{$#2$}}%
\newcommand*{\Resize}[2]{\resizebox{#1}{!}{$#2$}}%
\bear\nonumber 
&&\!\!\!\!\!\!\!\!\!\!\!\Scale[0.95]{\delta_\Gamma Q_2^\epsilon=Z(\epsilon)\,\int\limits_{-\infty}^{\infty}\,dt\,\delta x^\mu \dot{x}^\nu\,(\dd_\mu K_\nu-\dd_\nu K_\mu)\;-}\\ \nonumber \displaystyle
&&\!\!\!\!\!\!\!\!\!\!\!\Scale[0.95]{-{1\over 2}\,\int\limits_{-\infty}^{\infty}\,dt\,\delta x^\mu(t-\epsilon)\,\dot{x}^\alpha(t) [(\ast K)_\mu(t-\epsilon), (\ast K)_\alpha(t)]\;+}\\ \nonumber \displaystyle
&&\!\!\!\!\!\!\!\!\!\!\!\Scale[0.95]{+{1\over 2}\,\int\limits_{-\infty}^{\infty}\,dt\,\dot{x}^\mu(t) \delta x^\alpha(t+\epsilon) [(\ast K)_\mu(t), (\ast K)_\alpha(t+\epsilon)]\;+}\\ \nonumber \scriptstyle
&&\!\!\!\!\!\!\!\!\!\!\!\Scale[0.95]{+ {1\over 2}\,\int\limits_{-\infty}^{\infty}\,ds\,\int\limits_{-\infty}^{s-\epsilon} \,dt\, \left((\delta x^\mu(t) \dot{x}^\lambda(t)-\delta x^\lambda(t) \dot{x}^\mu(t)) [\dd_\lambda (\ast K)_\mu(t), (\ast K)_\alpha(s)]\dot{x}^{\alpha}(s)-(s\leftrightarrow t) \right)}\eear
The `bulk term' (the one in the last line) vanishes, since due to the antisymmetry of $\delta x^\mu(t) \dot{x}^\lambda(t)-\delta x^\lambda(t) \dot{x}^\mu(t)$ it turns out to be proportional to $\dd_\mu K_\mu=0$ (conservation of the current). The final result is therefore
\bear
&&\delta_\Gamma Q_2^\epsilon=\,\int\limits_{-\infty}^{\infty}\,dt\,\left[ Z(\epsilon)\,\delta x^\mu \dot{x}^\nu\,(\dd_\mu K_\nu-\dd_\nu K_\mu)+\right.\\ \nonumber && \label{deltaQ2}\left. +{1\over 2} \epsilonup_{\mu\nu}\epsilonup_{\alpha\beta}\,[K_\nu(t-\epsilon), K_\beta(t)]\,\left(\dot{x}^\mu(t-\epsilon) \delta x^\alpha(t) -\delta x^{\mu}(t-\epsilon) \dot{x}^\alpha(t)\right)\right]
\eear
Now in the last line we will use the OPE's of two currents obtained in Section~\ref{OPEsect1}. If, say, $\mu=z$ and $\alpha=\bar{z}$, one needs the OPE $[K_z(p+\epsilon), K_{\bar{z}}(p)]$, which, according to~(\ref{OPEleadfull}), is logarithmic in $\epsilon$. Therefore in this case in the bracket $\left(\dot{z}(t-\epsilon) \delta \bar{z}(t) -\delta z(t-\epsilon) \dot{\bar{z}}(t)\right)$ we can pass to the limit $\epsilon \to 0$. The logarithmic divergence itself is canceled by tuning the dependence on $\epsilon$ of the factor $Z(\epsilon)$.

\vspace{0.3cm}
If however $\mu=\alpha=z$, we have to consider the OPE $[K_{\bar{z}}(p+\epsilon), K_{\bar{z}}(p)]$, which according to~(\ref{antiKOPE}) has a pole term $\sim {1\over \epsilon}$ at the leading order. This term is multiplied by $\left(\dot{z}(t-\epsilon) \delta z(t) -\delta z(t-\epsilon) \dot{z}(t)\right)\sim \epsilon \left(\dot{\delta z}\,\dot{z}- \ddot{z}\, \delta z \right)$, producing an undesirable extra contribution in addition to (\ref{anomdef0}). In other words, if one assumes the definition~(\ref{Q2reg}), the anomaly is not given by a 2-form.

\section{The integral giving the anomaly at order $1\over \sqrt{N}$}\label{intsimpl}

In this Appendix we wish to consider the integral (\ref{anomint1}):
\bear\nonumber
&&\!\!\!\!\!\!\!\!\!\!I\!=\!\int\!\!\frac{d^2k}{(2\pi)^2}\!\left(\frac{k_{\bar{z}}(k_z+p_z)}{(k^2+m^2)((k+p)^2+m^2)}\,\left((2k_{\bar{z}}+p_{\bar{z}})\mathcal{A}_z(p)+(2k_{z}+p_{z})\mathcal{A}_{\bar{z}}(p)\right)\,-\right. \\  \nonumber  &&\left. \quad\quad\quad\quad\quad-\frac{k_{\bar{z}}}{k^2+m^2}\,\mathcal{A}_z(p)-\frac{k_z+p_z}{(p+k)^2+m^2}\,\mathcal{A}_{\bar{z}}(p) \right)\,e^{ik\epsilon}
\eear
and to bring it to the form (\ref{anomint2}). We start by rewriting the factors in the numerators of the first two summands as follows:
\bear\label{momdec1}
&&k_{\bar{z}}(k_z+p_z)={1\over 2}((k+p)^2+m^2)-p_{\bar{z}}(k_z+p_z)-{1\over 2}m^2\\ \label{momdec2}
&&k_{\bar{z}}(k_z+p_z)={1\over 2}(k^2+m^2)+k_{\bar{z}}p_z-{1\over 2}m^2
\eear
The first terms here allow cancelling one of the propagators each. The remaining terms are multiplied  by a factor $e^{ik\epsilon}\left(2k_\mu+p_\mu\over (k^2+m^2)((k+p)^2+m^2)\right)$, which in the limit $\epsilon \to 0$ is odd under the interchange $k\to-k-p$. As a result, we are free to add to (\ref{momdec1})-(\ref{momdec2}) arbitrary terms that are independent of the momentum $k$, since such terms will vanish in the integral in the limit $\epsilon \to 0$ due to anti-symmetry. We use this arbitrariness to bring the expressions (\ref{momdec1})-(\ref{momdec2}) to the form
\bear
&&k_{\bar{z}}(k_z+p_z)\to {1\over 2}((k+p)^2+m^2)-{1\over 2}p_{\bar{z}}(2k_z+p_z)\\
&&k_{\bar{z}}(k_z+p_z)\to{1\over 2}(k^2+m^2)+{1\over 2}(2k_{\bar{z}}+p_{\bar{z}})p_z\,.
\eear
When substituted into the integral, this gives
\bear\nonumber
&&I=\int\,\frac{d^2k}{(2\pi)^2}\,\left[p_z \mathcal{A}_{\bar{z}}\left({1\over 2}\frac{(2k_z+p_z)(2k_{\bar{z}}+p_{\bar{z}})}{(k^2+m^2)((k+p)^2+m^2)}-{1\over 2}{1\over (k+p)^2+m^2}\right) - \right. \\  \nonumber
&& \left. \quad\quad - p_{\bar{z}}\mathcal{A}_z\left({1\over 2}\frac{(2k_z+p_z)(2k_{\bar{z}}+p_{\bar{z}})}{(k^2+m^2)((k+p)^2+m^2)}-{1\over 2}{1\over k^2+m^2}\right) \right]\times e^{ik\epsilon}+\ldots
\eear
Again, up to terms that vanish in the limit $\epsilon\to 0$ (which arise due to the change of variables $k\to -k-p$ in the first bracket), we obtain:
\bea\nonumber
I\!=\!(p_z \mathcal{A}_{\bar{z}}-p_{\bar{z}}\mathcal{A}_z)\int\,\frac{d^2k}{(2\pi)^2}\,\left({1\over 2}\frac{(2k_z+p_z)(2k_{\bar{z}}+p_{\bar{z}})}{(k^2+m^2)((k+p)^2+m^2)}-{1\over 2}{1\over k^2+m^2}\right)\times e^{ik\epsilon}+\ldots
\eea
We now wish to relate this integral to the $\Gamma_{z\bar{z}}$ function entering the effective action~(\ref{effact}). The analogous function $\Gamma_{z\bar{z}}^{(M^2)}$ in the presence of a Pauli-Villars regulator of mass $M$ can be written as a difference of two convergent terms:
\bea
\Gamma_{z\bar{z}}^{(M^2)}=\int\,\frac{d^2k}{(2\pi)^2}\,\left({1\over k^2+m^2}-{1\over 2}\frac{(2k_z+p_z)(2k_{\bar{z}}+p_{\bar{z}})}{(k^2+m^2)((k+p)^2+m^2)}\right)-\left(m^2\to m^2+M^2\right)
\eea
The limit of the last  term $\left(m^2\to m^2+M^2\right)$ as $M\to \infty$ is $1\over 4\pi$. Therefore we can rewrite $I$ as follows:
\bea
I=(p_{\bar{z}}\mathcal{A}_z-p_z \mathcal{A}_{\bar{z}})\,\left(\Gamma_{z\bar{z}}+{1\over 4\pi}\right)-{1\over 2}(p_{\bar{z}}\mathcal{A}_z-p_z \mathcal{A}_{\bar{z}})\,D(\epsilon)+\ldots\,,
\eea
which is the expression (\ref{anomint2}).

\section{One more integral}\label{onemore}

Here we will prove the relation 
\bea\label{taugammarel}
\tau_{zz}(p|\epsilon)=-2\Gamma_{zz}-{1\over 2\pi}\,\frac{\bar{\epsilon}}{\epsilon}+\ldots\quad\quad\textrm{for}\quad\quad \epsilon\to 0
\eea
between the $\Gamma_{zz}$-component of the effective action  and the integral
\bea
\tau_{zz}(p|\epsilon):=\int\,\frac{d^2 k}{(2\pi)^2} \,e^{i(k\cdot\epsilon)}\,\frac{(2k+p)_z\,(2k+p)_z }{((k+p)^2+m^2)(k^2+m^2)}\,,
\eea
which we encountered in Section~\ref{ord1Nsect}. First, we formally introduce a factor $e^{ik\epsilon}$ in the definition of $\Gamma_{zz}$, which allows us to write:
\bea
\Gamma_{zz}=\underset{\epsilon\to 0}{\textrm{lim}} \left(-{1\over 2}\tau_{zz}(p|\epsilon)+\underbracket[0.6pt][0.6ex]{\int\,\frac{d^2k}{(2\pi)^2}\,\frac{2 k_z^2\, e^{i(k\cdot \epsilon)}}{(k^2+m^2)^2}}_{:=\mathcal{I}}\right)
\eea
Therefore we need to construct the asymptotics of the second (simpler) integral $\mathcal{I}$ as $\epsilon\to 0$. We recall that, by definition, $(k\cdot \epsilon)=k_z \epsilon+k_{\bar{z}} \bar{\epsilon}$. As a consequence of rotational symmetry $\epsilon\to e^{i\gamma} \epsilon$, the integral is equal to
\bea
\mathcal{I}=\frac{\bar{\epsilon}}{\epsilon}\,f(|\epsilon|)\,.
\eea
To calculate $f$, we may set $\epsilon$ to be real and positive. We pass to polar coordinates setting $k_z={\rho \over \sqrt{2}} e^{i\psi}$, then $f=\int\limits_{0}^\infty\,\frac{\rho d\rho}{2\pi}\,\frac{\rho^2}{(\rho^2+m^2)^2}\,\int\limits_{0}^{2\pi}\,\frac{d\psi}{2\pi}\,e^{i(2\psi+\rho\epsilon \cos(\psi))}=-\int\limits_{0}^\infty\,\frac{\rho d\rho}{2\pi}\,\frac{\rho^2\,J_2(\rho \epsilon)}{(\rho^2+m^2)^2}$. Rescaling $\rho\to {\rho\over \epsilon}$ and passing to the limit $\epsilon\to 0$ (which does not introduce divergences in the integral, thanks to the properties of the Bessel function at $0$ and $\infty$), we get $\underset{\epsilon\to 0}{\textrm{lim}} \,f(\epsilon)=-\int\limits_{0}^\infty\,\frac{ d\rho}{2\pi \rho}\,J_2(\rho)=-{1\over 4\pi}$. This leads to the relation~(\ref{taugammarel}).

\end{document}